\RequirePackage{fix-cm}
\documentclass[smallextended]{svjour3}       
\smartqed  
\usepackage{graphicx}
\usepackage{multirow}
\usepackage[cmex10]{amsmath}
\usepackage{color}
\begin{document}

\title{Fuzzy Overlapping Community Quality Metrics
}


\author{Mingming Chen  \and
        Boleslaw K. Szymanski 
}


\institute{Mingming Chen \at
              Department of Computer Science \\
              Rensselaer Polytechnic Institute \\
              110 8th Street, Troy, NY 12180 \\
              \email{chenm8@rpi.edu}           
           \and
           Boleslaw K. Szymanski \at
              Department of Computer Science \\
              Rensselaer Polytechnic Institute \\
              110 8th Street, Troy, NY 12180 \\
              \email{szymab@rpi.edu}
}

\date{Received: date / Accepted: date}

\maketitle

\begin{abstract}
Modularity is widely used to effectively measure the strength of the disjoint community structure found by community detection algorithms. Several overlapping extensions of modularity were proposed to measure the quality of overlapping community structure. However, all these extensions differ just in the way they define the belonging coefficient and belonging function. Yet, there is lack of systematic comparison of different extensions. To fill this gap, we overview overlapping extensions of modularity and generalize them with a uniform definition enabling application of different belonging coefficients and belonging functions to select the best. In addition, we extend localized modularity, modularity density, and eight local community quality metrics to enable their usages for overlapping communities. The experimental results on a large number of real networks and synthetic networks using overlapping extensions of modularity, overlapping modularity density, and local metrics show that the best results are obtained when the product of the belonging coefficients of two nodes is used as the belonging function. Moreover, the results may be used to guide researchers on which metrics to adopt when measuring the quality of overlapping community structure.
\end{abstract}

\section{Introduction}
Many networks, including Internet, citation networks, transportation networks, email networks, and social and biochemical networks, display community structure which identifies groups of nodes within which connections are denser than between them \cite{UWDModularity}. Detecting and characterizing such community structure, which is known as community detection, is one of the fundamental issues in network science. Community detection has been shown to reveal latent yet meaningful structure in networks such as groups in online and contact-based social networks, functional modules in protein-protein interaction networks, groups of customers with similar interests in online retailer user networks, groups of scientists in interdisciplinary collaboration networks, etc. \cite{CommunityReport}.

In the last decade, the most popular community detection method, proposed by Newman \cite{NewmanGreedy}, has been to maximize the quality metric known as modularity \cite{UWDModularity,PNASModularity} over all possible partitions of a network. This metric measures the difference between the fraction of all edges that are within the actual community and such a fraction of edges that would be inside the community in a randomized graph with the same number of nodes and the same degree sequence. It is widely used to measure the strength of community structures discovered by community detection algorithms.

Newman's modularity can only be used to measure the quality of disjoint communities. However, it is more realistic to expect that nodes in real networks belong to more than one community, resulting in overlapping communities \cite{Overlapping_survey}. Therefore, several overlapping extensions of modularity (\cite{fuzzy_cmeans_Qov,fuzzy_opt_Qov,Qov_num_coms,Qov_max_clique,Qov_node_strength,Qov_edge_degree,OverlappingQ_survey}) were proposed to measure the quality of overlapping community structure. Yet, to date no attempt has been made to systematically compare different overlapping extensions and propose metric selection criteria for different types of networks. In this paper, we consider several overlapping extensions of modularity and test their quality on real and synthetic networks. We also extend localized modularity \cite{NeighboringModularity}, modularity density \cite{QdsConference,QdsJournal}, and eight local community quality metrics for overlapping communities following the same principles used by the overlapping extensions of modularity.

We conducted experiments on a large number of real-world networks and synthetic networks using overlapping extensions of modularity, overlapping modularity density, and eight local metrics (the number of Intra-edges, Intra-density, Contraction, the number of Boundary-edges, Expansion, Conductance \cite{QdsConference,QdsJournal}, the Fitness function \cite{Fitness}, and the Average Modularity Degree \cite{AverageModularityDegree}). The results show that selecting the product of the belonging coefficients of two nodes as a belonging function for overlapping extensions yields better results on these networks than using other belonging functions. The experimental results also give a guidance to researchers on which metrics to choose when measuring the quality of overlapping community structure.

\textbf{Methodology}: Below we introduce in steps the methodology we use to evaluate overlapping extensions of modularity:
\vspace{-0.5em}
\begin{enumerate}
\item [(1)] We first give a generalized definition for existing overlapping extensions of modularity which covers four such extensions, each using one of the two different versions of belonging coefficient and one of the two different versions of belonging function.
\item [(2)] Next, we extend localized modularity, modularity density, and eight local community quality metrics to be applicable to overlapping community structures following the same principle as the overlapping extensions of modularity do.
\item [(3)] Then, for the generalized definitions of these metrics, we first determine which version of the belonging coefficient and which version of the belonging function perform best for each metric.
\item [(4)] Moreover, we determine which version of the belonging coefficient and which version of the belonging function scores the largest number of quality metrics consistent with each other on determining the best values of parameters of compared community detection algorithms: the threshold $r$ for SLPA \cite{SLPA2012}, the parameter $k$ for CFinder \cite{CPM}, and the  threshold $tr$ for SpeakEasy \cite{SpeakEasy} on the real and synthetic networks.
\item [(5)] Finally, we compare the performance of the overlapping metrics with the best combination of belonging coefficient and belonging function by looking at how many times each metric is among those that are consistent with each other on determining the best values of parameters for the same algorithms as used in step (4).
\end{enumerate}

This work is an extension of our previous paper \cite{QdsOv} which considered four real networks and ten community quality metrics for comparison. Also, it only adopted one overlapping community detection algorithm, SLPA \cite{SLPA2012}, as evaluation method. However, in this paper, we consider totally 23 real network datasets, including friendship network, collaboration networks, co-purchasing networks, biology networks, etc. Table~\ref{tab:real_networks} shows the basic properties of all these datasets. These networks have different numbers of nodes and different numbers of edges, varying from very small to very large networks. Edges in some networks have weights and directions. Besides real networks, we also consider LFR benchmark networks \cite{LFROv}, each of which is instantiated with a wide range of parameters in the experiments. Moreover, we consider two more community quality metrics, the Fitness function \cite{Fitness} and Average Modularity Degree \cite{AverageModularityDegree}, totally twelve metrics for comparison. In addition, we adopt three overlapping community detection algorithms, SLPA \cite{SLPA2012}, CFinder \cite{CPM}, and SpeakEasy \cite{SpeakEasy}, as testing methods.

\section{Modularity}
\label{sec:related_work}

\subsection{Newman's Modularity}
\label{subsec:modularity}
Newman's modularity \cite{UWDModularity,PNASModularity} for unweighted and undirected networks is defined as the difference between the fractions of the actual and expected (in a randomized graph with the same number of nodes and the same degree sequence) number of edges within the community. A larger value of modularity means a stronger community structure. For a given community partition of a network $G=(V,E)$ with $|E|$ edges, modularity ($Q$) \cite{UWDModularity} is given by:
\begin{equation}
\label{eq:uwdmodularity}
Q=\sum_{c \in C} \left[\frac{|E_{c}^{in}|}{|E|}-\left(\frac{2|E_{c}^{in}|+|E_{c}^{out}|}{2|E|}\right)^2\right],
\end{equation}
where $C$ is the set of all the communities, $c$ is a specific community in $C$, $|E_{c}^{in}|$ is the number of edges between nodes within community $c$, and $|E_{c}^{out}|$ is the number of edges from the nodes in community $c$ to the nodes outside $c$.

Modularity can also be expressed as \cite{PNASModularity}:
\begin{equation}
\label{eq:spectral_Q}
Q=\frac{1}{2|E|} \sum\limits_{ij} \left[A_{ij} - \frac{k_i k_j}{2|E|}\right] \delta_{{c_i}, {c_j}},
\end{equation}
where $k_i$ is the degree of node $i$, $A_{ij}$ is an element of the adjacency matrix between node $i$ and node $j$, $\delta_{{c_i}, {c_j}}$ is the Kronecker delta symbol, and $c_i$ is the label of the community to which node $i$ is assigned.

\subsection{Overlapping Definition of Modularity}
\label{sec:ov_Q}
Newman's modularity is used to measure the quality of disjoint community structure of a network. However, it is more realistic that nodes in networks belong to more than one community, resulting in overlapping communities \cite{Overlapping_survey}. For instance, a researcher may be active in several research areas, and a node in biological networks might have multiple functions. It is also quite common that people in social networks are naturally characterized by multiple community memberships depending on their families, friends, professional colleagues, neighbors, etc. For this reason, discovering overlapping communities became very popular in the last few years. Several overlapping extensions of modularity \cite{fuzzy_cmeans_Qov,fuzzy_opt_Qov,Qov_num_coms,Qov_max_clique,Qov_node_strength,Qov_edge_degree,OverlappingQ_survey} were proposed to measure the quality of overlapping community structure. These extensions are described below.

If communities overlap, each node can belong to multiple communities, although the strength of this connection can generally be different for different communities. Given a set of overlapping communities $C=\{c_1, c_2, ..., c, ..., c_{|C|}\}$ in which a node may belong to more than one of them, a vector of \textit{belonging coefficients} $(a_{i,c_1}, a_{i,c_2}, ..., a_{i,c}, ..., a_{i,c_{|C|}})$ \cite{fuzzy_opt_Qov,Qov_edge_degree} can be assigned to each node $i$ in the network. $|C|$ is the number of communities. The belonging coefficient $a_{i,c}$ measures the strength of association between node $i$ and community $c$. Without loss of generality, the following constraints are assumed to hold:
\begin{equation}
\begin{split}
&0 \le a_{i,c} \le 1~~\forall i \in V, \forall c \in C \\
&~~~~~~~~~~~~~~~~\text{and} \\
&~~~~~~~~\sum_{c \in C} a_{i,c} = 1.
\end{split}
\end{equation}
Therefore, the belonging strength is measured as a real value in the range of $[0 ,1]$ and the sum of all belonging coefficients, which is 1, is the same for all nodes in the network.

Zhang et al. \cite{fuzzy_cmeans_Qov} proposed an extended modularity which uses the average of the belonging coefficients of two nodes as belonging function to measure the quality of overlapping community structure:
\begin{equation}
\label{eq:Qov_Z}
Q_{ov}^Z=\sum_{c \in C}\left[ \frac{|E_{c}^{in}|}{|E|} - \left( \frac{2|E_{c}^{in}|+|E_{c}^{out}|}{2|E|} \right)^2 \right],
\end{equation}
where $|E_{c}^{in}|=\frac{1}{2}\sum_{i,j \in c}\frac{a_{i,c}+a_{j,c}}{2}A_{ij}$, $|E_{c}^{out}|=\sum_{i \in c, j \in V-c}\frac{a_{i,c}+(1-a_{j,c})}{2}A_{ij}$, and $|E|=\frac{1}{2}\sum_{ij}A_{ij}$. For the case of disjoint communities, $Q_{ov}^Z$ reduces exactly to Newman's modularity ($Q$) given by Equation~(\ref{eq:uwdmodularity}).

Nepusz et al. \cite{fuzzy_opt_Qov} considered the belonging coefficient $a_{i,c}$ as the probability that node $i$ is active in community $c$. Then, the probability that node $i$ is active in the same communities as node $j$ is the dot product of their membership vectors, denoted as $s_{ij}$:
\begin{equation}
s_{ij}=\sum_{c \in C} a_{i,c}a_{j,c}.
\end{equation}
The authors also adopted $s_{ij}$ as the similarity measure between nodes $i$ and $j$. By replacing $\delta_{c_i,c_j}$ in Equation~(\ref{eq:spectral_Q}) with the similarity measure $s_{ij}$ defined above, they proposed a fuzzified variant of modularity:
\begin{equation}
\label{eq:Qov_F}
\begin{split}
Q_{ov}^{F}&=\frac{1}{2|E|} \sum\limits_{ij} \left[A_{ij} - \frac{k_i k_j}{2|E|}\right] s_{ij} \\
&=\frac{1}{2|E|} \sum_{c \in C} \sum\limits_{i,j \in c} \left[A_{ij} - \frac{k_i k_j}{2|E|}\right]a_{i,c}a_{j,c}.
\end{split}
\end{equation}
In case communities are disjoint, there exists only one community $c$ for every node $i$ for which $a_{i,c}=1$. Then, the fuzzified modularity ($Q_{ov}^{F}$) reduces to exactly the original modularity ($Q$) described in Equation~(\ref{eq:spectral_Q}).

Shen et al. \cite{Qov_num_coms} proposed an extension of modularity for overlapping community structure using Equation~(\ref{eq:Qov_F}) but defined the belonging coefficients of node $i$ to be the reciprocal of the number of communities to which it belongs:
\begin{equation}
\label{eq:num_com_factor}
a_{i,c}=\frac{1}{O_i},
\end{equation}
where $O_i$ is the number of communities containing node $i$. Then, the extended modularity for overlapping community structure is given by:
\begin{equation}
\label{eq:Qov_E}
\begin{split}
Q_{ov}^{E}&=\frac{1}{2|E|} \sum_{c \in C} \sum\limits_{i,j \in c} \left[A_{ij} - \frac{k_i k_j}{2|E|}\right]a_{i,c}a_{j,c} \\
&=\frac{1}{2|E|} \sum_{c \in C} \sum\limits_{i,j \in c} \left[A_{ij} - \frac{k_i k_j}{2|E|}\right]\frac{1}{O_iO_j}.
\end{split}
\end{equation}
For disjoint community structure, $Q_{ov}^E$ reduces to the original modularity ($Q$) described in Equation~(\ref{eq:spectral_Q}).

Shen et al. \cite{Qov_max_clique} proposed another extension of modularity for overlapping communities also using Equation~(\ref{eq:Qov_F}). In this case, the coefficient of node $i$ belonging to community $c$ is defined as:
\begin{equation}
\label{eq:max_clique_factor}
a_{i,c}=\frac{1}{a_i}\sum_{k\in c} \frac{M_{ik}^c}{M_{ik}} A_{ik},
\end{equation}
where $M_{ik}$ denotes the number of maximal cliques in the network containing edge $(i,k)$, $M_{ik}^c$ is the number of maximal cliques in community $c$ that contains edge $(i,k)$, and $a_{i}$ is a normalization term defined as:
\begin{equation}
a_{i}=\sum_{c \in C} \sum_{k \in c} \frac{M_{ik}^c}{M_{ik}} A_{ik}.
\end{equation}
The maximal clique is a clique that is not a subset of any other cliques. Then, the extended modularity for overlapping community structure is given by:
\begin{equation}
\label{eq:Qov_C}
Q_{ov}^{C}=\frac{1}{2|E|} \sum_{c \in C} \sum\limits_{i,j \in c} \left[A_{ij} - \frac{k_i k_j}{2|E|}\right]a_{i,c}a_{j,c}.
\end{equation}
Note that for disjoint communities, this new extension also reduces to Newman's modularity shown in Equation~(\ref{eq:spectral_Q}).

Chen et al. \cite{Qov_node_strength} also proposed another extension of modularity with the same Equation~(\ref{eq:Qov_F}) but with the belonging coefficient defined as:
\begin{equation}
\label{eq:node_strength_factor}
a_{i,c}=\frac{\sum_{k \in c} A_{ik}}{\sum_{c' \in C_i} \sum_{k \in c'} A_{ik}},
\end{equation}
where $C_i$ is the set of communities to which node $i$ belongs. It measures how tightly node $i$ connects to community $c$. Consequently, the extended definition of modularity for overlapping community structure is given by:
\begin{equation}
\label{eq:Qov_O}
\begin{split}
&Q_{ov}^{O}=\frac{1}{2|E|} \sum_{c \in C} \sum\limits_{i,j \in c} \left[A_{ij} - \frac{k_i k_j}{2|E|}\right]a_{i,c}a_{j,c} \\
&=\frac{1}{2|E|} \sum_{c \in C} \sum\limits_{i,j \in c} \left[A_{ij} - \frac{k_i k_j}{2|E|}\right]\frac{\sum\limits_{k \in c} A_{ik}}{\sum\limits_{c' \in C_i} \sum\limits_{k \in c'} A_{ik}} \frac{\sum\limits_{k \in c} A_{jk}}{\sum\limits_{c' \in C_j} \sum\limits_{k \in c'} A_{jk}}.
\end{split}
\end{equation}
Still, for disjoint community structure, $Q_{ov}^{O}$ reduces to the original modularity given by Equation~(\ref{eq:spectral_Q}).

Unlike the node-based extensions of modularity presented above, Nicosia et al. \cite{Qov_edge_degree} proposed an edge-based extension of modularity for overlapping communities. In this case, the belonging coefficients represent how edges are assigned to communities. The coefficient for edge $l=(i,j)$ belonging to community $c$ is $\beta_{l(i,j),c}=F(a_{i,c}, a_{j,c})$, where $F(a_{i,c}, a_{j,c})$ could be any function (product, average, or maximum) of $a_{i,c}$ and $a_{j,c}$. After trying several different functions, the authors stated that the best $F$ is a two-dimensional logistic function:
\begin{equation}
\label{eq:Qov_L_bf}
F(a_{i,c}, a_{j,c}) = \frac{1}{(1+e^{-f(a_{i,c})})(1+e^{-f(a_{j,c})})},
\end{equation}
where $f(a_{i,c})$ is a simple linear scaling function $f(x)=2px-p, p \in R$. In papers \cite{Overlapping_survey,ov_lpa}, $p$ was selected to be $30$. Then, the expected belonging coefficient of any edge $l=(i,k)$ starting from node $i$ in community $c$ is given by $\beta_{l(i,k),c}^e=\frac{1}{|V|}\sum_{k \in V}\beta_{l(i,k),c}$ running over all nodes in the network. Accordingly, the expected belonging coefficient of any edge $l=(k,j)$ pointing to node $j$ in community $c$ is defined as $\beta_{l(k,j),c}^e=\frac{1}{|V|}\sum_{k \in V}\beta_{l(k,j),c}$. Then, the edge-based extension of modularity is given by:
\begin{equation}
\label{eq:Qov_edge}
\begin{split}
Q_{ov}^L&=\frac{1}{2|E|} \sum_{c \in C} \sum\limits_{i,j \in c} \left[r_{ijc}A_{ij} - s_{ijc}\frac{k_i k_j}{2|E|}\right] \\
&=\frac{1}{2|E|} \sum_{c \in C} \sum\limits_{i,j \in c} \left[\beta_{l(i,j),c}A_{ij} - \frac{\beta_{l(i,k),c}^e k_i \beta_{l(k,j),c}^e k_j}{2|E|}\right],
\end{split}
\end{equation}
where
\begin{equation}
r_{ijc}=\beta_{l(i,j),c}=F(a_{i,c}, a_{j,c})
\end{equation}
and
\begin{equation}
\begin{split}
s_{ijc}&=\beta_{l(i,k),c}^e \beta_{l(k,j),c}^e \\
&=\frac{\sum_{k \in V}\beta_{l(i,k),c}\sum_{k \in V}\beta_{l(k,j),c}}{|V|^2} \\
&=\frac{\sum_{k \in V}F(a_{i,c}, a_{k,c})\sum_{k \in V}F(a_{k,c}, a_{j,c})}{|V|^2}.
\end{split}
\end{equation}
In $Q_{ov}^L$, $r_{ijc}$ is used as the weight corresponding to the probability of the observed edge $l=(i,j)$, while $s_{ijc}$ is used as the weight of the probability of an edge from node $i$ to node $j$ in the null model. Note that although for disjoint communities $F(a_{i,c}, a_{j,c})$ is practically 1 when both $a_{i,c}$ and $a_{j,c}$ are equal to 1, $Q_{ov}^L$ does not exactly reduce to the original modularity given by Equation~(\ref{eq:spectral_Q}).

Generally, there are two categories of overlapping community structures: crisp (non-fuzzy) overlapping and fuzzy overlapping \cite{OverlappingQ_survey}. For crisp overlapping community structure, each node belongs to one or more communities but without the corresponding belonging coefficients. That is, the relationship between a node and a community is binary: a node either belongs to a community or it does not. For fuzzy overlapping community structure, each node can be a member of multiple communities, but in general the values of belonging coefficients are different. Fuzzy overlapping can be easily transformed to crisp overlapping with a threshold parameter. Namely, if the belonging coefficient of node $i$ to community $c$ is larger than the value of the threshold, then node $i$ stays in community $c$. Otherwise, node $i$ is deleted from community $c$. Crisp overlapping can be converted to fuzzy overlapping by calculating the value of the belonging coefficient using Equations~(\ref{eq:num_com_factor}), (\ref{eq:max_clique_factor}), or (\ref{eq:node_strength_factor}). However, calculating the belonging coefficient using Equation~(\ref{eq:max_clique_factor}) is computationally expensive since it needs to find all the maximal cliques of the network first. Hence, in this paper we only consider Equation~(\ref{eq:num_com_factor}) and Equation~(\ref{eq:node_strength_factor}) when converting crisp overlapping to fuzzy overlapping.

Now, we give two general definitions, $Q_{ov}$ and $Q_{ov}'$, for node-based extensions of modularity. First, $Q_{ov}$ is given by:
\begin{equation}
\label{eq:Qov}
Q_{ov}=\sum_{c \in C}\left[ \frac{|E_{c}^{in}|}{|E|} - \left( \frac{2|E_{c}^{in}|+|E_{c}^{out}|}{2|E|} \right)^2 \right],
\end{equation}
where $|E_{c}^{in}|=\frac{1}{2}\sum_{i,j \in c} f(a_{i,c}, a_{j,c})A_{ij}$, $|E_{c}^{out}|=\sum_{i \in c} \sum_{\substack{c' \in C \\ c' \ne c \\ j \in c' }}f(a_{i,c}, a_{j,c'})A_{ij}$, and $|E|=\frac{1}{2}\sum_{ij}A_{ij}$. The belonging function $f(a_{i,c}, a_{j,c})$ can be the average or product of $a_{i,c}$ and $a_{j,c}$. That is, $f(a_{i,c}, a_{j,c})=\frac{a_{i,c}+a_{j,c}}{2}$ or $f(a_{i,c}, a_{j,c})=a_{i,c}a_{j,c}$. Clearly, $Q_{ov}$ with $f(a_{i,c}, a_{j,c})=\frac{a_{i,c}+a_{j,c}}{2}$ is very similar to $Q_{ov}^Z$ in Equation~(\ref{eq:Qov_Z}). Second, $Q_{ov}'$ is given by:
\begin{equation}
Q_{ov}'=\frac{1}{2|E|} \sum_{c \in C} \sum\limits_{i,j \in c} \left[A_{ij} - \frac{k_i k_j}{2|E|}\right] f(a_{i,c}, a_{j,c}).
\end{equation}
where $f(a_{i,c}, a_{j,c})$ is the same as that in Equation~(\ref{eq:Qov}). It is worth noting that $Q_{ov}'$ with the belonging function $f(a_{i,c}, a_{j,c})=a_{i,c}a_{j,c}$ is actually the same as $Q_{ov}^{F}$ in Equation~(\ref{eq:Qov_F}), $Q_{ov}^{E}$ in Equation~(\ref{eq:Qov_E}), $Q_{ov}^{C}$ in Equation~(\ref{eq:Qov_C}), and $Q_{ov}^{O}$ in Equation~(\ref{eq:Qov_O}). The only difference between these formulas is how the value of $a_{i,c}$ is calculated.

It is easy to prove that $Q_{ov}$ is equivalent to $Q_{ov}'$ when $f(a_{i,c}, a_{j,c})=a_{i,c}a_{j,c}$. From the definition of $Q_{ov}$, we know that
$|E_{c}^{in}|=\frac{1}{2}\sum_{i,j \in c} a_{i,c}a_{j,c}A_{ij}$ which is in fact the same as the first term of $Q_{ov}'$. Moreover, it is easy to show that $\left( 2|E_{c}^{in}|+|E_{c}^{out}|\right)^2=\sum_{i,j \in c} k_ik_ja_{i,c}a_{j,c}$. Hence, the second term of $Q_{ov}$ is the same as the second term of $Q_{ov}'$. Similarly, it can be shown that $Q_{ov}$ is not equal to $Q_{ov}'$ when $f(a_{i,c}, a_{j,c})=\frac{a_{i,c}+a_{j,c}}{2}$.

\subsection{Localized Modularity Based on Community's Neighborhood}
Newman's modularity is a global measure which assumes that all pairs of nodes have equal probability to connect with each other, which reflects the connectivity among all communities. However, Muff et al. \cite{NeighboringModularity} argued that in many complex networks most communities are connected to only a small fraction of remaining communities, called local cluster connectivity property. Thus, they modified the definition of Newman's modularity by taking into account local cluster connectivity only to overcome global network dependency. The resulting measure is called localized modularity. We denote the localized modularity here as \textit{NQ} because it is based on the neighborhood of a community. \textit{NQ} for unweighted and undirected  networks is given by:
\begin{equation}
\label{eq:NQ}
NQ=\sum_{c \in C} \left[\frac{|E_{c}^{in}|}{|E_c^{neighb}|}-\left(\frac{2|E_{c}^{in}|+|E_{c}^{out}|}{2|E_c^{neighb}|}\right)^2\right],
\end{equation}
where $|E_c^{neighb}|$ is the total number of edges in the subnetwork containing the community $c$ and all its neighboring communities, i.e. the neighborhood of community $c$. It means that the contribution of each community $c$ to \textit{NQ} is calculated based on the neighborhood of $c$.

Unlike the traditional modularity ($Q$), the localized version of modularity ($NQ$) is not bounded above by 1. The more locally connected communities a network has, the bigger its $NQ$ can grow. In a network where all communities are connected to each other, $NQ$ yields the same value as $Q$. $NQ$ considers individual communities and their neighbors, and therefore provides a measure of community quality that is not dependent on other parts of the network.

Similar to the general node-based overlapping definition of modularity $Q_{ov}$ in Equation~(\ref{eq:Qov}), we extend $NQ$ for overlapping community structure as $NQ_{ov}$. The formula for $NQ_{ov}$ is exactly the same with $NQ$ in Equation~(\ref{eq:NQ}), while the difference is that $|E_{c}^{in}|$, $|E_{c}^{out}|$, and $|E_c^{neighb}|$ in $NQ_{ov}$ should consider the belonging coefficients of nodes and also the belonging function between pairs of nodes. For disjoint communities, $NQ_{ov}$ reduces exactly to $NQ$.

\section{Modularity Density}
\subsection{Modularity Density for Disjoint Communities}
Chen et al. \cite{QdsConference,QdsJournal,fine_tuned} proposed modularity density which simultaneously addresses two opposite yet coexisting problems of Newman's modularity: in some cases, it tends to favor small communities over large ones while in others, large communities over small ones. The latter tendency is known in the literature as the resolution limit problem of modularity \cite{ResolutionLimit}. Modularity density mixes two additional components, split penalty and the community density, into Newman's modularity given in Equation~(\ref{eq:uwdmodularity}). Split penalty is the fraction of edges that connect nodes of different communities. Community density includes internal community density and pair-wise community density. The definition of modularity density ($Q_{ds}$) for unweighted and undirected networks is given by:
\begin{equation}
\label{eq:Qds}
\begin{split}
& Q_{ds}=\sum_{c \in C} \left[\frac{|E_{c}^{in}|}{|E|}d_{c}-\left(\frac{2|E_{c}^{in}|+|E_{c}^{out}|}{2|E|}d_{c}\right)^2-\sum_{\substack{c' \in C \\ c' \ne c}}\frac{|E_{c,c'}|}{2|E|}d_{c,c'} \right], \\
&~~~~d_{c}=\frac{2|E_{c}^{in}|}{|c|(|c|-1)}, \\
&~~~~d_{c,c'}=\frac{|E_{c,c'}|}{|c||c'|},
\end{split}
\end{equation}
where $d_{c}$ is the internal density of community $c$, and $d_{c,c'}$ is the pair-wise density between community $c$ and community $c'$.

\subsection{Modularity Density for Overlapping Communities}
According to $Q_{ov}$ in Equation~(\ref{eq:Qov}), we extend $Q_{ds}$ for overlapping community structure as:
\begin{equation}
\label{eq:Qov_ds}
\begin{split}
&Q_{ds}^{ov}=\sum_{c \in C} \left[\frac{|E_{c}^{in}|}{|E|}d_{c}-\left(\frac{2|E_{c}^{in}|+|E_{c}^{out}|}{2|E|}d_{c}\right)^2-\sum_{\substack{c' \in C \\ c' \ne c}}\frac{|E_{c,c'}|}{2|E|}d_{c,c'}\right], \\
&~~~~d_c = \frac{2|E_{c}^{in}|}{\sum_{i,j \in c, i \ne j} f(a_{i,c}, a_{j,c})}, \\
&~~~~d_{c,c'}=\frac{|E_{c,c'}|}{\sum_{i \in c, j \in c'} f(a_{i,c}, a_{j,c'})},
\end{split}
\end{equation}
where $|E_{c}^{in}|=\frac{1}{2}\sum_{i,j \in c} f(a_{i,c}, a_{j,c})A_{ij}$, $|E_{c}^{out}|=\sum\limits_{i \in c} \sum\limits_{\substack{c' \in C \\ c' \ne c \\ j \in c' }}f(a_{i,c}, a_{j,c'})A_{ij}$, $|E_{c,c'}|=\sum_{i \in c, j \in c'} f(a_{i,c}, a_{j,c'})A_{ij}$, and $|E|=\frac{1}{2}\sum\limits_{ij}A_{ij}$. The belonging function $f(a_{i,c}, a_{j,c})$ can be the product or average of $a_{i,c}$ and $a_{j,c}$. For disjoint communities, $Q_{ds}^{ov}$ reduces exactly to $Q_{ds}$ given by Equation~(\ref{eq:Qds}). Notice that we do not extend modularity density based on $Q_{ov}^L$ since it is too complicated and far from intuitive.

\section{Evaluation and Analysis}
\label{sec:evaluation}

From Subsection~\ref{sec:ov_Q}, we know that all node-based overlapping extensions of modularity can be expressed with $Q_{ov}$ in Equation~(\ref{eq:Qov}) using the belonging function $f(a_{i,c}, a_{j,c})=\frac{a_{i,c}+a_{j,c}}{2}$ or $f(a_{i,c}, a_{j,c})=a_{i,c}a_{j,c}$. For the edge-based overlapping extension of modularity ($Q_{ov}^L$), the belonging function is given by Equation~(\ref{eq:Qov_L_bf}). Also, the overlapping extension of the localized modularity $NQ_{ov}$ has the belonging function $f(a_{i,c}, a_{j,c})=\frac{a_{i,c}+a_{j,c}}{2}$ or $f(a_{i,c}, a_{j,c})=a_{i,c}a_{j,c}$. For the overlapping extension of modularity density $(Q_{ds}^{ov})$, the belonging function $f(a_{i,c}, a_{j,c})$ can also be the average or the product of $a_{i,c}$ and $a_{j,c}$. Thus, there are two versions of the belonging function for $Q_{ov}$, $NQ_{ov}$, and $Q_{ds}^{ov}$. Therefore, we have $Q_{ov}(average)$ with $f(a_{i,c}, a_{j,c})=\frac{a_{i,c}+a_{j,c}}{2}$, $Q_{ov}(product)$ with $f(a_{i,c}, a_{j,c})=a_{i,c}a_{j,c}$, $Q_{ov}^L$ in Equation~(\ref{eq:Qov_edge}), $NQ_{ov}(average)$ with $f(a_{i,c}, a_{j,c})=\frac{a_{i,c}+a_{j,c}}{2}$, $NQ_{ov}(product)$ with $f(a_{i,c}, a_{j,c})=a_{i,c}a_{j,c}$, $Q_{ds}^{ov}(average)$ with $f(a_{i,c}, a_{j,c})=\frac{a_{i,c}+a_{j,c}}{2}$, and $Q_{ds}^{ov}(product)$ with $f(a_{i,c}, a_{j,c})=a_{i,c}a_{j,c}$. For fuzzy overlapping community structures, $a_{i,c}$ is given for each node $i$ to each community $c$ to which it belongs. For crisp overlapping community structures, we can adopt Equation~(\ref{eq:num_com_factor}) and Equation~(\ref{eq:node_strength_factor}) to calculate $a_{i,c}$. Consequently, two versions of the belonging coefficient can be used to convert crisp overlapping to fuzzy overlapping.

We also consider eight local community quality metrics: the number of \textit{Intra-edges}, \textit{Intra-density}, \textit{Contraction}, the number of \textit{Boundary-edges}, \textit{Expansion}, \textit{Conductance} \cite{QdsConference,QdsJournal}, the \textit{Fitness} function \cite{Fitness}, and the \textit{Average Modularity Degree} \cite{AverageModularityDegree}. These metrics describe how the connectivity structure of a given set of nodes resembles a community. All of them rely on the intuition that communities are sets of nodes with many edges inside them and few edges outside of them. We also extend these metrics to be applicable to fuzzy overlapping community structures in which nodes are assigned probability of belonging to each community of which they are part. Two versions of the belonging coefficient and two versions of the belonging function are considered for each metric. For fuzzy overlapping community structure, we define the size of a community $c$ as $|c|=\sum_{i \in c}a_{i,c}$.\\
\textbf{The number of \textit{Intra-edges} (\textit{IE}):} $|E_c^{in}|$; it is the total number of edges in $c$. A large value of this metric is better than a small value in terms of the community quality. \\
\textbf{\textit{Intra-density} (\textit{ID}):} $d_c$ in Equation~(\ref{eq:Qov_ds}). The larger the value of this metric, the higher the quality of the communities. \\
\textbf{\textit{Contraction} (\textit{CNT}):} $2|E_c^{in}|/|c|$; it measures the average number of edges per node inside the community $c$. A larger value of contraction means a better community quality. \\
\textbf{The number of \textit{Boundary-edges} (BE):} $|E_c^{out}|$; it is the total number of edges on the boundary of $c$. A small value of this metric is better than a large value in terms of the community quality.  \\
\textbf{\textit{Expansion} (\textit{EXP}):} $|E_c^{out}|/|c|$; it measures the average number of edges (per node) that point outside the community $c$. A smaller value of expansion corresponds to a better community structure. \\
\textbf{\textit{Conductance} (\textit{CND}):} $\frac{|E_c^{out}|}{2|E_c^{in}|+|E_c^{out}|}$; it measures the fraction of the total number of edges that point outside the community. A smaller value of conductance means a better community quality. \\
\textbf{The \textit{Fitness} (\textit{F}) function:} $\frac{|E_c^{in}|}{|E_c^{in}|+|E_c^{out}|}$; it is the ratio between the internal degree and the total degree of a community $c$. A larger value of $F$ indicates a better community quality. \\
\textbf{\textit{Average Modularity Degree} (\textit{D}):} $\sum_{c \in C}\frac{2|E_c^{in}|-|E_c^{out}|}{|c|}$; it is the summation of the average modularity degree of each community. The average modularity degree of a community ($\frac{2|E_c^{in}|-|E_c^{out}|}{|c|}$) equals to the average inner degree ($\frac{2|E_c^{in}|}{|c|}$) minus the average outer degree ($\frac{|E_c^{out}|}{|c|}$). The larger the value of $D$, the higher the quality of the community structure.

In this section, we compare different choices of the belonging coefficient and the belonging function to be used for $Q_{ov}$, $Q_{ov}^L$, $NQ_{ov}$, $Q_{ds}^{ov}$, and the eight local community quality metrics in order to see which version of the belonging coefficient and which version of the belonging function are better. Then, we try to determine which of the three overlapping extensions of modularity (two kinds of node-based extensions of modularity and the edge-based extension of modularity) is the best. In addition, we compare the performance of all these overlapping metrics with the best combination of belonging coefficient and belonging function to recommend which metrics to select when measuring the quality of overlapping community structures.

The experiments are done with three community detection algorithms, Speaker-listener Label Propagation Algorithm (SLPA) \cite{SLPA2012}, Clique Percolation Method (CFinder) \cite{CPM}, and SpeakEasy \cite{SpeakEasy} which is a label propagation algorithm specialized for biology networks, on a large number of real networks and synthetic networks. We vary the threshold parameter $r$ of SLPA \cite{SLPA2012} from 0.05 to 0.5 with step 0.05. SLPA gets crisp overlapping communities when $r<0.5$ and gets disjoint communities when $r=0.5$ (for $r>0.5$ SLPA generates the same disjoint communities as for $r=0.5$). For each value of threshold $r$, we adopt 10 running samples since the community detection result of SLPA is not deterministic. We vary the parameter $k$ of CFinder from 3 to 20 with step 1 but only when such $k$-clique-community is available. It is usually the case that some nodes are not in the final discovered $k$-clique-communities so we consider each of these nodes forming a community of itself. The threshold parameter $tr$ of SpeakEasy is varied from 0.05 to 1 with step 0.05. SpeakEasy gets crisp overlapping community structures when $tr<1$ and gets disjoint community structures when $tr=1$.

Then, for the community detection results of SLPA with different values of threshold $r$, the results of CFinder with different values of $k$, and the results of SpeakEasy with different thresholds $tr$ on each of these networks, we calculate the values of $Q_{ov}$, $Q_{ov}^L$, $NQ_{ov}$ $Q_{ds}^{ov}$, and the eight local community quality metrics (twelve metrics in total) with two versions of the \textbf{belonging coefficient (BC)} and two versions of the \textbf{belonging function (BF)}. For each $r$ of SLPA, the values of all the metrics are calculated as the average of the 10 runs. For convenience, we denote Equation~(\ref{eq:num_com_factor}) and Equation~(\ref{eq:node_strength_factor}) as the first and the second version of the belonging coefficient, respectively. We also denote the belonging function $f(a_{i,c}, a_{j,c})=\frac{a_{i,c}+a_{j,c}}{2}$ as the first version of the belonging function and $f(a_{i,c}, a_{j,c})=a_{i,c}a_{j,c}$ as the second version of the belonging function. We determine which version of the belonging coefficient and which version of the belonging function are better based on the largest number of quality metrics consistent with each other on determining the best value of threshold $r$ for SLPA, the best value of parameter $k$ for CFinder, and the best value of threshold $tr$ for SpeakEasy on all these networks. Finally, we compare the performance of these overlapping metrics with the best combination of belonging coefficient and belonging function by looking at how many times each metric is among those that are consistent with each other on determining the best values of parameters for the three adopted community detection algorithms.

\subsection{Real Network Datasets}

\begin{table}
\centering
\caption{Basic properties of all real network datasets used in the experiments.}
\label{tab:real_networks}       
\setlength{\tabcolsep}{0.5pt}
\begin{tabular}{c||c|c|c|c}
\hline \hline
Name & \#Nodes & \#Edges & Type & Description \\
\hline
Celegans & 453 & 2025 & Unweighted \& Undirected & Metabolic network of C. elegans \cite{ExtremalQ} \\
\hline
Dolphin & 62 & 159 & Unweighted \& Undirected & Dolphin social network \cite{Dolphin} \\
\hline
Email & 1133 & 5451 & Unweighted \& Undirected & Email network \cite{Email} \\
\hline
Football & 115 & 613 & Unweighted \& Undirected & American college football network \cite{football} \\
\hline
Jazz & 198 & 2742 & Unweighted \& Undirected & Jazz musicians network \cite{Jazz} \\
\hline
Karate & 34 & 78 & Unweighted \& Undirected & Zachary's karate club network \cite{karate} \\
\hline
Lesmis & 77 & 254 & Weighted \& Undirected & Characters coappearance network \cite{Lesmis} \\
\hline
Netscience & 1461 & 2742 & Weighted \& Undirected & Coauthorship network \cite{EigenvectorCommunity} \\
\hline
PGP & 10680 & 24316 & Unweighted \& Undirected & PGP network \cite{PGPNetwork} \\
\hline
Polblogs & 1224 & 19022 & Unweighted \& Directed & Political blogs network \cite{Polblogs} \\
\hline
Polbooks & 105 & 441 & Unweighted \& Undirected & Network of books about US politics \cite{Polbooks} \\
\hline
Railway & 297 & 1213 & Unweighted \& Undirected & Indian railway network \cite{Railway} \\
\hline
Santafe & 118 & 200 & Unweighted \& Undirected & Collaboration network of scientists \cite{football} \\
\hline
Collins\_cyc & 1097 & 6392 & Unweighted \& Undirected & \multirow{8}{*}{Protein-protein interaction networks \cite{SpeakEasy}} \\
\cline{1-4}
Collins\_cyc\_w & 1097 & 6392 & Weighted \& Undirected & \\
\cline{1-4}
Collins\_mips & 734 & 4778 & Unweighted \& Undirected & \\
\cline{1-4}
Collins\_sgd & 809 & 2955 & Unweighted \& Undirected & \\
\cline{1-4}
Gavin\_cyc & 997 & 4031 & Unweighted \& Undirected & \\
\cline{1-4}
Gavin\_cyc\_w & 997 & 4031 & Weighted \& Undirected & \\
\cline{1-4}
Gavin\_mips & 701 & 2695 & Unweighted \& Undirected & \\
\cline{1-4}
Gavin\_sgd & 747 & 2639 & Unweighted \& Undirected & \\
\hline
Amazon & 319948 & 880215 & Unweighted \& Undirected & Amazon product network \cite{Leskovec_dataset} \\
\hline
DBLP & 260998 & 950059 & Unweighted \& Undirected & DBLP collaboration network \cite{Leskovec_dataset} \\

\hline \hline
\end{tabular}
\end{table}

We consider totally 23 real network datasets, including friendship network, collaboration networks, co-purchasing networks, biology networks, etc. Table~\ref{tab:real_networks} shows the basic properties of all these datasets. It can be seen that these networks have different numbers of nodes and different numbers of edges, varying from very small networks to very large networks. Moreover, edges in some networks have weights and directions.

\textbf{Celegans} is a metabolic network of C. elegans \cite{ExtremalQ}. \textbf{Dolphin} is an social network of frequent associations between 62 dolphins in a community living off Doubtful Sound, New Zealand \cite{Dolphin}. \textbf{Email} is a network of email interchanges between members of the Univeristy Rovira i Virgili (Tarragona) \cite{Email}. \textbf{Football} is a network that represents the schedule of games between college football teams in a single season \cite{football}. \textbf{Jazz} is a network of collaborations between jazz musicians \cite{Jazz}. \textbf{Karate} is a network representing the friendships between $34$ members of a karate club at a US university during two years \cite{karate}. \textbf{Lesmis} is a coappearance network of characters in the novel Les Miserables \cite{Lesmis}. \textbf{Netscience} is a coauthorship network of scientists working on network theory and experiment \cite{EigenvectorCommunity}. \textbf{PGP} is the giant component of the network of users of the Pretty-Good-Privacy (PGP) algorithm for secure information interchange \cite{PGPNetwork}. \textbf{Polblogs} is a directed network of hyperlinks between weblogs on US politics, recorded in 2005 by Adamic and Glance \cite{Polblogs}. \textbf{Polbooks} is a network of books about US politics published around the time of the 2004 presidential election and sold by the online bookseller Amazon.com \cite{Polbooks}. \textbf{Railway} is a network with nodes representing Indian railway stations, where two stations are connected by an edge if there exists at least one train-route such that both stations are scheduled stops on that route \cite{Railway}. \textbf{Santafe} is the largest connected component of the collaboration network of scientists at the Santa Fe Institute during years 1999 and 2000 \cite{football}. \textbf{Collins\_cyc}, \textbf{Collins\_cyc\_w}, \textbf{Collins\_mips}, \textbf{Collins\_sgd}, \textbf{Gavin\_cyc}, \textbf{Gavin\_cyc\_w}, \textbf{Gavin\_mips}, and \textbf{Gavin\_sgd} are two kinds (referred as Collins \cite{Collins} and Gavin \cite{Gavin} here) of popular high throughput protein-protein interaction networks derived from measurements obtained by affinity purification and mass spectrometry (AP-MS) techniques \cite{SpeakEasy}. These two kinds of networks are further refined with three gold-standards for protein complexes, including the classic Munich Information Center for Protein Sequences (MIPS) \cite{MIPS} and the more recent Saccharomyces Genome Database (SGD) \cite{SGD}. The complete MIPS dataset as well as partial information from SGD are incorporated into a third protein complex list known as CYC2008 \cite{CYC}. Thus, we have \textbf{Collins\_cyc}, \textbf{Collins\_mips}, \textbf{Collins\_sgd}, \textbf{Gavin\_cyc}, \textbf{Gavin\_mips}, and \textbf{Gavin\_sgd}, respectively. \textbf{Collins\_cyc\_w} and \textbf{Gavin\_cyc\_w} are respectively the weighted versions of \textbf{Collins\_cyc} and \textbf{Gavin\_cyc}, in which the weight is proportional to the probability a given interaction pair truly exists. \textbf{Amazon} is a product co-purchasing network of the Amazon website \cite{Leskovec_dataset}. The nodes of the network represent products and edges link commonly copurchased products. \textbf{DBLP} is a scientific collaboration network where nodes represent authors and edges connect authors that have co-authored a paper \cite{Leskovec_dataset}.

\begin{table}
\centering
\caption{The best value of threshold $r$ for SLPA and the corresponding number of community quality metrics (out of twelve) that are consistent with each other on determining this best $r$ for the four combinations of two versions of belonging coefficient and two versions of belonging function on all real network datasets. The best value in each row is marked by red italic font.}
\label{tab:real_networks:slpa}       
\setlength{\tabcolsep}{4pt}
\begin{tabular}{c||c|c|c|c}
\hline \hline
 & \multicolumn{4}{c}{SLPA} \\
\hline
Datasets & (BC,BF)=(1,1) & (BC,BF)=(1,2) & (BC,BF)=(2,1) & (BC,BF)=(2,2) \\
\hline
Celegans & \textcolor{red}{\textbf{\emph{0.5} (\emph{6})}} & 0.05 (4) & \textcolor{red}{\textbf{\emph{0.5} (\emph{6})}} & 0.4 (5) \\
\hline
Dolphin & 0.5 (5) & \textcolor{red}{\textbf{\emph{0.4} (\emph{8})}} & 0.5 (6) & \{0.05,0.4\} (5) \\
\hline
Email & 0.5 (8) & \textcolor{red}{\textbf{\emph{0.5} (\emph{10})}} & 0.5 (7) & 0.5 (8) \\
\hline
Football & \{0.45,0.5\} (9) & \textcolor{red}{\textbf{\{\emph{0.45},\emph{0.5}\} (\emph{11})}} & \{0.45,0.5\} (8) & 0.25 (7) \\
\hline
Jazz & \textcolor{red}{\textbf{\emph{0.5} (\emph{10})}} & \textcolor{red}{\textbf{\emph{0.5} (\emph{10})}} & 0.5 (9) & 0.5 (9) \\
\hline
Karate & 0.5 (8) & \textcolor{red}{\textbf{\emph{0.45} (\emph{10})}} & 0.5 (7) & \textcolor{red}{\textbf{\emph{0.45} (\emph{10})}} \\
\hline
Lesmis & \{0.25,0.5\} (4) & \textcolor{red}{\textbf{\emph{0.25} (\emph{6})}} & 0.5 (4) & 0.15 (5) \\
\hline
Netscience & 0.35 (4) & 0.5 (4) & \textcolor{red}{\textbf{\emph{0.35} (\emph{5})}} & \{0.35,0.5\} (3) \\
\hline
PGP & 0.5 (8) & \textcolor{red}{\textbf{\emph{0.5} (\emph{10})}} & 0.5 (7) & 0.5 (8) \\
\hline
Polblogs & 0.5 (7) & \textcolor{red}{\textbf{\emph{0.5} (\emph{8})}} & 0.5 (6) & 0.5 (4) \\
\hline
Polbooks & \textcolor{red}{\textbf{\emph{0.5} (\emph{7})}} & 0.2 (5) & \textcolor{red}{\textbf{\emph{0.5} (\emph{7})}} & 0.2 (4) \\
\hline
Railway & 0.5 (8) & \textcolor{red}{\textbf{\emph{0.5} (\emph{9})}} & 0.5 (7) & 0.5 (7) \\
\hline
Santafe & 0.4 (6) & \textcolor{red}{\textbf{\emph{0.4} (\emph{7})}} & 0.4 (5) & 0.4 (5) \\
\hline
Collins\_cyc & \textcolor{red}{\textbf{\emph{0.5} (\emph{8})}} & 0.5 (5) & \textcolor{red}{\textbf{\emph{0.5} (\emph{8})}} & 0.5 (6) \\
\hline
Collins\_cyc\_w & 0.05 (7) & 0.05 (7) & 0.05 (6) & \textcolor{red}{\textbf{\emph{0.05} (\emph{8})}} \\
\hline
Collins\_mips & 0.45 (5) & \textcolor{red}{\textbf{\emph{0.45} (\emph{6})}} & \{0.05,0.45\} {4} & \{0.05,0.45\} (4) \\
\hline
Collins\_sgd & 0.5 (4) & 0.1 (4) & 0.5 (4) & \textcolor{red}{\textbf{\emph{0.1} (\emph{6})}} \\
\hline
Gavin\_cyc & 0.45 (5) & 0.45 (6) & 0.45 (5) & \textcolor{red}{\textbf{\emph{0.45} (\emph{7})}} \\
\hline
Gavin\_cyc\_w & 0.35 (5) & \textcolor{red}{\textbf{\emph{0.3} (\emph{6})}} & 0.05 (4) & 0.3 (5) \\
\hline
Gavin\_mips & 0.5 (9) & \textcolor{red}{\textbf{\emph{0.5} (\emph{11})}} & 0.5 (8) & 0.5 (6) \\
\hline
Gavin\_sgd & 0.5 (8) & \textcolor{red}{\textbf{\emph{0.5} (\emph{9})}} & 0.5 (7) & 0.5 (7) \\
\hline
Amazon & 0.5 (8) & \textcolor{red}{\textbf{\emph{0.5} (\emph{10})}} & 0.5 (6) & 0.5 (9) \\
\hline
DBLP & 0.5 (7) & \textcolor{red}{\textbf{\emph{0.5} (\emph{10})}} & 0.5 (6) & 0.5 (8) \\

\hline \hline
\end{tabular}
\end{table}

Tables~\ref{tab:real_networks:slpa}-\ref{tab:real_networks:speak_easy} show the best value of threshold $r$ for SLPA, the best value of parameter $k$ for CFinder, and the best value of threshold $tr$ for SpeakEasy, respectively, along with the corresponding number of community quality metrics (out of twelve) that are consistent with each other on determining this best $r$, this best $k$, and this best $tr$ for the four combinations of two versions of belonging coefficient and two versions of belonging function on all 23 real network datasets. For instance, entry \textbf{0.5 (5)} in row \textbf{Dolphin} and column \textbf{(BC,BF)=(1,1)} in Table~\ref{tab:real_networks:slpa} means that there are totally five out of twelve community quality metrics showing that the best value of threshold $r$ for SLPA is 0.5 when adopting the first version of belonging coefficient and the first version of belonging function. The red italic font in each dataset row of these tables denotes the best combination of the two versions of belonging coefficient and two versions of belonging function for each dataset. For example, \textcolor{red}{\textbf{\emph{0.4} (\emph{8})}} in row \textbf{Dolphin} and column \textbf{(BC,BF)=(1,2)} in Table~\ref{tab:real_networks:slpa} indicates that the twelve metrics with the first version of belonging coefficient and the second version of belonging function is the best since the number of metrics (which is eight here) consistent with each other on determining the best value of $r$ is the largest among the four combinations.

\begin{table}
\centering
\caption{The best value of parameter $k$ for CFinder and the corresponding number of community quality metrics (out of twelve) that are consistent with each other on determining this best $k$ for the four combinations of two versions of belonging coefficient and two versions of belonging function on all real network datasets. The best value in each row is marked by red italic font.}
\label{tab:real_networks:cfinder}       
\setlength{\tabcolsep}{5pt}
\begin{tabular}{c||c|c|c|c}
\hline \hline
 & \multicolumn{4}{c}{CFinder} \\
\hline
Datasets & (BC,BF)=(1,1) & (BC,BF)=(1,2) & (BC,BF)=(2,1) & (BC,BF)=(2,2) \\
\hline
Celegans & 3 (6) & \textcolor{red}{\textbf{\emph{3} (\emph{7})}} & 3 (5) & \{3,9\} (4) \\
\hline
Dolphin & 3 (11) & \textcolor{red}{\textbf{\emph{3} (\emph{12})}} & 3 (7) & 3 (7) \\
\hline
Email & \textcolor{red}{\textbf{\emph{3} (\emph{10})}} & \textcolor{red}{\textbf{\emph{3} (\emph{10})}} & 4 (4) & \{3,4,9-12\} (3) \\
\hline
Football & \textcolor{red}{\textbf{\emph{4} (\emph{6})}} & \{3,4\} (5) & 4 (5) & 4 (5) \\
\hline
Jazz & \textcolor{red}{\textbf{\emph{3} (\emph{8})}} & \textcolor{red}{\textbf{\emph{3} (\emph{8})}} & 3 (6) & 3 (5) \\
\hline
Karate & \textcolor{red}{\textbf{\emph{3} (\emph{11})}} & \textcolor{red}{\textbf{\emph{3} (\emph{11})}} & 3 (8) & 3 (7) \\
\hline
Lesmis & 3 (7) & \textcolor{red}{\textbf{\emph{3} (\emph{8})}} & 6 (4) & \{3,6\} (4) \\
\hline
Netscience & 3 (11) & \textcolor{red}{\textbf{\emph{3} (\emph{12})}} & 3 (9) & 3 (9) \\
\hline
PGP & 3 (10) & \textcolor{red}{\textbf{\emph{3} (\emph{12})}} & 3 (6) & 3 (8) \\
\hline
Polblogs & N/A & N/A & N/A & N/A \\
\hline
Polbooks & \textcolor{red}{\textbf{\emph{3} (\emph{10})}} & \textcolor{red}{\textbf{\emph{3} (\emph{10})}} & 3 (7) & 3 (6) \\
\hline
Railway & \textcolor{red}{\textbf{\emph{3} (\emph{9})}} & \textcolor{red}{\textbf{\emph{3} (\emph{9})}} & 3 (4) & 3 (4) \\
\hline
Santafe & 3 (10) & \textcolor{red}{\textbf{\emph{3} (\emph{11})}} & 3 (7) & 3 (8) \\
\hline
Collins\_cyc & \textcolor{red}{\textbf{\emph{3} (\emph{11})}} & \textcolor{red}{\textbf{\emph{3} (\emph{11})}} & 3 (7) & 3 (8) \\
\hline
Collins\_cyc\_w & N/A & N/A & N/A & N/A \\
\hline
Collins\_mips & \textcolor{red}{\textbf{\emph{3} (\emph{10})}} & \textcolor{red}{\textbf{\emph{3} (\emph{10})}} & 3 (6) & 3 (6) \\
\hline
Collins\_sgd & \textcolor{red}{\textbf{\emph{3} (\emph{12})}} & \textcolor{red}{\textbf{\emph{3} (\emph{12})}} & 3 (8) & 3 (8) \\
\hline
Gavin\_cyc & 3 (11) & \textcolor{red}{\textbf{\emph{3} (\emph{12})}} & 3 (6) & 3 (6) \\
\hline
Gavin\_cyc\_w & N/A & N/A & N/A & N/A \\
\hline
Gavin\_mips & \textcolor{red}{\textbf{\emph{3} (\emph{12})}} & \textcolor{red}{\textbf{\emph{3} (\emph{12})}} & 3 (7) & 3 (6) \\
\hline
Gavin\_sgd & \textcolor{red}{\textbf{\emph{3} (\emph{11})}} & \textcolor{red}{\textbf{\emph{3} (\emph{11})}} & 3 (6) & 3 (6) \\
\hline
Amazon & \textcolor{red}{\textbf{\emph{3} (\emph{11})}} & \textcolor{red}{\textbf{\emph{3} (\emph{11})}} & 3 (6) & 3 (6) \\
\hline
DBLP & 3 (9) & \textcolor{red}{\textbf{\emph{3} (\emph{10})}} & 3 (5) & 3 (5) \\

\hline \hline
\end{tabular}
\end{table}

It can be observed from Table~\ref{tab:real_networks:slpa} that almost all the networks, except \textbf{Celegans}, \textbf{Netscience}, \textbf{Polbooks}, \textbf{Collins\_cyc}, \textbf{Collins\_cyc\_w}, \textbf{Collins\_sgd}, and \textbf{Gavin\_cyc}, imply that (BC,BF)=(1,2) is the best among the four possible combinations when using SLPA. Table~\ref{tab:real_networks:cfinder} shows that (BC,BF)=(1,2) is the best on all the networks, except \textbf{Football}, when using CFinder. Results for CFinder on \textbf{Polblogs}, \textbf{Collins\_cyc\_w}, and \textbf{Gavin\_cyc\_w} are not provided because it has not finished running on these three networks for more than two months processing many potential $k$-cliques. Similarly, Table~\ref{tab:real_networks:speak_easy} indicates that (BC,BF)=(1,2) is the best on all the networks, except \textbf{Jazz}, when using SpeakEasy. We can observe from the three tables that for each network there are at least two out of three algorithms (SLPA, CFinder, and SpeakEasy) supporting conclusion that (BC,BF)=(1,2) is the best. Thus, we determined that the first version of the belonging coefficient is better than the alternative. It means that to convert crisp overlapping to fuzzy overlapping the belonging coefficient of a node to a community should be the reciprocal of the number of communities of which this node is a part. When the relationship between a node and the communities to which it belongs is binary, there is no information about the strength of the membership. In this case, it is intuitive and reasonable to assign a node to its communities using equal belonging coefficients. We also determined that the second version of the belonging function is better than the alternative. It means that the probability of the event that two nodes belong to the same community should be the product, not the average, of their belonging coefficients to that community. In addition, $Q_{ov}=Q_{ov}'$ when $f(a_{i,c}, a_{j,c})=a_{i,c}a_{j,c}$ as proved in Subsection~\ref{sec:ov_Q}, which is another way of showing that the second version of the belonging function is much more suitable for use in the metric than the first. Therefore, we conclude that the overlapping community quality metrics with the first version of the belonging coefficient and the second version of the belonging function are the best among the four possible combinations on all the real network datasets. Tables 1-23 in Supplementary Materials contain more detailed results on these real network datasets.

\begin{table}
\centering
\caption{The best value of threshold $tr$ for SpeakEasy and the corresponding number of community quality metrics (out of twelve) that are consistent with each other on determining this best $tr$ for the four combinations of two versions of belonging coefficient and two versions of belonging function on all real network datasets. The best value in each row is marked by red italic font.}
\label{tab:real_networks:speak_easy}       
\setlength{\tabcolsep}{0.5pt}
\begin{tabular}{c||c|c|c|c}
\hline \hline
 & \multicolumn{4}{c}{SpeakEasy} \\
\hline
Datasets & (BC,BF)=(1,1) & (BC,BF)=(1,2) & (BC,BF)=(2,1) & (BC,BF)=(2,2) \\
\hline
Celegans & 0.75 (6) & \textcolor{red}{\textbf{\emph{0.75} (\emph{7})}} & 0.75 (6) & 0.75 (6) \\
\hline
Dolphin & 0.4 (3) & \textcolor{red}{\textbf{\emph{0.4} (\emph{4})}} & \textcolor{red}{\textbf{\emph{0.15} (\emph{4})}} & \textcolor{red}{\textbf{\emph{0.7} (\emph{4})}} \\
\hline
Email & \textcolor{red}{\textbf{\emph{0.9} (\emph{3})}} & \textcolor{red}{\textbf{\emph{1} (\emph{3})}} & \{0.05,0.9\} (2) & \textcolor{red}{\textbf{\{\emph{0.5},\emph{1}\} (\emph{3})}} \\
\hline
Football & \textcolor{red}{\textbf{\emph{0.6} (\emph{10})}} & \textcolor{red}{\textbf{\emph{0.6} (\emph{10})}} & \textcolor{red}{\textbf{\emph{0.6} (\emph{10})}} &  \textcolor{red}{\textbf{\emph{0.6} (\emph{10})}} \\
\hline
Jazz & 0.75 (5) & 0.75 (5) & 0.75 (5) & \textcolor{red}{\textbf{\emph{0.75} (\emph{6})}} \\
\hline
Karate & 0.45 (5) & \textcolor{red}{\textbf{\emph{0.45} (\emph{6})}} & 0.45 (5) & \textcolor{red}{\textbf{\emph{0.45} (\emph{6})}} \\
\hline
Lesmis & 0.85 (5) & \textcolor{red}{\textbf{\emph{0.85} (\emph{6})}} & 0.85 (4) & 0.85 (4) \\
\hline
Netscience & 0.7 (4) & \textcolor{red}{\textbf{\emph{0.25} (\emph{10})}} & 0.05 (3) & \{0.15,0.2,0.25,0.7\} (2) \\
\hline
PGP & 0.85 (5) & \textcolor{red}{\textbf{\emph{0.85} (\emph{7})}} & \{0.05,0.75,0.85\} (3) & 0.85 (4) \\
\hline
Polblogs & \textcolor{red}{\textbf{\emph{0.7} (\emph{4})}} & \textcolor{red}{\textbf{\emph{0.7} (\emph{4})}} & 0.45 (3) & \{0.5,0.7,0.8,0.9\} (2) \\
\hline
Polbooks & 0.95 (5) & \textcolor{red}{\textbf{\emph{0.95} (\emph{6})}} & \{0.5,0.95\} (3) & 0.95 (4) \\
\hline
Railway & 0.8 (4) & \textcolor{red}{\textbf{\emph{0.8} (\emph{5})}} & 0.8 (3) & 0.8 (4) \\
\hline
Santafe & \textcolor{red}{\textbf{\emph{0.9} (\emph{5})}} & \textcolor{red}{\textbf{\emph{0.9} (\emph{5})}} & 0.9 (4) & \{0.65,0.9\} (4) \\
\hline
Collins\_cyc & 0.9 (8) & \textcolor{red}{\textbf{\emph{0.9} (\emph{10})}} & 0.9 (4) & 0.9 (6) \\
\hline
Collins\_cyc\_w & 0.55 (6) & \textcolor{red}{\textbf{\emph{0.55} (\emph{9})}} & \{0.1,0.55\} (3) & 0.55 (4) \\
\hline
Collins\_mips & 0.4 (7) & \textcolor{red}{\textbf{\emph{0.4} (\emph{10})}} & 0.4 (5) & \{0.25,0.4,0.6\} (3) \\
\hline
Collins\_sgd & 0.8 (7) & \textcolor{red}{\textbf{\emph{0.8} (\emph{10})}} & \{0.5,0.8\} (4) & \{0.5,0.8\} (5) \\
\hline
Gavin\_cyc & 0.7 (6) & \textcolor{red}{\textbf{\emph{0.7} (\emph{8})}} & 0.7 (5) &  0.7 (6) \\
\hline
Gavin\_cyc\_w & 0.7 (4) & \textcolor{red}{\textbf{\emph{0.7} (\emph{5})}} & \{0.05,0.5,0.7,0.95\} (2) & 0.7 (3) \\
\hline
Gavin\_mips & 0.9 (6) & \textcolor{red}{\textbf{\emph{0.9} (\emph{8})}} & 0.9 (5) & 0.9 (5) \\
\hline
Gavin\_sgd & 0.8 (5) & \textcolor{red}{\textbf{\emph{0.8} (\emph{6})}} & \textcolor{red}{\textbf{\emph{0.8} (\emph{6})}} & \textcolor{red}{\textbf{\emph{0.8} (\emph{6})}}  \\
\hline
Amazon & 1 (9) & \textcolor{red}{\textbf{\emph{1} (\emph{11})}} & 1 (8) & \textcolor{red}{\textbf{\emph{1} (\emph{11})}}  \\
\hline
DBLP & 1 (9) & \textcolor{red}{\textbf{\emph{1} (\emph{10})}} & 1 (8) & \textcolor{red}{\textbf{\emph{1} (\emph{10})}} \\

\hline \hline
\end{tabular}
\end{table}

\begin{table}
\centering
\caption{The number of times (maximum three since there are totally three community detection algorithms adopted) that the community quality metric is among those consistent with each other on determining the best value of threshold $r$ for SLPA, the best value of parameter $k$ for CFinder, and the best value of threshold $tr$ for SpeakEasy on each real network with (BC,BF)=(1,2). The best value in the last row is marked by red italic font.}
\label{tab:real_networks:all_metrics_1_2}       
\setlength{\tabcolsep}{0.5pt}
\begin{tabular}{c||c|c|c|c|c|c|c|c|c|c|c|c}
\hline \hline
Datasets & $Q_{ov}$ & $NQ_{ov}$ & $Q_{ov}^L$ & $Q_{ds}^{ov}$ & $IE$ & $ID$ & $CNT$ & $BE$ & $EXP$ & $CND$ & $F$ & $D$ \\
\hline
Celegans & 1 & 1 & 1 & 1 & 3 & 0 & 2 & 0 & 2 & 2 & 3 & 2 \\
\hline
Dolphin & 3 & 2 & 3 & 2 & 2 & 2 & 1 & 1 & 2 & 2 & 1 & 3 \\
\hline
Email & 2 & 2 & 3 & 2 & 3 & 1 & 1 & 1 & 2 & 2 & 2 & 2 \\
\hline
Football & 2 & 1 & 3 & 3 & 3 & 2 & 3 & 2 & 3 & 3 & 3 & 3 \\
\hline
Jazz & 2 & 2 & 2 & 2 & 2 & 1 & 1 & 3 & 2 & 2 & 2 & 2 \\
\hline
Karate & 3 & 3 & 3 & 3 & 2 & 1 & 2 & 1 & 3 & 2 & 2 & 2 \\
\hline
Lesmis & 0 & 2 & 1 & 0 & 2 & 2 & 3 & 0 & 2 & 3 & 3 & 2 \\
\hline
Netscience & 3 & 2 & 3 & 2 & 3 & 1 & 1 & 3 & 2 & 2 & 2 & 2 \\
\hline
PGP & 3 & 2 & 3 & 2 & 2 & 1 & 3 & 2 & 3 & 3 & 2 & 3 \\
\hline
Polblogs & 1 & 1 & 1 & 1 & 1 & 0 & 1 & 1 & 1 & 1 & 1 & 2 \\
\hline
Polbooks & 2 & 1 & 2 & 1 & 2 & 2 & 1 & 0 & 2 & 3 & 3 & 2 \\
\hline
Railway & 1 & 1 & 1 & 1 & 2 & 1 & 3 & 2 & 2 & 3 & 3 & 3 \\
\hline
Santafe & 2 & 2 & 2 & 1 & 1 & 2 & 3 & 0 & 2 & 3 & 3 & 2 \\
\hline
Collins\_cyc & 3 & 2 & 3 & 1 & 2 & 3 & 2 & 2 & 2 & 2 & 2 & 2 \\
\hline
Collins\_cyc\_w & 2 & 1 & 2 & 0 & 2 & 0 & 1 & 1 & 2 & 2 & 2 & 1 \\
\hline
Collins\_mips & 2 & 3 & 3 & 1 & 2 & 2 & 2 & 1 & 3 & 2 & 2 & 3 \\
\hline
Collins\_sgd & 2 & 2 & 2 & 1 & 3 & 2 & 3 & 1 & 2 & 3 & 3 & 2 \\
\hline
Gavin\_cyc & 2 & 2 & 2 & 2 & 2 & 1 & 1 & 2 & 3 & 3 & 3 & 3 \\
\hline
Gavin\_cyc\_w & 0 & 0 & 0 & 1 & 1 & 0 & 2 & 0 & 1 & 2 & 2 & 2 \\
\hline
Gavin\_mips & 3 & 2 & 3 & 2 & 2 & 2 & 3 & 2 & 3 & 3 & 3 & 3 \\
\hline
Gavin\_sgd & 2 & 3 & 2 & 1 & 2 & 2 & 3 & 1 & 2 & 3 & 3 & 2 \\
\hline
Amazon & 3 & 3 & 3 & 3 & 3 & 2 & 3 & 1 & 3 & 2 & 3 & 3 \\
\hline
DBLP & 2 & 2 & 3 & 3 & 3 & 2 & 2 & 1 & 3 & 3 & 3 & 3 \\
\hline
     & \shortstack{~~~ \\ 8 \\ (2)} & \shortstack{4 \\ (1.83)} & \shortstack{12 \\ (2.22)} & \shortstack{4 \\ (1.57)} & \shortstack{8 \\ (2.17)} & \shortstack{1 \\ (1.39)} & \shortstack{10 \\ (2.04)} & \shortstack{2 \\ (1.22)} & \shortstack{9 \\ (2.26)} & \shortstack{13 \\ \textcolor{red}{\textbf{(\emph{2.43})}}} & \shortstack{~~ \\ \textcolor{red}{\textbf{\emph{14}}} \\ \textcolor{red}{\textbf{(\emph{2.43})}}} & \shortstack{11 \\ (2.35)} \\ [1ex]
\hline \hline
\end{tabular}
\end{table}

Table~\ref{tab:real_networks:all_metrics_1_2} shows the number of times (maximum three since there are totally three community detection algorithms adopted) that the community quality metric is among the quality metrics that are consistent with each other on determining the best value of threshold $r$ for SLPA, the best value of parameter $k$ for CFinder, and the best value of threshold $tr$ for SpeakEasy on each real network with (BC,BF)=(1,2). The last row in the table presents on how many networks this metric got support from all three algorithms and how many times on average (maximum three; but for \textbf{Polblogs}, \textbf{Collins\_cyc\_w}, and \textbf{Gavin\_cyc\_w} the maximum value is two because CFinder has no result for these networks) this metric got support from each of the 23 real networks. This table indicates that the edge-based overlapping definition is the best overlapping extension for modularity among the three extensions (two kinds of node-based extensions of modularity and the edge-based extension of modularity). The fitness function gets the largest values in the last row and generally local metrics are better than global metrics here which may imply that local community quality metrics are more applicable for measuring the quality of overlapping community structures. Note that eight out of twelve metrics are local metrics, so they are majority here which may impact the selection of the most consistent metrics. In addition, even though we have already shown in \cite{QdsConference,QdsJournal,fine_tuned} that modularity density simultaneously solves two opposite yet coexisting problems of modularity, $Q_{ds}^{ov}$ is not highly consistent with majority of other metrics. This might be because only one other metric consider community density. We will investigate if adding more community density based metrics can change the above results.

\begin{table}
\centering
\caption{The best value of threshold $r$ for SLPA and the corresponding number of community quality metrics (out of twelve) that are consistent with each other on determining this best $r$ for the four combinations of two versions of belonging coefficient and two versions of belonging function on LFR benchmark networks with $(\alpha,\beta)=(1,2)$ and $\mu=0.3,0.35,0.4$. The best value in each row is marked by red italic font.}
\label{tab:lfr:slpa}       
\setlength{\tabcolsep}{5pt}
\begin{tabular}{c|c||c|c|c|c}
\hline \hline
\multicolumn{2}{c||}{} & \multicolumn{4}{c}{SLPA} \\
\hline
$\mu$ & $O_m$ & (BC,BF)=(1,1) & (BC,BF)=(1,2) & (BC,BF)=(2,1) & (BC,BF)=(2,2) \\
\hline
\multirow{5}{*}{0.3} & 1 & 0.5 (7) & \textcolor{red}{\textbf{\emph{0.5} (\emph{8})}} & 0.5 (7) & 0.5 (6) \\
\cline{2-6}
& 2 & \textcolor{red}{\textbf{\emph{0.5} (\emph{8})}} & 0.5 (5) & 0.5 (7) & 0.3 (4) \\
\cline{2-6}
& 4 & 0.5 (9) & \textcolor{red}{\textbf{\emph{0.5} (\emph{10})}} & 0.5 (8) & 0.5 (5) \\
\cline{2-6}
& 6 & 0.5 (9) & \textcolor{red}{\textbf{\emph{0.5} (\emph{10})}} & 0.5 (8) & 0.5 (6) \\
\cline{2-6}
& 8 & 0.5 (9) & \textcolor{red}{\textbf{\emph{0.5} (\emph{10})}} & 0.5 (8) & 0.5 (9) \\
\hline

\multirow{5}{*}{0.35} & 1 & 0.25 (5) & \textcolor{red}{\textbf{\emph{0.25} (\emph{6})}} & 0.25 (4) & 0.25 (5) \\
\cline{2-6}
& 2 & \textcolor{red}{\textbf{\emph{0.5} (\emph{7})}} & \textcolor{red}{\textbf{\emph{0.45} (\emph{7})}} & \textcolor{red}{\textbf{\emph{0.5} (\emph{7})}} & 0.45 (6) \\
\cline{2-6}
& 4 & 0.5 (8) & \textcolor{red}{\textbf{\emph{0.5} (\emph{9})}} & 0.5 (7) & 0.5 (6) \\
\cline{2-6}
& 6 & 0.5 (8) & \textcolor{red}{\textbf{\emph{0.5} (\emph{9})}} & 0.5 (7) & 0.5 (8) \\
\cline{2-6}
& 8 & 0.5 (7) & \textcolor{red}{\textbf{\emph{0.5} (\emph{9})}} & 0.5 (6) & 0.5 (8) \\
\hline

\multirow{5}{*}{0.4} & 1 & \textcolor{red}{\textbf{\emph{0.5} (\emph{6})}} & 0.25 (5) & 0.5 (4) & \{0.2,0.25,0.45\} (3) \\
\cline{2-6}
& 2 & 0.5 (8) & \textcolor{red}{\textbf{\emph{0.5} (\emph{9})}} & 0.5 (7) & 0.5 (5) \\
\cline{2-6}
& 4 & 0.5 (9) & \textcolor{red}{\textbf{\emph{0.5} (\emph{10})}} & 0.5 (8) & 0.5 (8) \\
\cline{2-6}
& 6 & 0.5 (9) & \textcolor{red}{\textbf{\emph{0.5} (\emph{10})}} & 0.5 (9) & 0.5 (9) \\
\cline{2-6}
& 8 & 0.5 (8) & \textcolor{red}{\textbf{\emph{0.5} (\emph{9})}} & 0.5 (7) & 0.5 (8) \\
\hline \hline
\end{tabular}
\end{table}

\subsection{LFR Benchmark Networks}
LFR (named after the initials of names of authors) benchmark networks \cite{LFROv} have become a standard in the evaluation of the performance of community detection algorithms. The LFR benchmark network that we used here has $1000$ nodes with average degree $15$ and maximum degree $50$. The exponent $\gamma$ for the degree sequence varies from $2$ to $3$. The exponent $\beta$ for the community size distribution ranges from $1$ to $2$. Then, four pairs of the exponents $(\gamma, \beta) = (2, 1), (2, 2), (3, 1), \text{and}~(3, 2)$ are chosen in order to explore the widest spectrum of graph structures. The mixing parameter $\mu$ is varied from $0.05$ to $0.95$. It means that each node shares a fraction $(1- \mu)$ of its edges with the other nodes in its community and shares a fraction $\mu$ of its edges with the nodes outside its community. Thus, low mixing parameters indicate strong community structure. The degree of overlap is determined by two parameters. $O_n$ is the number of overlapping nodes, and $O_m$ is the number of communities to which each overlapping node belongs. $O_n$ here is set to 10\% of the total number of nodes. Instead of fixing $O_m$, we allow it to vary from 1 to 8 indicating the overlapping diversity of overlapping nodes. By increasing the value of $O_m$, we create harder detection tasks. Also, we generate $10$ network instances for each configuration of these parameters. Hence, each metric value for a certain configuration of LFR represents the average metric values of all $10$ instances. Since the experimental results are similar for all four pairs of exponents $(\gamma, \beta) = (2, 1), (2, 2), (3, 1), \text{and}~(3, 2)$, for the sake of brevity, we only present the results for $(\gamma, \beta) = (2, 1)$ here. In addition, these results are similar for different values of $\mu$ and we cannot show all the results in one page as can be observed from Tables 24-26 in the Supplementary Materials, so here we only show the results for $\mu=0.3,0.35,\text{and~} 0.4$. We choose $\mu=0.3,0.35,\text{and~} 0.4$ to better illustrate the results since with $\mu=0.3,0.35,\text{and~} 0.4$ the community structures generated by LFR are around the boundary of well-separated communities and well-connected communities. For each node, $\mu=0.5$ means that the number of its edges with other nodes in its communities is equal to the number of its edges with nodes outside its community, which makes the community structure difficult to discover.

\begin{table}
\centering
\caption{The best value of parameter $k$ for CFinder and the corresponding number of community quality metrics (out of twelve) that are consistent with each other on determining this best $k$ for the four combinations of two versions of belonging coefficient and two versions of belonging function on LFR benchmark networks with $(\alpha,\beta)=(1,2)$ and $\mu=0.3,0.35,0.4$. The best value in each row is marked by red italic font.}
\label{tab:lfr:cfinder}       
\setlength{\tabcolsep}{5pt}
\begin{tabular}{c|c||c|c|c|c}
\hline \hline
\multicolumn{2}{c||}{} & \multicolumn{4}{c}{CFinder} \\
\hline
$\mu$ & $O_m$ & (BC,BF)=(1,1) & (BC,BF)=(1,2) & (BC,BF)=(2,1) & (BC,BF)=(2,2) \\
\hline
\multirow{5}{*}{0.3} & 1 & \textcolor{red}{\textbf{\emph{4} (\emph{9})}} & \textcolor{red}{\textbf{\emph{4} (\emph{9})}} & 4 (7) & 4 (6) \\
\cline{2-6}
& 2 & \{3,4\} (5) & \textcolor{red}{\textbf{\emph{4} (\emph{6})}} & 3 (5) & 4 (5) \\
\cline{2-6}
& 4 & \textcolor{red}{\textbf{\emph{3} (\emph{6})}} & \textcolor{red}{\textbf{\emph{3} (\emph{6})}} & 3 (4) & 3 (4) \\
\cline{2-6}
& 6 & 3 (5) & \textcolor{red}{\textbf{\emph{4} (\emph{7})}} & \{3,4,11\} (3) & 4 (4) \\
\cline{2-6}
& 8 & 3 (5) & \textcolor{red}{\textbf{\emph{4} (\emph{7})}} & \{3,4,12\} (3) & 4 (4) \\
\hline

\multirow{5}{*}{0.35} & 1 & \textcolor{red}{\textbf{\emph{4} (\emph{8})}} & \textcolor{red}{\textbf{\emph{4} (\emph{8})}} & 4 (6) & 4 (5) \\
\cline{2-6}
& 2 & \textcolor{red}{\textbf{\emph{3} (\emph{6})}} & \textcolor{red}{\textbf{\emph{3} (\emph{6})}} & \{3,8\} (4) & 4 (4) \\
\cline{2-6}
& 4 & \textcolor{red}{\textbf{\emph{3} (\emph{6})}} & \textcolor{red}{\textbf{\emph{4} (\emph{6})}} & 3 (4) & 4 (4) \\
\cline{2-6}
& 6 & 4 (5) & \textcolor{red}{\textbf{\emph{4} (\emph{7})}} & 4 (5) & 4 (5) \\
\cline{2-6}
& 8 & 4 (6) & \textcolor{red}{\textbf{\emph{4} (\emph{7})}} & 4 (4) & 4 (5) \\
\hline

\multirow{5}{*}{0.4} & 1 & \textcolor{red}{\textbf{\emph{4} (\emph{6})}} & \textcolor{red}{\textbf{\emph{3} (\emph{6})}} & \{3,4,8\} (4) & 4 (4) \\
\cline{2-6}
& 2 & \textcolor{red}{\textbf{\emph{3} (\emph{6})}} & \textcolor{red}{\textbf{\emph{3} (\emph{6})}} & \{3,9\} (4) & 4 (4) \\
\cline{2-6}
& 4 & \textcolor{red}{\textbf{\emph{3} (\emph{6})}} & \textcolor{red}{\textbf{\emph{4} (\emph{6})}} & \{3,9\} (4) & 4 (4) \\
\cline{2-6}
& 6 & 4 (7) & \textcolor{red}{\textbf{\emph{4} (\emph{8})}} & 4 (5) & 4 (5) \\
\cline{2-6}
& 8 & 4 (6) & \textcolor{red}{\textbf{\emph{4} (\emph{7})}} & 4 (5) & 4 (5) \\
\hline \hline
\end{tabular}
\end{table}

\begin{table}
\centering
\caption{The best value of threshold $tr$ for SpeakEasy and the corresponding number of quality metrics (out of twelve) that are consistent with each other on determining this best $tr$ for the four combinations of two versions of belonging coefficient and two versions of belonging function on LFR benchmark networks with $(\alpha,\beta)=(1,2)$ and $\mu=0.3,0.35,0.4$. The best value in each row is marked by red italic font.}
\label{tab:lfr:speak_easy}       
\setlength{\tabcolsep}{4pt}
\begin{tabular}{c|c||c|c|c|c}
\hline \hline
\multicolumn{2}{c||}{} & \multicolumn{4}{c}{SpeakEasy} \\
\hline
$\mu$ & $O_m$ & (BC,BF)=(1,1) & (BC,BF)=(1,2) & (BC,BF)=(2,1) & (BC,BF)=(2,2) \\
\hline
\multirow{5}{*}{0.3} & 1 & \textcolor{red}{\textbf{\emph{0.75} (\emph{9})}} & \textcolor{red}{\textbf{\emph{0.75} (\emph{9})}} & \textcolor{red}{\textbf{\emph{0.75} (\emph{9})}} & \textcolor{red}{\textbf{\emph{0.75} (\emph{9})}} \\
\cline{2-6}
& 2 & 0.8 (7) & \textcolor{red}{\textbf{\emph{0.8} (\emph{8})}} & 0.8 (6) & 0.8 (7) \\
\cline{2-6}
& 4 & \textcolor{red}{\textbf{\emph{1} (\emph{4})}} & \textcolor{red}{\textbf{\emph{1} (\emph{4})}} & \textcolor{red}{\textbf{\emph{1} (\emph{4})}} & \textcolor{red}{\textbf{\emph{1} (\emph{4})}} \\
\cline{2-6}
& 6 & \textcolor{red}{\textbf{\emph{0.95} (\emph{4})}} & \textcolor{red}{\textbf{\emph{0.95} (\emph{4})}} & 0.05 (3) & \textcolor{red}{\textbf{\{\emph{0.35},\emph{0.7}\} (\emph{4})}} \\
\cline{2-6}
& 8 & 0.05 (3) & \textcolor{red}{\textbf{\{\emph{0.4},\emph{0.6}\} (\emph{4})}} & \{0.05,0.85\} (3) & \{0.4,0.85\} (3) \\
\hline

\multirow{5}{*}{0.35} & 1 & \textcolor{red}{\textbf{\emph{0.8} (\emph{9})}} & \textcolor{red}{\textbf{\emph{0.8} (\emph{9})}} & \textcolor{red}{\textbf{\emph{0.8} (\emph{9})}} & \textcolor{red}{\textbf{\emph{0.8} (\emph{9})}} \\
\cline{2-6}
& 2 & 0.95 (5) & \textcolor{red}{\textbf{\emph{0.95} (\emph{6})}} & 0.95 (4) & \{0.35,0.95\} (3) \\
\cline{2-6}
& 4 & \{0.05,0.95\} (3) & \textcolor{red}{\textbf{\emph{0.75} (\emph{4})}} & \{0.05,0.95\} (3) & 0.95 (3) \\
\cline{2-6}
& 6 & \textcolor{red}{\textbf{\emph{0.85} (\emph{5})}} & 0.85 (4) & 0.85 (4) & 0.45 (3) \\
\cline{2-6}
& 8 & \{0.05,0.85\} (3) & \textcolor{red}{\textbf{\emph{0.9} (\emph{5})}} & \{0.05,0.85,1\} (3) & \{0.85,0.9\} (3) \\
\hline

\multirow{5}{*}{0.4} & 1 & \textcolor{red}{\textbf{\emph{0.15} (\emph{9})}} & \textcolor{red}{\textbf{\emph{0.15} (\emph{9})}} & \textcolor{red}{\textbf{\emph{0.15} (\emph{9})}} & \textcolor{red}{\textbf{\emph{0.15} (\emph{9})}} \\
\cline{2-6}
& 2 & \{0.8,0.85,0.95\} (3) & 0.95 (4) & \{0.05,0.85,0.95\} (3) & \textcolor{red}{\textbf{\emph{0.95} (\emph{5})}} \\
\cline{2-6}
& 4 & 0.95 (5) & \textcolor{red}{\textbf{\emph{0.95} (\emph{6})}} & 0.95 (4) & 0.95 (4) \\
\cline{2-6}
& 6 & 0.95 (4) & \{0.65,0.75\} (3) & 0.75 (4) & \textcolor{red}{\textbf{\emph{0.75} (\emph{6})}} \\
\cline{2-6}
& 8 & \{0.05,0.9,1\} (3) & \textcolor{red}{\textbf{\emph{0.4} (\emph{5})}} & \{0.05,0.95,1\} (3) & \{0.4,0.5\} (3) \\
\hline \hline
\end{tabular}
\end{table}

Tables~\ref{tab:lfr:slpa}-\ref{tab:lfr:speak_easy} show the best value of threshold $r$ for SLPA, the best value of parameter $k$ for CFinder, and the best value of threshold $tr$ for SpeakEasy, respectively, along with the corresponding number of community quality metrics (out of twelve) that are consistent with each other on determining this best $r$, this best $k$, and this best $tr$ for the four possible combinations of two versions of belonging coefficient and two versions of belonging function on LFR benchmark networks with $(\alpha,\beta)=(1,2)$ and $\mu=0.3,0.35,0.4$. Table~\ref{tab:lfr:slpa} implies that (BC,BF)=(1,2) is the best among the four possible combinations on all configurations of LFR networks except $\mu=0.3,O_m=2$ and $\mu=0.4,O_m=1$ when using SLPA. Table~\ref{tab:lfr:cfinder} demonstrates that (BC,BF)=(1,2) is the best on all configurations of LFR networks when using CFinder. Table~\ref{tab:lfr:speak_easy} indicates that (BC,BF)=(1,2) is the best among the four combinations on all configurations of LFR networks except $\mu=0.35,O_m=6$ and $\mu=0.4,O_m=2,4$ when using SpeakEasy. Consequently, we could conclude that the overlapping community quality metrics with the first version of belonging coefficient and the second version of the belonging function are the best among the four possible combinations on LFR networks. Please refer to Tables 24-26 in Supplementary Materials for more detailed results on LFR benchmark networks.

\begin{table}
\centering
\caption{The number of times (maximum three since there are totally three community detection algorithms adopted) that the community quality metric is among those consistent with each other on determining the best value of threshold $r$ for SLPA, the best value of parameter $k$ for CFinder, and the best value of threshold $tr$ for SpeakEasy on each configuration of LFR benchmark networks with $(\alpha,\beta)=(1,2)$ and $\mu=0.3,0.35,0.4$ when (BC,BF)=(1,2). The best value in the last row is marked by red italic font.}
\label{tab:lfr:all_metrics_1_2}       
\setlength{\tabcolsep}{1.5pt}
\begin{tabular}{c|c||c|c|c|c|c|c|c|c|c|c|c|c}
\hline \hline
$\mu$ & $O_m$ & $Q_{ov}$ & $NQ_{ov}$ & $Q_{ov}^L$ & $Q_{ds}^{ov}$ & $IE$ & $ID$ & $CNT$ & $BE$ & $EXP$ & $CND$ & $F$ & $D$ \\
\hline
\multirow{5}{*}{0.3} &	1	& 3	& 2	& 3	& 3	& 1	& 0	& 2	& 0	& 3	& 3	& 3	& 3 \\
\cline{2-14}
&	2	& 3	& 1	& 3	& 2	& 1	& 1	& 2	& 0	& 2	& 1	& 1	& 2 \\
\cline{2-14}
&	4	& 1	& 2	& 1	& 1	& 1	& 3	& 2	& 2	& 1	& 2	& 2	& 2 \\
\cline{2-14}
&	6	& 2	& 1	& 2	& 2	& 0	& 1	& 2	& 0	& 3	& 3	& 3	& 2 \\
\cline{2-14}
&	8	& 3	& 1	& 3	& 2	& 1	& 1	& 3	& 0	& 3	& 3	& 3	& 2 \\
\hline
\multirow{5}{*}{0.35}	& 1	& 3	& 1	& 3	& 2	& 1	& 0	& 2	& 0	& 3	& 2	& 3	& 3 \\
\cline{2-14}
&	2	& 2	& 0	& 2	& 1	& 1	& 2	& 2	& 1	& 1	& 2	& 2	& 3 \\
\cline{2-14}
&	4	& 2	& 2	& 2	& 2	& 0	& 0	& 3	& 0	& 2	& 2	& 2	& 2 \\
\cline{2-14}
&	6	& 2	& 1	& 2	& 2	& 0	& 0	& 2	& 0	& 3	& 3	& 3	& 2 \\
\cline{2-14}
&	8	& 3	& 1	& 3	& 2	& 1	& 0	& 2	& 0	& 3	& 2	& 2	& 2 \\
\hline
\multirow{5}{*}{0.4}	& 1	& 1	& 0	& 1	& 1	& 2	& 1	& 3	& 0	& 2	& 3	& 3	& 3 \\
\cline{2-14}
&	2	& 2	& 1	& 2	& 1	& 1	& 2	& 2	& 0	& 1	& 2	& 2	& 3 \\
\cline{2-14}
&	4	& 3	& 2	& 3	& 2	& 0	& 1	& 3	& 0	& 2	& 2	& 2	& 2 \\
\cline{2-14}
&	6	& 3	& 2	& 3	& 3	& 0	& 1	& 3	& 0	& 2	& 3	& 3	& 1 \\
\cline{2-14}
&	8	& 2	& 1	& 2	& 2	& 1	& 0	&3	& 0	& 2	& 3	& 3	& 2 \\
\hline
\multicolumn{2}{c||}{} &	\shortstack{~~~ \\ 7 \\ (2.33)}	& \shortstack{0 \\ (1.2)}	& \shortstack{7 \\ (2.33)}	& \shortstack{2 \\ (1.87)}	& \shortstack{0 \\ (0.73)}	& \shortstack{1 \\ (0.87)}	& \shortstack{6 \\ (2.4)}	& \shortstack{0 \\ (0.2)}	& \shortstack{6	\\ (2.2)} & \shortstack{7 \\ (2.4)}	& \shortstack{\textcolor{red}{\textbf{\emph{8}}} \\ \textcolor{red}{\textbf{(\emph{2.47})}}}	& \shortstack{5 \\ (2.27)} \\ [1ex]
\hline \hline
\end{tabular}
\end{table}

Table~\ref{tab:lfr:all_metrics_1_2} shows the number of times (maximum three since there are totally three community detection algorithms adopted) that the community quality metric is among the metrics that are consistent with each other on determining the best value of threshold $r$ for SLPA, the best value of parameter $k$ for CFinder, and the best value of threshold $tr$ for SpeakEasy on each configuration of LFR benchmark networks with $(\alpha,\beta)=(1,2)$ and $\mu=0.3,0.35,0.4$ when (BC,BF)=(1,2). The last row in the table presents how many different configurations of LFR networks support this metric across all three algorithms and how many times on average (maximum three) this metric got support from each of the 15 different configurations. This table shows results similar with those presented in Table~\ref{tab:real_networks:all_metrics_1_2}. Node-based overlapping extension and edge-based overlapping extension of modularity perform equally well on LFR networks. The last row shows that the fitness function has the largest values and generally local metrics are better than global metrics, and still $Q_{ds}^{ov}$ performs poor. The reason is the same as given for the results in Table~\ref{tab:real_networks:all_metrics_1_2}.

\section{Conclusion}
\label{sec:conclusion}
In this paper, we determined which versions of the belonging coefficient and the belonging function are better for measuring quality of fuzzy overlapping community structures. We found that the first version of the belonging coefficient is better than the second one, which means that the coefficient of a node belonging to a community should be the reciprocal of the number of communities to which this node belongs. In addition, we found that the second version of the belonging function is better than the first version, meaning that the probability that two nodes belong to the same community should be the product, not the average, of their belonging coefficients. Moreover, we proposed overlapping extensions for localized modularity, modularity density, and eight local community quality metrics analogous to such extension of modularity. Based on the experimental results, we recommend using the edge-based overlapping extension of modularity with the first version of belonging coefficient and with its own belonging function. We also recommend using the node-based overlapping extension of modularity and overlapping extension of modularity density with the first version of belonging coefficient and the second version of belonging function as the metrics of the global quality of overlapping community structures.

In the future, we plan to explore local community quality metrics for overlapping community structures and investigate more community quality metrics incorporating community density to see whether putting community density into these metrics will make them perform better.

\section*{Acknowledgment}
This work was supported in part by the ARL under Cooperative Agreement W911NF-09-2-0053, the Office of Naval Research Grant N00014-09-1-0607 and by the EU's 7FP Grant Agreement 316097. The views and conclusions contained in this paper are those of the authors and should not be interpreted as representing the official policies either expressed or implied of the Army Research Laboratory or the U.S. Government.





\begin{thebibliography}{10}
\providecommand{\url}[1]{{#1}}
\providecommand{\urlprefix}{URL }
\expandafter\ifx\csname urlstyle\endcsname\relax
  \providecommand{\doi}[1]{DOI~\discretionary{}{}{}#1}\else
  \providecommand{\doi}{DOI~\discretionary{}{}{}\begingroup
  \urlstyle{rm}\Url}\fi

\bibitem{Polblogs}
Adamic, L.A., Glance, N.: The political blogosphere and the 2004 u.s. election:
  Divided they blog.
\newblock In: Proceedings of the 3rd International Workshop on Link Discovery,
  LinkKDD '05, pp. 36--43. ACM, New York, NY, USA (2005)

\bibitem{PGPNetwork}
Bogu\~{n}\'{a}, M., Pastor-Satorras, R., D\'{i}az-Guilera, A., Arenas, A.:
  Models of social networks based on social distance attachment.
\newblock Phys. Rev. E \textbf{70}, 056,122 (2004)

\bibitem{Railway}
Chakraborty, T., Srinivasan, S., Ganguly, N., Mukherjee, A., Bhowmick, S.: On
  the permanence of vertices in network communities.
\newblock In: Proceedings of the 20th ACM SIGKDD International Conference on
  Knowledge Discovery and Data Mining, KDD '14, pp. 1396--1405. ACM, New York,
  NY, USA (2014)

\bibitem{Qov_node_strength}
Chen, D., Shang, M., Lv, Z., Fu, Y.: Detecting overlapping communities of
  weighted networks via a local algorithm.
\newblock Physica A: Statistical Mechanics and its Applications
  \textbf{389}(19), 4177 -- 4187 (2010)

\bibitem{fine_tuned}
Chen, M., Kuzmin, K., Szymanski, B.: Community detection via maximization of
  modularity and its variants.
\newblock IEEE Transactions on Computational Social Systems \textbf{1}(1),
  46--65 (2014)

\bibitem{QdsOv}
Chen, M., Kuzmin, K., Szymanski, B.: Extension of modularity density for
  overlapping community structure.
\newblock In: Advances in Social Networks Analysis and Mining (ASONAM), 2014
  IEEE/ACM International Conference on, pp. 856--863 (2014)

\bibitem{QdsJournal}
Chen, M., Nguyen, T., Szymanski, B.K.: A new metric for quality of network
  community structure.
\newblock ASE Human Journal \textbf{2}(4), 226--240 (2013)

\bibitem{QdsConference}
Chen, M., Nguyen, T., Szymanski, B.K.: On measuring the quality of a network
  community structure.
\newblock In: Proceedings of ASE/IEEE International Conference on Social
  Computing, pp. 122--127. Washington, DC, USA (2013)

\bibitem{Collins}
Collins, S.R., Kemmeren, K.P., chu Zhao, F.X., Greenblatt, G.J.F., Spencer, F.,
  et~al.: Toward a comprehensive atlas of the physical interactome of
  saccharomyces cerevisiae.
\newblock Molecular \& Cellular Proteomic \textbf{6}, 439--450 (2007)

\bibitem{ExtremalQ}
Duch, J., Arenas, A.: Community detection in complex networks using extremal
  optimization.
\newblock Phys. Rev. E \textbf{72}, 027,104 (2005).
\newblock \doi{10.1103/PhysRevE.72.027104}

\bibitem{CommunityReport}
Fortunato, S.: Community detection in graphs.
\newblock Physics Reports \textbf{486}, 75--174 (2010).
\newblock \doi{10.1016/j.physrep.2009.11.002}

\bibitem{ResolutionLimit}
Fortunato, S., Barth\'{e}lemy, M.: Resolution limit in community detection.
\newblock Proceedings of the National Academy of Sciences \textbf{104}(1),
  36--41 (2007).
\newblock \doi{10.1073/pnas.0605965104}

\bibitem{SpeakEasy}
Gaiteri, C., Chen, M., Szymanski, B.K., Kuzmin, K., Xie, J., Lee, C., Blanche,
  T., Neto, E.C., Huang, S.C., Grabowski, T., Madhyastha, T., Komashko, V.:
  Identifying robust clusters and multi-community nodes by combining top-down
  and bottom-up approaches to clustering.
\newblock http://arxiv.org/abs/1501.04709  (2015)

\bibitem{Gavin}
Gavin, A.C., Aloy, P., Grandi, P., et~al.: Proteome survey reveals modularity
  of the yeast cell machinery.
\newblock Nature \textbf{440}(7084), 631--636 (2006)

\bibitem{football}
Girvan, M., Newman, M.E.J.: Community structure in social and biological
  networks.
\newblock Proceedings of the National Academy of Sciences \textbf{99}(12),
  7821--7826 (2002)

\bibitem{Jazz}
Gleiser, P., Danon, L.: Community structure in jazz.
\newblock Advances in Complex Systems \textbf{06}(04), 565--573 (2003)

\bibitem{ov_lpa}
Gregory, S.: Finding overlapping communities in networks by label propagation.
\newblock New Journal of Physics \textbf{12}(10), 103,018 (2010)

\bibitem{OverlappingQ_survey}
Gregory, S.: Fuzzy overlapping communities in networks.
\newblock Journal of Statistical Mechanics: Theory and Experiment
  \textbf{2011}(02), P02,017 (2011)

\bibitem{Email}
Guimer\`{a}, R., Danon, L., D\'{i}az\-Guilera, A., Giralt, F., Arenas, A.:
  Self-similar community structure in a network of human interactions.
\newblock Phys. Rev. E \textbf{68}, 065,103 (2003)

\bibitem{SGD}
Hong, E.L., Balakrishnan, R., Dong, Q., et~al.: Gene ontology annotations at
  \uppercase{SGD}: new data sources and annotation methods.
\newblock Nucleic Acids Research \textbf{36}, D577–--D581 (2008)

\bibitem{Lesmis}
Knuth, D.E.: The Stanford GraphBase: A Platform for Combinatorial Computing.
\newblock Addison-Wesley, Reading, MA (1993)

\bibitem{Polbooks}
Krebs, V.: http://www.orgnet.com/

\bibitem{LFROv}
Lancichinetti, A., Fortunato, S.: Benchmarks for testing community detection
  algorithms on directed and weighted graphs with overlapping communities.
\newblock Phys. Rev. E \textbf{80}, 016,118 (2009)

\bibitem{Fitness}
Lancichinetti, A., Fortunato, S., Kert\'{e}sz, J.: Detecting the overlapping
  and hierarchical community structure in complex networks.
\newblock New Journal of Physics \textbf{11}(3), 033,015 (2009)

\bibitem{AverageModularityDegree}
Li, Z., Zhang, S., Wang, R.S., Zhang, X.S., Chen, L.: Quantitative function for
  community detection.
\newblock Phys. Rev. E \textbf{77}, 036,109 (2008)

\bibitem{Dolphin}
Lusseau, D., Schneider, K., Boisseau, O., Haase, P., Slooten, E., Dawson, S.:
  The bottlenose dolphin community of doubtful sound features a large
  proportion of long-lasting associations.
\newblock Behavioral Ecology and Sociobiology \textbf{54}(4), 396--405 (2003)

\bibitem{MIPS}
Mewes, H.W., Amid, C., Arnold, R., et~al.: \uppercase{MIPS}: analysis and
  annotation of proteins from whole genomes.
\newblock Nucleic Acids Res. \textbf{32}, D41--D44 (2004)

\bibitem{NeighboringModularity}
Muff, S., Rao, F., Caflisch, A.: Local modularity measure for network
  clusterizations.
\newblock Phys. Rev. E \textbf{72}, 056,107 (2005)

\bibitem{fuzzy_opt_Qov}
Nepusz, T., Petr\'oczi, A., N\'egyessy, L., Bazs\'o, F.: Fuzzy communities and
  the concept of bridgeness in complex networks.
\newblock Phys. Rev. E \textbf{77}, 016,107 (2008)

\bibitem{NewmanGreedy}
Newman, M.E.J.: Fast algorithm for detecting community structure in networks.
\newblock Phys. Rev. E \textbf{69}, 066,133 (2004).
\newblock \doi{10.1103/PhysRevE.69.066133}

\bibitem{EigenvectorCommunity}
Newman, M.E.J.: Finding community structure in networks using the eigenvectors
  of matrices.
\newblock Phys. Rev. E \textbf{74}, 036,104 (2006)

\bibitem{PNASModularity}
Newman, M.E.J.: Modularity and community structure in networks.
\newblock Proceedings of the National Academy of Sciences \textbf{103}(23),
  8577--8582 (2006).
\newblock \doi{10.1073/pnas.0601602103}

\bibitem{UWDModularity}
Newman, M.E.J., Girvan, M.: Finding and evaluating community structure in
  networks.
\newblock Phys. Rev. E \textbf{69}, 026,113 (2004).
\newblock \doi{10.1103/PhysRevE.69.026113}

\bibitem{Qov_edge_degree}
Nicosia, V., Mangioni, G., Carchiolo, V., Malgeri, M.: Extending the definition
  of modularity to directed graphs with overlapping communities.
\newblock Journal of Statistical Mechanics: Theory and Experiment
  \textbf{2009}(03), P03,024 (2009)

\bibitem{CPM}
Palla, G., Der\'{e}nyi, I., Farkas, I., Vicsek, T.: Uncovering the overlapping
  community structure of complex networks in nature and society.
\newblock Nature \textbf{435}(7043), 814--818 (2005).
\newblock \doi{10.1038/nature03607}

\bibitem{CYC}
Pu, S., Wong, J., Turner, B., Cho, E., Wodak, S.J.: Up-to-date catalogues of
  yeast protein complexes.
\newblock Nucleic Acids Research \textbf{37}(3), 825--831 (2009)

\bibitem{Qov_num_coms}
Shen, H., Cheng, X., Cai, K., Hu, M.B.: Detect overlapping and hierarchical
  community structure in networks.
\newblock Physica A: Statistical Mechanics and its Applications
  \textbf{388}(8), 1706 -- 1712 (2009)

\bibitem{Qov_max_clique}
Shen, H.W., Cheng, X.Q., Guo, J.F.: Quantifying and identifying the overlapping
  community structure in networks.
\newblock Journal of Statistical Mechanics: Theory and Experiment
  \textbf{2009}(07), P07,042 (2009)

\bibitem{Overlapping_survey}
Xie, J., Kelley, S., Szymanski, B.K.: Overlapping community detection in
  networks: The state-of-the-art and comparative study.
\newblock ACM Comput. Surv. \textbf{45}(4), 43:1--43:35 (2013)

\bibitem{SLPA2012}
Xie, J., Szymanski, B.K.: Towards linear time overlapping community detection
  in social networks.
\newblock In: The 16th Pacific-Asia Conference on Knowledge Discovery and Data
  Mining (PAKDD), pp. 25--36 (2012)

\bibitem{Leskovec_dataset}
Yang, J., Leskovec, J.: Defining and evaluating network communities based on
  ground-truth.
\newblock In: Proceedings of the ACM SIGKDD Workshop on Mining Data Semantics,
  MDS '12, pp. 3:1--3:8. ACM, New York, NY, USA (2012)

\bibitem{karate}
Zachary, W.: An information flow model for conflict and fission in small
  groups.
\newblock Journal of Anthropological Research \textbf{33}, 452--473 (1977)

\bibitem{fuzzy_cmeans_Qov}
Zhang, S., Wang, R.S., Zhang, X.S.: Identification of overlapping community
  structure in complex networks using fuzzy c-means clustering.
\newblock Physica A: Statistical Mechanics and its Applications
  \textbf{374}(1), 483 -- 490 (2007)

\end{thebibliography}


\end{document}


\mainmatter  

\title{Supplementary Materials\\Fuzzy Overlapping Community Quality Metrics}


%
%
\author{Mingming Chen \and Boleslaw K. Szymanski}
%

\institute{Department of Computer Science \\ Rensselaer Polytechnic Institute \\ 110 8th Street, Troy, NY 12180
\\ chenm8@rpi.edu and szymab@rpi.edu
}

\maketitle

\section{Real Network Datasets}
\subsection{C. elegans Metabolic Network}
This is the metabolic network of C. elegans \cite{ExtremalQ} with 453 nodes and 2025 edges. Table~\ref{tab:celegans:global} shows the best value of threshold $r$ for SLPA \cite{SLPA2012}, the best value of parameter $k$ for CFinder \cite{CPM,CPMw}, and the best value of threshold $tr$ for SpeakEasy \cite{SpeakEasy} determined by the twelve community quality metrics with four possible combinations of the two versions of belonging coefficient and two version of belonging function for this network. The last column in this table (and all the following tables) is the best value of threshold $r$ for SLPA, the best value of parameter $k$ for CFinder, or the best value of the threshold $tr$ for SpeakEasy along with the corresponding number of community quality metrics (out of twelve) that are consistent with each other on determining this best $r$, this best $k$, and this best $tr$ for each combination of belonging coefficient and belonging function. The table shows that the first version of belonging function is better than the second version of belonging function when using SLPA. For CFinder and SpeakEasy, it implies that (BC,BF)=(1,2) is the best among the four possible combinations of two versions of belonging coefficient and two versions of belonging function. In conclusion, two out of three algorithms show that (BC,BF)=(1,2) is the best on C. elegans metabolic network.

\begin{table}
\vspace{1em}
\centering
\caption{The best value of threshold $r$ for SLPA, the best value of parameter $k$ for CFinder, and the best value of threshold $tr$ for SpeakEasy determined by the twelve community quality metrics with four possible combinations of the two versions of belonging coefficient and two version of belonging function on C. elegans metabolic network. The best value for subcolumn of the last column is marked by red italic font.}
\label{tab:celegans:global}       
\vspace{-0.8em}
\setlength{\tabcolsep}{3pt}
\begin{tabular}{c||c|c|c|c|c|c|c|c|c|c|c|c|c||c}
\hline \hline
Algorithm & (BC,BF) & $Q_{ov}$ & $NQ_{ov}$ & $Q_{ov}^L$ & $Q_{ds}^{ov}$ & $IE$ & $ID$ & $CNT$ & $BE$ & $EXP$ & $CND$ & $F$ & $D$ & \\
\hline
\multirow{4}{*}{SLPA} & (1,1) & 0.5	& 0.3	& 0.3	& 0.35	& 0.05	& 0.5	& 0.05	& 0.5	& 0.5	& 0.5	& 0.05	& 0.5 & \textcolor{red}{\textbf{\emph{0.5} (\emph{6})}} \\
\cline{2-15}
& (1,2) & 0.3	& 0.3	& 0.3	& 0.4	& 0.05	& 0.5	& 0.05	& 0.5	& 0.4	& 0.05	& 0.05	& 0.4 & 0.05 (4) \\
\cline{2-15}
& (2,1) & 0.5	& 0.3	& 0.45	& 0.4	& 0.05	& 0.5	& 0.05	& 0.5	& 0.5	& 0.5	& 0.05	& 0.5 & \textcolor{red}{\textbf{\emph{0.5} (\emph{6})}} \\
\cline{2-15}
& (2,2) & 0.4	& 0.4	& 0.45	& 0.4	& 0.05	& 0.35	& 0.05	& 0.35	& 0.4	& 0.05	& 0.05	& 0.4 & 0.4 (5) \\
\cline{1-15} \cline{1-15}

\multirow{4}{*}{CFinder} & (1,1) & 9	& 3	& 4	& 3	& 3	& 4	& 4	& 7	& 7	& 3	& 3	& 3 & 3 (6)\\
\cline{2-15}
& (1,2) & 4	& 3	& 4	& 4	& 3	& 4	& 3	& 7	& 3	& 3	& 3	& 3 & \textcolor{red}{\textbf{\emph{3} (\emph{7})}} \\
\cline{2-15}
& (2,1) & 7	& 3	& 4	& 3	& 3	& 4	& 4	& 9	& 9	& 3	& 3	& 9 & 3 (5) \\
\cline{2-15}
& (2,2) & 5	& 9	& 4	& 4	& 3	& 4	& 3	& 9	& 9	& 9	& 3	& 3 & \{3,9\} (4)\\
\cline{1-15} \cline{1-15}

\multirow{4}{*}{SpeakEasy} & (1,1) & 0.75	& 0.8	& 0.75	& 0.75	& 0.05	& 0.8	& 0.05	& 1	& 0.75	& 0.9 & 0.75 & 0.75 & 0.75 (6) \\
\cline{2-15}
& (1,2) & 0.75	& 0.8	& 0.75	& 0.75 & 0.75	& 0.8	& 0.9	& 0.35	& 0.75	& 0.9	& 0.75	& 0.75 & \textcolor{red}{\textbf{\emph{0.75} (\emph{7})}} \\
\cline{2-15}
& (2,1) & 0.75	& 0.8	& 0.75	& 0.75	& 0.05	& 0.8	& 0.05	& 1	& 0.75	& 0.9	& 0.75	& 0.75 & 0.75 (6) \\
\cline{2-15}
& (2,2) & 0.75	& 0.8	& 0.75	& 0.75	& 0.75	& 0.8	& 0.9	& 0.35	& 0.75	& 0.5	& 0.9	& 0.75 & 0.75 (6) \\
\hline \hline
\end{tabular}
\end{table}

\begin{table}
\centering
\caption{The best value of threshold $r$ for SLPA, the best value of parameter $k$ for CFinder, and the best value of threshold $tr$ for SpeakEasy determined by the twelve community quality metrics with four possible combinations of the two versions of belonging coefficient and two version of belonging function on dolphin social network. The best value for subcolumn of the last column is marked by red italic font.}
\label{tab:dolphin:global}       
\vspace{-0.8em}
\setlength{\tabcolsep}{2.5pt}
\begin{tabular}{c||c|c|c|c|c|c|c|c|c|c|c|c|c||c}
\hline \hline
Algorithm & (BC,BF) & $Q_{ov}$ & $NQ_{ov}$ & $Q_{ov}^L$ & $Q_{ds}^{ov}$ & $IE$ & $ID$ & $CNT$ & $BE$ & $EXP$ & $CND$ & $F$ & $D$ & \\
\hline
\multirow{4}{*}{SLPA} & (1,1) & 0.5	& 0.4	& 0.4	& 0.4 & 0.05	& 0.5	& 0.05	& 0.5	& 0.5	& 0.45	& 0.5	& 0.45 & 0.5 (5) \\
\cline{2-15}
& (1,2) & 0.4	& 0.4	& 0.4	& 0.4	& 0.05	& 0.4	& 0.05	& 0.5	& 0.4	& 0.4	& 0.05	& 0.4 & \textcolor{red}{\textbf{\emph{0.4} (\emph{8})}} \\
\cline{2-15}
& (2,1) & 0.5	& 0.4	& 0.05	& 0.4	& 0.05	& 0.5	& 0.05	& 0.5	& 0.5	& 0.5	& 0.5	& 0.45 & 0.5 (6) \\
\cline{2-15}
& (2,2) & 0.4	& 0.4	& 0.05	& 0.4	& 0.05	& 0.3	& 0.05	& 0.3	& 0.4	& 0.05	& 0.05	& 0.4 & \{0.05,0.4\} (5)\\
\cline{1-15} \cline{1-15}

\multirow{4}{*}{CFinder} & (1,1) & 3	& 3	& 3	& 3	& 3	& 3	& 3	& 4	& 3	& 3	& 3	& 3 & 3 (11) \\
\cline{2-15}
& (1,2) & 3	& 3	& 3	& 3& 3	& 3	& 3	& 3	& 3	& 3	& 3	& 3 & \textcolor{red}{\textbf{\emph{3} (\emph{12})}} \\
\cline{2-15}
& (2,1) & 3	& 3	& 3	& 4	& 3	& 3	& 3	& 5	& 5	& 5	& 3	& 4 & 3 (7) \\
\cline{2-15}
& (2,2) & 3	& 3	& 3	& 4	& 3	& 3	& 3	& 5	& 5	& 5	& 3	& 4 & 3 (7) \\
\cline{1-15} \cline{1-15}

\multirow{4}{*}{SpeakEasy} & (1,1) & 0.4	& 0.55	& 0.4	& 0.2	& 0.15	& 0.85	& 0.05	& 0.7	& 1	& 0.45	& 0.45	& 0.4 & 0.4 (3) \\
\cline{2-15}
& (1,2) & 0.4	& 0.2	& 0.4	& 0.2	& 0.4	& 0.85	& 0.45	& 0.7	& 1	& 0.45	& 0.45	& 0.4 & \textcolor{red}{\textbf{\emph{0.4} (\emph{4})}} \\
\cline{2-15}
& (2,1) & 0.15	& 0.8	& 0.15	& 0.8	& 0.15	& 0.85	& 0.15	& 0.7	& 0.7	& 0.45	& 0.45	& 0.7  & \textcolor{red}{\textbf{\emph{0.15} (\emph{4})}} \\
\cline{2-15}
& (2,2) & 0.4	& 0.8	& 0.15	& 0.8	& 0.4	& 0.85	& 0.45	& 0.7	& 0.7	& 0.7	& 0.45	& 0.7 & \textcolor{red}{\textbf{\emph{0.7} (\emph{4})}} \\
\hline \hline
\end{tabular}
\vspace{1.5em}
\end{table}

\vspace{1.5em}
\subsection{Dolphin Social Network}
This is a social network of frequent associations between 62 dolphins in a community living off Doubtful Sound, New Zealand \cite{Dolphin}. There are 62 nodes and 159 edges. Table~\ref{tab:dolphin:global} shows the best value of threshold $r$ for SLPA, the best value of parameter $k$ for CFinder, and the best value of threshold $tr$ for SpeakEasy determined by the twelve community quality metrics with four possible combinations of the two versions of belonging coefficient and two version of belonging function on dolphin social network. We could learn from the table that all the three algorithms demonstrate that (BC,BF)=(1,2) is the best among the four combinations of belonging coefficient and belonging function.

\vspace{2.5em}
\subsection{Email network}
This network represents email interchanges between members of the Univeristy Rovira i Virgili (Tarragona) \cite{Email}. It has 1133 nodes and 5451 edges. Table~\ref{tab:email:global} shows the best value of threshold $r$ for SLPA, the best value of parameter $k$ for CFinder, and the best value of threshold $tr$ for SpeakEasy determined by the twelve community quality metrics with four possible combinations of the two versions of belonging coefficient and two version of belonging function on email network. It can be observed from this table that all three algorithms demonstrate that (BC,BF)=(1,2) is the best among the four combinations of belonging coefficient and belonging function.

\begin{table}
\vspace{1em}
\centering
\caption{The best value of threshold $r$ for SLPA, the best value of parameter $k$ for CFinder, and the best value of threshold $tr$ for SpeakEasy determined by the twelve community quality metrics with four possible combinations of the two versions of belonging coefficient and two version of belonging function on email network. The best value for subcolumn of the last column is marked by red italic font.}
\label{tab:email:global}       
\vspace{-0.8em}
\setlength{\tabcolsep}{2.5pt}
\begin{tabular}{c||c|c|c|c|c|c|c|c|c|c|c|c|c||c}
\hline \hline
Algorithm & (BC,BF) & $Q_{ov}$ & $NQ_{ov}$ & $Q_{ov}^L$ & $Q_{ds}^{ov}$ & $IE$ & $ID$ & $CNT$ & $BE$ & $EXP$ & $CND$ & $F$ & $D$ & \\
\hline
\multirow{4}{*}{SLPA} & (1,1) & 0.5	& 0.5	& 0.5	& 0.35	& 0.1	& 0.4	& 0.05	& 0.5	& 0.5	& 0.5	& 0.5	& 0.5 & 0.5 (8) \\
\cline{2-15}
& (1,2) & 0.5	& 0.5	& 0.5	& 0.5	& 0.5	& 0.4	& 0.5	& 0.35	& 0.5	& 0.5	& 0.5	& 0.5 & \textcolor{red}{\textbf{\emph{0.5} (\emph{10})}} \\
\cline{2-15}
& (2,1) & 0.5	& 0.5	& 0.3	& 0.4	& 0.1	& 0.4	& 0.05	& 0.5	& 0.5	& 0.5	& 0.5	& 0.5 & 0.5 (7) \\
\cline{2-15}
& (2,2) & 0.5	& 0.5	& 0.3	& 0.5	& 0.5	& 0.4	& 0.5	& 0.35	& 0.4	& 0.5	& 0.5	& 0.5 & 0.5 (8) \\
\cline{1-15} \cline{1-15}

\multirow{4}{*}{CFinder} & (1,1) & 4	& 3	& 3	& 3	& 3	& 3	& 4	& 3	& 3	& 3	& 3	& 3 & \textcolor{red}{\textbf{\emph{3} (\emph{10})}} \\
\cline{2-15}
& (1,2) & 4	& 3	& 3	& 3	& 3	& 3	& 4	& 3	& 3	& 3	& 3	& 3 & \textcolor{red}{\textbf{\emph{3} (\emph{10})}} \\
\cline{2-15}
& (2,1) & 4	& 5	& 3	& 4	& 3	& 3	& 4	& 10-12 & 	9 &	10-12 & 	4	& 8 & 4 (4) \\
\cline{2-15}
& (2,2) & 4	& 7	& 3	& 5	& 3	& 3	& 4	& 9-12 &	9-12	& 9-12 &	4	& 7 & \{3,4,9-12\} (3) \\
\cline{1-15} \cline{1-15}

\multirow{4}{*}{SpeakEasy} & (1,1) & 0.3	& 0.85	& 1	& 0.8	& 0.05	& 0.75	& 0.05	& 0.7	& 0.55	& 0.9	& 0.9	& 0.9 & \textcolor{red}{\textbf{\emph{0.9} (\emph{3})}} \\
\cline{2-15}
& (1,2) & 1	& 0.85	& 1	& 0.8	& 1	& 0.75	& 0.85	& 0.7	& 0.5	& 0.9	& 0.9	& 0.5 & \textcolor{red}{\textbf{\emph{1} (\emph{3})}} \\
\cline{2-15}
& (2,1) & 0.3	& 0.85	& 1	& 0.8	& 0.05	& 0.75	& 0.05	& 0.7	& 0.95	& 0.9	& 0.9	& 0.5 & \{0.05,0.9\} (2) \\
\cline{2-15}
& (2,2) & 1	& 0.85	& 1	& 0.8	& 1	& 0.75	& 0.85	& 0.7	& 0.5	& 0.5	& 0.9	& 0.5 & \textcolor{red}{\textbf{\{\emph{0.5},\emph{1}\} (\emph{3})}} \\
\hline \hline
\end{tabular}
\end{table}

\begin{table}
\scriptsize
\centering
\caption{The best value of threshold $r$ for SLPA, the best value of parameter $k$ for CFinder, and the best value of threshold $tr$ for SpeakEasy determined by the twelve quality metrics with four possible combinations of the two versions of belonging coefficient and two version of belonging function on American college football network. The best value for subcolumn of the last column is marked by red italic font.}
\label{tab:football:global}       
\vspace{-0.8em}
\setlength{\tabcolsep}{0.2pt}
\begin{tabular}{c||c|c|c|c|c|c|c|c|c|c|c|c|c||c}
\hline \hline
Algorithm & (BC,BF) & $Q_{ov}$ & $NQ_{ov}$ & $Q_{ov}^L$ & $Q_{ds}^{ov}$ & $IE$ & $ID$ & $CNT$ & $BE$ & $EXP$ & $CND$ & $F$ & $D$ & \\
\hline
\multirow{4}{*}{SLPA} & (1,1) & 0.4	& 0.45,0.5	& 0.45,0.5	& 0.45,0.5	& 0.05	& 0.45,0.5	& 0.05	& 0.45,0.5	& 0.45,0.5	& 0.45,0.5	& 0.45,0.5	& 0.45,0.5 & \{0.45,0.5\} (9) \\
\cline{2-15}
& (1,2) & 0.4	& 0.45,0.5	& 0.45,0.5	& 0.45,0.5	& 0.45,0.5	& 0.45,0.5	& 0.45,0.5	& 0.45,0.5	& 0.45,0.5	& 0.45,0.5	& 0.45,0.5	& 0.45,0.5 & \textcolor{red}{\textbf{\{\emph{0.45},\emph{0.5}\} (\emph{11})}} \\
\cline{2-15}
& (2,1) & 0.4	& 0.45,0.5	& 0.3	& 0.45,0.5	& 0.05	& 0.45,0.5	& 0.05	& 0.45,0.5	& 0.45,0.5	& 0.45,0.5	& 0.45,0.5	& 0.45,0.5 & \{0.45,0.5\} (8)\\
\cline{2-15}
& (2,2) & 0.4	& 0.45,0.5	& 0.3	& 0.45,0.5	& 0.25	& 0.45,0.5	& 0.25	& 0.25	& 0.25	& 0.25	& 0.25	& 0.25 & 0.25 (7) \\
\cline{1-15} \cline{1-14}

\multirow{4}{*}{CFinder} & (1,1) & 4	& 5	& 4	& 4	& 3	& 4	& 3	& 9	& 4	& 3	& 3	& 4 & \textcolor{red}{\textbf{\emph{4} (\emph{6})}} \\
\cline{2-15}
& (1,2) & 4	& 5	& 4	& 4	& 3	& 4	& 3	& 9	& 3	& 3	& 3	& 4 & \{3,4\} (5) \\
\cline{2-15}
& (2,1) & 4	& 6	& 4	& 4	& 3	& 4	& 3	& 9	& 9	& 9	& 3	& 4 & 4 (5) \\
\cline{2-15}
& (2,2) & 4	& 6	& 4	& 4	& 3	& 4	& 3	& 9	& 9	& 9	& 3	& 4 & 4 (5) \\
\cline{1-15} \cline{1-15}

\multirow{4}{*}{SpeakEasy} & (1,1) & 0.1,0.6	& 0.75	& 0.6	& 0.1,0.6	& 0.1,0.6	& 0.75	& 0.1,0.6	& 0.1,0.6	& 0.1,0.6	& 0.1,0.6	& 0.6	& 0.1,0.6 & \textcolor{red}{\textbf{\emph{0.6} (\emph{10})}} \\
\cline{2-15}
& (1,2) & 0.1,0.6	& 0.75	& 0.6	& 0.1,0.6	& 0.1,0.6	& 0.75	& 0.1,0.6	& 0.1,0.6	& 0.1,0.6	& 0.1,0.6	& 0.6	& 0.1,0.6 & \textcolor{red}{\textbf{\emph{0.6} (\emph{10})}} \\
\cline{2-15}
& (2,1) & 0.1,0.6	& 0.75	& 0.6	& 0.1,0.6	& 0.1,0.6	& 0.75	& 0.1,0.6	& 0.1,0.6	& 0.1,0.6	& 0.1,0.6	& 0.6	& 0.1,0.6 & \textcolor{red}{\textbf{\emph{0.6} (\emph{10})}} \\
\cline{2-15}
& (2,1) & 0.1,0.6	& 0.75	& 0.6	& 0.1,0.6	& 0.1,0.6	& 0.75	& 0.1,0.6	& 0.1,0.6	& 0.1,0.6	& 0.1,0.6	& 0.6	& 0.1,0.6 & \textcolor{red}{\textbf{\emph{0.6} (\emph{10})}} \\
\hline \hline
\end{tabular}
\vspace{2em}
\end{table}

\vspace{1em}
\subsection{American College Football Network}
The network represents the schedule of games between college football teams in a single season \cite{football}. There are 115 nodes and 613 edges. Table~\ref{tab:football:global} shows the best value of threshold $r$ for SLPA, the best value of parameter $k$ for CFinder, and the best value of threshold $tr$ for SpeakEasy determined by the twelve community quality metrics with four possible combinations of the two versions of belonging coefficient and two version of belonging function on American college football network. It can be observed from this table that SLPA performs best with (BC,BF)=(1,2), CFinder implies that (BC,BF)=(1,1) is the best, while SpeakEasy has no preferences.

\vspace{1.5em}
\subsection{Jazz Musicians Network}
This is a network with 198 nodes and 2742 edges of collaborations between jazz musicians \cite{Jazz}. Table~\ref{tab:jazz:global} shows the best value of threshold $r$ for SLPA, the best value of parameter $k$ for CFinder, and the best value of threshold $tr$ for SpeakEasy determined by the twelve community quality metrics with four possible combinations of the two versions of belonging coefficient and two version of belonging function on jazz musicians network. From this table, we could see that the first version of belonging coefficient is better than the second version when using SLPA and CFinder, while SpeakEasy demonstrates that (BC,BF)=(2,2) is the best among the four combinations. In summary, two of the three algorithms support that the first version of belonging coefficient is better than the second one on jazz musicians network.

\begin{table}
\vspace{1.5em}
\centering
\caption{The best value of threshold $r$ for SLPA, the best value of parameter $k$ for CFinder, and the best value of threshold $tr$ for SpeakEasy determined by the twelve community quality metrics with four possible combinations of the two versions of belonging coefficient and two version of belonging function on jazz musicians network. The best value for subcolumn of the last column is marked by red italic font.}
\label{tab:jazz:global}       
\vspace{-0.8em}
\setlength{\tabcolsep}{3pt}
\begin{tabular}{c||c|c|c|c|c|c|c|c|c|c|c|c|c||c}
\hline \hline
Algorithm & (BC,BF) & $Q_{ov}$ & $NQ_{ov}$ & $Q_{ov}^L$ & $Q_{ds}^{ov}$ & $IE$ & $ID$ & $CNT$ & $BE$ & $EXP$ & $CND$ & $F$ & $D$ & \\
\hline
\multirow{4}{*}{SLPA} & (1,1) & 0.5	& 0.5	& 0.5	& 0.5	& 0.05	& 0.5	& 0.05	& 0.5	& 0.5	& 0.5	& 0.5	& 0.5 & \textcolor{red}{\textbf{\emph{0.5} (\emph{10})}} \\
\cline{2-15}
& (1,2) & 0.5	& 0.5	& 0.5	& 0.5	& 0.1	& 0.5	& 0.1	& 0.5	& 0.5	& 0.5	& 0.5	& 0.5 & \textcolor{red}{\textbf{\emph{0.5} (\emph{10})}} \\
\cline{2-15}
& (2,1) & 0.5	& 0.5	& 0.4	& 0.5	& 0.05	& 0.5	& 0.05	& 0.5	& 0.5	& 0.5	& 0.5	& 0.5 &  0.5 (9) \\
\cline{2-15}
& (2,2) & 0.5	& 0.5	& 0.4	& 0.5	& 0.1	& 0.5	& 0.1	& 0.5	& 0.5	& 0.5	& 0.5	& 0.5 & 0.5 (9) \\
\cline{1-15} \cline{1-15}

\multirow{4}{*}{CFinder} & (1,1) & 14	& 3	& 10	& 10	& 3	& 8	& 3	& 3	& 3	& 3	& 3	& 3 & \textcolor{red}{\textbf{\emph{3} (\emph{8})}} \\
\cline{2-15}
& (1,2) & 10	& 3	& 10	& 10	& 3	& 8	& 3	& 3	& 3	& 3	& 3	& 3 & \textcolor{red}{\textbf{\emph{3} (\emph{8})}} \\
\cline{2-15}
& (2,1) & 14	& 10	& 8	& 10	& 3	& 8	& 3	& 3	& 3	& 18	& 3	& 3 & 3 (6) \\
\cline{2-15}
& (2,2) & 10	& 17	& 8	& 10	& 3	& 8	& 3	& 3	& 19,20	& 19,20 &	3	& 3 & 3 (5) \\
\cline{1-15} \cline{1-15}

\multirow{4}{*}{SpeakEasy} & (1,1) & 0.75	& 0.5	& 0.75	& 0.75	& 0.1	& 0.85	& 0.1	& 0.75	& 0.75	& 0.55	& 0.55	& 0.8 & 0.75 (5) \\
\cline{2-15}
& (1,2) & 0.75	& 0.5	& 0.75	& 0.75	& 0.75	& 0.85	& 0.7	& 0.75	& 0.5	& 0.55	& 0.55	& 0.5 & 0.75 (5) \\
\cline{2-15}
& (2,1) & 0.75	& 0.5	& 0.75	& 0.75	& 0.1	& 0.85	&0.1	& 0.75	& 0.75	& 0.55	& 0.55	& 0.8 & 0.75 (5) \\
\cline{2-15}
& (2,2) & 0.75	& 0.5	& 0.75	& 0.75	& 0.75	& 0.85	& 0.7	& 0.75	& 0.75	& 0.8	& 0.55	& 0.8 & \textcolor{red}{\textbf{\emph{0.75} (\emph{6})}} \\
\hline \hline
\end{tabular}
\end{table}

\vspace{2em}
\subsection {Zachary's Karate Club Network}
This network represents the friendships between $34$ members of a karate club at a US university during two years \cite{karate}. It has 34 nodes and 78 edges. Table~\ref{tab:karate:global} shows the best value of threshold $r$ for SLPA, the best value of parameter $k$ for CFinder, and the best value of threshold $tr$ for SpeakEasy determined by the twelve community quality metrics with four possible combinations of the two versions of belonging coefficient and two version of belonging function on Zachary's karate club network. It can be observed from this table that all three algorithms show that (BC,BF)=(1,2) is the best among the four combinations of belonging coefficient and belonging function.

\begin{table}
\centering
\caption{The best value of threshold $r$ for SLPA, the best value of parameter $k$ for CFinder, and the best value of threshold $tr$ for SpeakEasy determined by the twelve quality metrics with four possible combinations of the two versions of belonging coefficient and two version of belonging function on Zachary's karate club network. The best value for subcolumn of the last column is marked by red italic font.}
\label{tab:karate:global}       
\vspace{-0.8em}
\setlength{\tabcolsep}{2.5pt}
\begin{tabular}{c||c|c|c|c|c|c|c|c|c|c|c|c|c||c}
\hline \hline
Algorithm & (BC,BF) & $Q_{ov}$ & $NQ_{ov}$ & $Q_{ov}^L$ & $Q_{ds}^{ov}$ & $IE$ & $ID$ & $CNT$ & $BE$ & $EXP$ & $CND$ & $F$ & $D$ & \\
\hline
\multirow{4}{*}{SLPA} & (1,1) & 0.5	& 0.5	& 0.45	& 0.5	& 0.1	& 0.35	& 0.05	& 0.5	& 0.5	& 0.5	& 0.5	& 0.5 & 0.5 (8) \\
\cline{2-15}
& (1,2) & 0.45	& 0.45	& 0.45	& 0.45	& 0.15	& 0.35	& 0.45	& 0.45	& 0.45	& 0.45 &	0.45	& 0.45 & \textcolor{red}{\textbf{\emph{0.45} (\emph{10})}} \\
\cline{2-15}
& (2,1) & 0.5	& 0.5	& 0.45	& 0.5	& 0.1	& 0.35	& 0.05	& 0.5	& 0.5	& 0.5	& 0.5	& 0.45 & 0.5 (7) \\
\cline{2-15}
& (2,2) & 0.45	& 0.45	& 0.45	& 0.45	& 0.15	& 0.35	& 0.45	& 0.45	& 0.45	& 0.45	& 0.45	& 0.45 & \textcolor{red}{\textbf{\emph{0.45} (\emph{10})}} \\
\cline{1-15} \cline{1-15}

\multirow{4}{*}{CFinder} & (1,1) & 3	& 3	& 3	& 3	& 3 &	3 &	3 &	5	& 3	& 3 &	3	& 3 & \textcolor{red}{\textbf{\emph{3} (\emph{11})}} \\
\cline{2-15}
& (1,2) & 3	& 3	& 3	& 3 & 3	& 3	& 3	& 5	& 3	& 3	& 3	& 3 & \textcolor{red}{\textbf{\emph{3} (\emph{11})}} \\
\cline{2-15}
& (2,1) & 4	& 3	& 3	& 3	& 3	& 3	& 3	& 5	& 5	& 5	& 3	& 3 & 3 (8) \\
\cline{2-15}
& (2,2) & 4	& 4	& 3	& 3	& 3	& 3	& 3	& 5	& 5	& 5	& 3	& 3 & 3 (7) \\
\cline{1-15} \cline{1-15}

\multirow{4}{*}{SpeakEasy} & (1,1) & 0.45	& 0.45	& 0.45	& 0.45	& 0.2	& 0.95	& 0.05	& 0.7,0.75	& 0.45	& 0.65	& 0.65	& 0.65 & 0.45 (5) \\
\cline{2-15}
& (1,2) & 0.45	& 0.45	& 0.45	& 0.45	& 0.45	& 0.95	& 0.65	& 0.15	& 0.45	& 0.65	& 0.65 & 	0.65 & \textcolor{red}{\textbf{\emph{0.45} (\emph{6})}} \\
\cline{2-15}
& (2,1) & 0.45	& 0.45	& 0.15	& 0.45	& 0.2	& 0.95	& 0.05	& 0.9	& 0.45	& 0.65	& 0.65	& 0.45 & 0.45 (5) \\
\cline{2-15}
& (2,2) & 0.45	& 0.45	& 0.15	& 0.45	& 0.2	& 0.35	& 0.65	& 0.9	& 0.45	& 0.45	& 0.85	& 0.45 & \textcolor{red}{\textbf{\emph{0.45} (\emph{6})}} \\
\hline \hline
\end{tabular}
\end{table}

\begin{table}
\centering
\caption{The best value of threshold $r$ for SLPA, the best value of parameter $k$ for CFinder, and the best value of threshold $tr$ for SpeakEasy determined by the twelve community quality metrics with four possible combinations of the two versions of belonging coefficient and two version of belonging function on Les Miserables network. The best value for subcolumn of the last column is marked by red italic font.}
\label{tab:lesmis:global}       
\vspace{-0.8em}
\setlength{\tabcolsep}{0.5pt}
\begin{tabular}{c||c|c|c|c|c|c|c|c|c|c|c|c|c||c}
\hline \hline
Algorithm & (BC,BF) & $Q_{ov}$ & $NQ_{ov}$ & $Q_{ov}^L$ & $Q_{ds}^{ov}$ & $IE$ & $ID$ & $CNT$ & $BE$ & $EXP$ & $CND$ & $F$ & $D$ & \\
\hline
\multirow{4}{*}{SLPA} & (1,1) & 0.35	& 0.5	& 0.25	& 0.35	& 0.05	& 0.5	& 0.05	& 0.5	& 0.5	& 0.25	& 0.25	& 0.25 & \{0.25,0.5\} (4) \\
\cline{2-15}
& (1,2) & 0.35	& 0.5	& 0.25	& 0.35	& 0.15	& 0.5	& 0.25	& 0.35	& 0.25	& 0.25	& 0.25	& 0.25 & \textcolor{red}{\textbf{\emph{0.25} (\emph{6})}} \\
\cline{2-15}
& (2,1) & 0.35	& 0.5	& 0.15	& 0.4	& 0.05	& 0.5	& 0.05	& 0.5	& 0.5	& 0.25	& 0.25	& 0.25 & 0.5 (4) \\
\cline{2-15}
& (2,2) & 0.35	& 0.5	& 0.15	& 0.35	& 0.15	& 0.5	& 0.15	& 0.35	& 0.25	& 0.15	& 0.15	& 0.25 & 0.15 (5) \\
\cline{1-15} \cline{1-15}

\multirow{4}{*}{CFinder} & (1,1) & 6	& 3	& 4	& 4	& 3	& 3	& 5	& 8	& 3 &	3	& 3	& 3 & 3 (7) \\
\cline{2-15}
& (1,2) & 5	& 3	& 4	& 5	& 3	& 3	& 3	& 4	& 3	& 3	& 3	& 3 & \textcolor{red}{\textbf{\emph{3} (\emph{8})}} \\
\cline{2-15}
& (2,1) & 6	& 6	& 4	& 6	& 3 & 3	& 5	& 9	& 9	& 9	& 3	& 6 & 6 (4) \\
\cline{2-15}
& (2,2) & 6	& 6	& 4	& 6	& 3	& 3	& 3	& 9	& 9	& 9	& 3 &	6 & \{3,6\} (4) \\
\cline{1-15} \cline{1-15}

\multirow{4}{*}{SpeakEasy} & (1,1) & 0.65	& 0.85 & 	0.65 &	0.6	& 0.85 &	0.85 &	0.1	& 0.55,0.95 & 	0.55 &	0.85	& 0.85	& 0.6 & 0.85 (5) \\
\cline{2-15}
& (1,2) & 0.65	& 0.85	& 0.65	& 0.6	& 0.85	& 0.85	& 0.85	& 0.55,0.95	& 0.55	& 0.85	& 0.85 &	0.6 & \textcolor{red}{\textbf{\emph{0.85} (\emph{6})}} \\
\cline{2-15}
& (2,1) & 0.65	& 0.7	& 0.65,0.8	& 0.6	& 0.85	& 0.85	& 0.1	& 0.95	& 0.55	& 0.85	& 0.85	& 0.6 & 0.85 (4) \\
\cline{2-15}
& (2,2) & 0.65,0.8	& 0.7	& 0.65,0.8	& 0.45,0.6	& 0.85	& 0.85	& 0.85	& 0.45,0.95	& 0.55	& 0.95	& 0.85	& 0.45,0.6 & 0.85 (4) \\
\hline \hline
\end{tabular}
\end{table}

\begin{table}
\centering
\caption{The best value of threshold $r$ for SLPA, the best value of parameter $k$ for CFinder, and the best value of threshold $tr$ for SpeakEasy determined by the twelve quality metrics with four possible combinations of the two versions of belonging coefficient and two version of belonging function Network Science coauthorship network. The best value for subcolumn of the last column is marked by red italic font.}
\label{tab:netscience:global}       
\vspace{-0.8em}
\setlength{\tabcolsep}{1.5pt}
\begin{tabular}{c||c|c|c|c|c|c|c|c|c|c|c|c|c||c}
\hline \hline
Algorithm & (BC,BF) & $Q_{ov}$ & $NQ_{ov}$ & $Q_{ov}^L$ & $Q_{ds}^{ov}$ & $IE$ & $ID$ & $CNT$ & $BE$ & $EXP$ & $CND$ & $F$ & $D$ & \\
\hline
\multirow{4}{*}{SLPA} & (1,1) & 0.05	& 0.4	& 0.5	& 0.35	& 0.05	& 0.4	& 0.05	& 0.5	& 0.5	& 0.35	& 0.35	& 0.35 & 0.35 (4) \\
\cline{2-15}
& (1,2) & 0.5	& 0.4	& 0.5	& 0.4	& 0.5	& 0.4	& 0.25	& 0.5	& 0.35	& 0.2	& 0.15	& 0.35 & 0.5 (4) \\
\cline{2-15}
& (2,1) & 0.05	& 0.4	& 0.35	& 0.35	& 0.05	& 0.4	& 0.05	& 0.5	& 0.5	& 0.35	& 0.35	& 0.35 & \textcolor{red}{\textbf{\emph{0.35} (\emph{5})}} \\
\cline{2-15}
& (2,2) & 0.5	& 0.4	& 0.35	& 0.35	& 0.15	& 0.4	& 0.25	& 0.5	& 0.5	& 0.2	& 0.15	& 0.35 & \{0.35,0.5\} (3) \\
\cline{1-15} \cline{1-15}

\multirow{4}{*}{CFinder} & (1,1) & 3	& 3	& 3	& 3	& 3	& 3	& 3	& 7	& 3	& 3	& 3	& 3 & 3 (11) \\
\cline{2-15}
& (1,2) & 3	& 3	& 3	& 3	& 3	& 3	& 3	& 3	& 3	& 3	& 3	& 3 & \textcolor{red}{\textbf{\emph{3} (\emph{12})}} \\
\cline{2-15}
& (2,1) & 3	& 3	& 3	& 3 & 3	& 3	& 3 &	11-20	& 11-20 &	11-20 &	3	& 3 & 3 (9) \\
\cline{2-15}
& (2,2) & 3	& 3	& 3	& 3	& 3	& 3	& 3	& 9-20 &	9-20 &	9-20 &	3	& 3 & 3 (9) \\
\cline{1-15} \cline{1-15}

\multirow{4}{*}{SpeakEasy} & (1,1) & 0.05	& 0.25	& 0.25	& 0.25	& 0.05	& 0.7	& 0.05	& 1	& 1	& 0.7	& 0.7	& 0.7 & 0.7 (4) \\
\cline{2-15}
& (1,2) & 0.25	& 0.25	& 0.25	& 0.25	& 0.25	& 0.7	& 0.7	& 0.25	& 0.25	& 0.25	& 0.25	& 0.25 & \textcolor{red}{\textbf{\emph{0.25} (\emph{10})}} \\
\cline{2-15}
& (2,1) & 0.05	& 0.3	& 0.15	& 0.25	& 0.05	& 0.7	& 0.05	& 0.85	& 1	& 0.25	& 0.7	& 1 & 0.05 (3) \\
\cline{2-15}
& (2,2) & 0.15	& 0.3	& 0.15	& 0.25	& 0.25	& 0.7	& 0.7	& 0.4	& 0.2	& 0.2	& 0.9	& 1 & \{0.15,0.2,0.25,0.7\} (2) \\
\hline \hline
\end{tabular}
\end{table}

\vspace{2em}
\subsection{Les Miserables Network}
This is a coappearance network of characters in the novel Les Miserables \cite{Lesmis}. It has 77 nodes and 254 edges. Table~\ref{tab:lesmis:global} shows the best value of threshold $r$ for SLPA, the best value of parameter $k$ for CFinder, and the best value of threshold $tr$ for SpeakEasy determined by the twelve community quality metrics with four possible combinations of the two versions of belonging coefficient and two version of belonging function on Les Miserables network. The table shows that for all three algorithms (BC,BF)=(1,2) is the best among the four combinations of belonging coefficient and belonging function.

\subsection{Network Science Coauthorship Network}
This is a coauthorship network of scientists working on network theory and experiment \cite{EigenvectorCommunity}. There are 1461 nodes and 2742 edges. Table~\ref{tab:netscience:global} shows the best value of threshold $r$ for SLPA, the best value of parameter $k$ for CFinder, and the best value of threshold $tr$ for SpeakEasy determined by the twelve community quality metrics with four possible combinations of the two versions of belonging coefficient and two version of belonging function on Network Science coauthorship network. We can observe from this table that SLPA shows that (BC,BF)=(2,1) is the best among the four combinations, while CFinder and SpeakEasy indicate that (BC,BF)=(1,2) is the best. Thus, we could conclude that (BC,BF)=(1,2) is the best on Network Science coauthorship network.

\vspace{2em}
\subsection{PGP Network}
This is the largest connected component of the network of users of the Pretty-Good-Privacy (PGP) algorithm for secure information interchange \cite{PGPNetwork}. It has 10680 nodes and 24316 edges in total. Table~\ref{tab:pgp:global} shows the best value of threshold $r$ for SLPA, the best value of parameter $k$ for CFinder, and the best value of threshold $tr$ for SpeakEasy determined by the twelve community quality metrics with four possible combinations of the two versions of belonging coefficient and two version of belonging function on PGP Network. It can be seen from the table that all three algorithms show that (BC,BF)=(1,2) is the best among the four combinations of belonging coefficient and belonging function.

\begin{table}
\vspace{1em}
\centering
\caption{The best value of threshold $r$ for SLPA, the best value of parameter $k$ for CFinder, and the best value of threshold $tr$ for SpeakEasy determined by the twelve community quality metrics with four possible combinations of the two versions of belonging coefficient and two version of belonging function on PGP network. The best value for subcolumn of the last column is marked by red italic font.}
\label{tab:pgp:global}       
\vspace{-0.8em}
\setlength{\tabcolsep}{1.5pt}
\begin{tabular}{c||c|c|c|c|c|c|c|c|c|c|c|c|c||c}
\hline \hline
Algorithm & (BC,BF) & $Q_{ov}$ & $NQ_{ov}$ & $Q_{ov}^L$ & $Q_{ds}^{ov}$ & $IE$ & $ID$ & $CNT$ & $BE$ & $EXP$ & $CND$ & $F$ & $D$ & \\
\hline
\multirow{4}{*}{SLPA} & (1,1) & 0.05	& 0.5	& 0.5	& 0.5	& 0.05	& 0.45	& 0.05	& 0.5	& 0.5	& 0.5	& 0.5	& 0.5 & 0.5 (8) \\
\cline{2-15}
& (1,2) & 0.5	& 0.5	& 0.5	& 0.5	& 0.05	& 0.45	& 0.5	& 0.5	& 0.5	& 0.5	& 0.5	& 0.5 & \textcolor{red}{\textbf{\emph{0.5} (\emph{10})}} \\
\cline{2-15}
& (2,1) & 0.05	& 0.5	& 0.2	& 0.5	& 0.05	& 0.45	& 0.05	& 0.5	& 0.5	& 0.5	& 0.5	& 0.5 & 0.5 (7) \\
\cline{2-15}
& (2,2) & 0.45	& 0.5	& 0.2	& 0.5	& 0.05	& 0.45	& 0.5	& 0.5	& 0.5	& 0.5	& 0.5	& 0.5 & 0.5 (8) \\
\cline{1-15} \cline{1-15}

\multirow{4}{*}{CFinder} & (1,1) & 4	& 3	& 3	& 3	& 3	& 3	& 3	& 13	& 3	& 3	& 3	& 3 & 3 (10) \\
\cline{2-15}
& (1,2) & 3	& 3	& 3	& 3	& 3	& 3	& 3	& 3	& 3	& 3	& 3	& 3 & \textcolor{red}{\textbf{\emph{3} (\emph{12})}} \\
\cline{2-15}
& (2,1) & 4	& 3	& 3	& 6	& 3	& 3	& 3	& 18	& 18	& 19	& 3	& 13 & 3 (6) \\
\cline{2-15}
& (2,2) & 3	& 3	& 3	& 6	& 3	& 3	& 3	& 18	& 18	& 14,15	 & 3	& 3 & 3 (8) \\
\cline{1-15} \cline{1-15}

\multirow{4}{*}{SpeakEasy} & (1,1) & 0.05	& 0.25	& 0.85	& 0.9	& 0.05	& 0.7	& 0.05	& 0.75	& 0.85	& 0.85	& 0.85	& 0.85 & 0.85 (5) \\
\cline{2-15}
& (1,2) & 0.85	& 0.25	& 0.85	& 0.9	& 0.85	& 0.7	& 0.85	& 0.75	& 0.85	& 0.85	& 0.25	& 0.85 & \textcolor{red}{\textbf{\emph{0.85} (\emph{7})}} \\
\cline{2-15}
& (2,1) & 0.05	& 0.8	& 0.85	& 0.9	& 0.05	& 0.7	& 0.05	& 0.75	& 0.75	& 0.85	& 0.85	& 0.75 & \{0.05,0.75,0.85\} (3) \\
\cline{2-15}
& (2,2) & 0.85	& 0.8	& 0.85	& 0.9	& 0.85	& 0.7	& 0.85	& 0.6	& 0.75	& 0.75	& 0.2	& 0.75 & 0.85 (4) \\
\hline \hline
\end{tabular}
\end{table}

\begin{table}
\centering
\caption{The best value of threshold $r$ for SLPA and the best value of threshold $tr$ for SpeakEasy determined by the twelve community quality metrics with four possible combinations of the two versions of belonging coefficient and two version of belonging function on political blogs network. The best value for subcolumn of the last column is marked by red italic font.}
\label{tab:polblogs:global}       
\vspace{-0.8em}
\setlength{\tabcolsep}{2pt}
\begin{tabular}{c||c|c|c|c|c|c|c|c|c|c|c|c|c||c}
\hline \hline
Algorithm & (BC,BF) & $Q_{ov}$ & $NQ_{ov}$ & $Q_{ov}^L$ & $Q_{ds}^{ov}$ & $IE$ & $ID$ & $CNT$ & $BE$ & $EXP$ & $CND$ & $F$ & $D$ & \\
\hline
\multirow{4}{*}{SLPA} & (1,1) & 0.5	& 0.5	& 0.5	& 0.4	& 0.5	& 0.35	& 0.3	& 0.4	& 0.45	& 0.5	& 0.5	& 0.5 & 0.5 (7) \\
\cline{2-15}
& (1,2) & 0.5	& 0.5	& 0.5	& 0.4	& 0.5	& 0.4	& 0.5	& 0.15	& 0.4	& 0.5	& 0.5	& 0.5 & \textcolor{red}{\textbf{\emph{0.5} (\emph{8})}} \\
\cline{2-15}
& (2,1) & 0.5	& 0.5	& 0.05	& 0.4	& 0.5	& 0.4	& 0.3	& 0.4	& 0.45	& 0.5	& 0.5	& 0.5 & 0.5 (6) \\
\cline{2-15}
& (2,2) & 0.25	& 0.25	& 0.05	& 0.25	& 0.5	& 0.4	& 0.5	& 0.05	& 0.35	& 0.5	& 0.5	& 0.3 & 0.5 (4) \\
\cline{1-15} \cline{1-15}

\multirow{4}{*}{SpeakEasy} & (1,1) & 0.85	& 0.45	& 0.85	& 0.7	& 0.9	& 0.35	& 0.05	& 0.7	& 0.7	& 0.25	& 0.45	& 0.7 & \textcolor{red}{\textbf{\emph{0.7} (\emph{4})}} \\
\cline{2-15}
& (1,2) & 0.85	& 0.45	& 0.85	& 0.7	& 0.9	& 0.35	& 0.9	& 0.7	& 0.7	& 0.25	& 0.45	& 0.7 & \textcolor{red}{\textbf{\emph{0.7} (\emph{4})}} \\
\cline{2-15}
& (2,1) & 0.85	& 0.45	& 0.45	& 0.7	& 0.9	& 0.35	& 0.05	& 0.7	& 0.5	& 0.25	& 0.45	& 0.5 & 0.45 (3) \\
\cline{2-15}
& (2,2) & 0.8	& 0.8	& 0.45	& 0.7	& 0.9	& 0.35	& 0.9	& 0.7	& 0.5	& 0.4	& 0.6	& 0.5 & \{0.5,0.7,0.8,0.9\} (2) \\
\hline \hline
\end{tabular}
\end{table}

\vspace{1.5em}
\subsection{Political Blogs Network}
This is a directed network of hyperlinks between weblogs on US politics, recorded in 2005 by Adamic and Glance \cite{Polblogs}. There are 1224 nodes and 19022 edges. Table~\ref{tab:polblogs:global} shows the best value of threshold $r$ for SLPA and the best value of threshold $tr$ for SpeakEasy determined by the twelve community quality metrics with four possible combinations of the two versions of belonging coefficient and two version of belonging function on political blogs network. Results for CFinder are not provided because it has not finished running on this network for more than two months processing many potential $k$-cliques resulting from dense connections. It can be seen from the table that both SLPA and SpeakEasy imply that (BC,BF)=(1,2) is the best among the four combinations of belonging coefficient and belonging function.

\vspace{2em}
\subsection{Political Books Network}
 This is a network of books about US politics published around the time of the 2004 presidential election and sold by the online bookseller Amazon.com \cite{Polbooks}. It has 105 nodes and 441 edges in total. Edges between books indicate frequent copurchasing of books by the same buyers. Table~\ref{tab:polbooks:global} shows the best value of threshold $r$ for SLPA, the best value of parameter $k$ for CFinder, and the best value of threshold $tr$ for SpeakEasy determined by the twelve community quality metrics with four possible combinations of the two versions of belonging coefficient and two version of belonging function on political books network. We could learn from the table that SLPA shows that the first version of belonging function is better than the second version, CFinder implies that the first version of belonging coefficient is better than the second one, and SpeakEasy indicates that (BC,BF)=(1,2) is the best among all four combinations. In summary, there are two out of three algorithms support that (BC,BF)=(1,2) is the best on political books network.

\begin{table}
\vspace{1em}
\centering
\caption{The best value of threshold $r$ for SLPA, the best value of parameter $k$ for CFinder, and the best value of threshold $tr$ for SpeakEasy determined by the twelve community quality metrics with four possible combinations of the two versions of belonging coefficient and two version of belonging function on political books network. The best value for subcolumn of the last column is marked by red italic font.}
\label{tab:polbooks:global}       
\vspace{-0.8em}
\setlength{\tabcolsep}{2.5pt}
\begin{tabular}{c||c|c|c|c|c|c|c|c|c|c|c|c|c||c}
\hline \hline
Algorithm & (BC,BF) & $Q_{ov}$ & $NQ_{ov}$ & $Q_{ov}^L$ & $Q_{ds}^{ov}$ & $IE$ & $ID$ & $CNT$ & $BE$ & $EXP$ & $CND$ & $F$ & $D$ & \\
\hline
\multirow{4}{*}{SLPA} & (1,1) & 0.5	& 0.5	& 0.2	& 0.4	& 0.05	& 0.4	& 0.05	& 0.5	& 0.5	& 0.5	& 0.5	& 0.5 & \textcolor{red}{\textbf{\emph{0.5} (\emph{7})}} \\
\cline{2-15}
& (1,2) & 0.5	& 0.5	& 0.2	& 0.4	& 0.1	& 0.4	& 0.1	& 0.4	& 0.2	& 0.2	& 0.2	& 0.2 & 0.2 (5) \\
\cline{2-15}
& (2,1) & 0.5	& 0.5	& 0.1	& 0.4	& 0.05	& 0.4	& 0.05	& 0.5	& 0.5	& 0.5	& 0.5	& 0.5 & \textcolor{red}{\textbf{\emph{0.5} (\emph{7})}} \\
\cline{2-15}
& (2,2) & 0.5	& 0.5	& 0.1	& 0.4	& 0.1	& 0.4	& 0.1	& 0.4	&0.2	& 0.2	& 0.2	& 0.2 & 0.2 (4) \\
\cline{1-15} \cline{1-15}

\multirow{4}{*}{CFinder} & (1,1) & 4	& 3	& 3	& 3	& 3	& 3	& 3	& 6	& 3	& 3	& 3	& 3 & \textcolor{red}{\textbf{\emph{3} (\emph{10})}} \\
\cline{2-15}
& (1,2) & 3	& 3	& 3	& 4	& 3	& 3	& 3	& 6	& 3	& 3	& 3	& 3 & \textcolor{red}{\textbf{\emph{3} (\emph{10})}} \\
\cline{2-15}
& (2,1) & 4	& 4	& 3	& 4	& 3	& 3	& 3	& 6	& 6	& 3	& 3	& 3 & 3 (7) \\
\cline{2-15}
& (2,2) & 4	& 4	& 3	& 4	& 3	& 3	& 3	& 6	& 6	& 6	& 3	& 3 & 3 (6) \\
\cline{1-15} \cline{1-15}

\multirow{4}{*}{SpeakEasy} & (1,1) & 0.95	& 1	& 0.5	& 0.95	& 0.25	& 0.95	& 0.05	& 0.5	& 0.9	& 0.95	& 0.95	& 0.9 & 0.95 (5) \\
\cline{2-15}
& (1,2) & 0.95	& 0.55	& 0.5	& 0.95	& 0.95	& 0.95	& 0.9	& 0.5	& 0.9	& 0.95	& 0.95	& 0.9 & \textcolor{red}{\textbf{\emph{0.95} (\emph{6})}} \\
\cline{2-15}
& (2,1) & 0.5	& 1	& 0.5	& 0.95	& 0.25	& 0.95	& 0.05	& 0.5	& 0.85	& 0.9	& 0.95	& 0.9 & \{0.5,0.95\} (3) \\
\cline{2-15}
& (2,2) & 0.5	& 0.55 &	0.5 &	0.95	& 0.95	& 0.95	& 0.9	& 0.5	& 0.85	& 0.85	& 0.95	& 0.9 & 0.95 (4) \\
\hline \hline
\end{tabular}
\end{table}

\begin{table}
\centering
\caption{The best value of threshold $r$ for SLPA, the best value of parameter $k$ for CFinder, and the best value of threshold $tr$ for SpeakEasy determined by the twelve community quality metrics with four possible combinations of the two versions of belonging coefficient and two version of belonging function on Indian railway network. The best value for subcolumn of the last column is marked by red italic font.}
\label{tab:railway:global}       
\vspace{-0.8em}
\setlength{\tabcolsep}{3pt}
\begin{tabular}{c||c|c|c|c|c|c|c|c|c|c|c|c|c||c}
\hline \hline
Algorithm & (BC,BF) & $Q_{ov}$ & $NQ_{ov}$ & $Q_{ov}^L$ & $Q_{ds}^{ov}$ & $IE$ & $ID$ & $CNT$ & $BE$ & $EXP$ & $CND$ & $F$ & $D$ & \\
\hline
\multirow{4}{*}{SLPA} & (1,1) & 0.15	& 0.5	& 0.5	& 0.5	& 0.05	& 0.45	& 0.05	& 0.5	& 0.5	& 0.5	& 0.5	& 0.5 & 0.5 (8) \\
\cline{2-15}
& (1,2) & 0.5	& 0.35	& 0.5	& 0.5	& 0.05	& 0.45	& 0.5	& 0.5	& 0.5	& 0.5	& 0.5	& 0.5 & \textcolor{red}{\textbf{\emph{0.5} (\emph{9})}} \\
\cline{2-15}
& (2,1) & 0.05	& 0.5	& 0.15	& 0.5	& 0.05	& 0.45	& 0.05	& 0.5	& 0.5	& 0.5	& 0.5	& 0.5 & 0.5 (7) \\
\cline{2-15}
& (2,2) & 0.5	& 0.35	& 0.15	& 0.5	& 0.05	& 0.45	& 0.5	& 0.35	& 0.5	& 0.5	& 0.5	& 0.5 & 0.5 (7) \\
\cline{1-15} \cline{1-15}

\multirow{4}{*}{CFinder} & (1,1) & 6	& 3	& 4	& 5	& 3	& 3	& 3	& 3	& 3	& 3	& 3	& 3 & \textcolor{red}{\textbf{\emph{3} (\emph{9})}} \\
\cline{2-15}
& (1,2) & 6	& 3	& 4	& 5	& 3	& 3	& 3	& 3	& 3	& 3	& 3	& 3 & \textcolor{red}{\textbf{\emph{3} (\emph{9})}} \\
\cline{2-15}
& (2,1) & 6	& 6	& 4	& 5	& 3	& 3	& 3	& 10	& 10	& 10	& 3	& 4 & 3 (4) \\
\cline{2-15}
& (2,2) & 6	& 6	& 4	& 6	& 3	& 3	& 3	& 10	& 10	& 10	& 3	& 4 & 3 (4) \\
\cline{1-15} \cline{1-15}

\multirow{4}{*}{SpeakEasy} & (1,1) & 0.55	& 0.8	& 0.45	& 1	& 0.05	& 0.7	& 0.05	& 0.95	& 0.95	& 0.8	& 0.8	& 0.8 & 0.8 (4) \\
\cline{2-15}
& (1,2) & 0.45	& 0.7	& 0.45	& 1	& 0.8	& 0.7	& 0.8	& 0.95	& 0.55	& 0.8	& 0.8	& 0.8 & \textcolor{red}{\textbf{\emph{0.8} (\emph{5})}} \\
\cline{2-15}
& (2,1) & 0.55	& 0.8	& 0.45	& 1	& 0.05	& 0.7	& 0.05	& 0.95	& 0.95	& 0.8	& 0.8	& 0.55 & 0.8 (3) \\
\cline{2-15}
& (2,2) & 0.45	& 0.8	& 0.45	& 1	& 0.8	& 0.7	& 0.8	& 0.95	& 0.55	& 0.55	& 0.8	& 0.55 & 0.8 (4) \\
\hline \hline
\end{tabular}
\end{table}

\vspace{1.5em}
\subsection{Indian Railway Network}
This network consists of nodes representing Indian railway stations, where two stations are connected by an edge if there exists at least one train-route such that both stations are scheduled stops on that route \cite{Railway}. There are 297 nodes and 1213 edges. Table~\ref{tab:railway:global} shows the best value of threshold $r$ for SLPA, the best value of parameter $k$ for CFinder, and the best value of threshold $tr$ for SpeakEasy determined by the twelve community quality metrics with four possible combinations of the two versions of belonging coefficient and two version of belonging function on Indian railway network. It can be seen from the table that all three algorithms show that (BC,BF)=(1,2) is the best among the four combinations of belonging coefficient and belonging function.

\vspace{2em}
\subsection{Santa Fe Institute Collaboration Network}
This is the largest connected component of the collaboration network of scientists at the Santa Fe Institute, an interdisciplinary research center in Santa Fe, New Mexico \cite{football}. It has 118 nodes and 200 edges. Nodes in this network represent scientists in residence at the Santa Fe Institute during any part of calendar year 1999 or 2000 and their collaborators. An edge is drawn between a pair of scientists if they coauthored one or more articles during the same time period. The network includes all journal and book publications by the scientists involved, along with all papers that appeared in the institute's technical reports series. Table~\ref{tab:santafe:global} shows the best value of threshold $r$ for SLPA, the best value of parameter $k$ for CFinder, and the best value of threshold $tr$ for SpeakEasy determined by the twelve community quality metrics with four possible combinations of the two versions of belonging coefficient and two version of belonging function on Santa Fe Institute collaboration network. We could learn that all three algorithms show that (BC,BF)=(1,2) is the best among the four combinations of belonging coefficient and belonging function.

\begin{table}
\vspace{1em}
\centering
\caption{The best value of threshold $r$ for SLPA, the best value of parameter $k$ for CFinder, and the best value of threshold $tr$ for SpeakEasy determined by the twelve quality metrics with four combinations of the two versions of belonging coefficient and two version of belonging function on Santa Fe Institute collaboration network. The best value for subcolumn of the last column is marked by red italic font.}
\label{tab:santafe:global}       
\vspace{-0.8em}
\setlength{\tabcolsep}{2.5pt}
\begin{tabular}{c||c|c|c|c|c|c|c|c|c|c|c|c|c||c}
\hline \hline
Algorithm & (BC,BF) & $Q_{ov}$ & $NQ_{ov}$ & $Q_{ov}^L$ & $Q_{ds}^{ov}$ & $IE$ & $ID$ & $CNT$ & $BE$ & $EXP$ & $CND$ & $F$ & $D$ & \\
\hline
\multirow{4}{*}{SLPA} & (1,1) & 0.4	& 0.5	& 0.4	& 0.5	& 0.05	& 0.5	& 0.05	& 0.5	& 0.4	& 0.4	& 0.4	& 0.4 & 0.4 (6) \\
\cline{2-15}
& (1,2) & 0.4	& 0.5	& 0.4	& 0.5	& 0.1	& 0.5	& 0.4	& 0.5	& 0.4	& 0.4	& 0.4	& 0.4 & \textcolor{red}{\textbf{\emph{0.4} (\emph{7})}} \\
\cline{2-15}
& (2,1) & 0.4	& 0.5	& 0.1	& 0.5	& 0.05	& 0.5	& 0.05	& 0.5	& 0.4	& 0.4	& 0.4	& 0.4 & 0.4 (5) \\
\cline{2-15}
& (2,2) & 0.4	& 0.5	& 0.1	& 0.5	& 0.1	& 0.5	& 0.1	& 0.5	& 0.4	& 0.4	& 0.4	& 0.4 & 0.4 (5) \\
\cline{1-15} \cline{1-15}

\multirow{4}{*}{CFinder} & (1,1) & 3	& 3	& 3	& 3	& 3	& 3	& 3	& 5	& 4	& 3	& 3	& 3 & 3 (10) \\
\cline{2-15}
& (1,2) & 3	& 3	& 3	& 3	& 3	& 3	& 3	& 4	& 3	& 3	& 3	& 3 & \textcolor{red}{\textbf{\emph{3} (\emph{11})}} \\
\cline{2-15}
& (2,1) & 3	& 3	& 3	& 4	& 3	& 3	& 3	& 5	& 5	& 5	& 3	& 5 & 3 (7) \\
\cline{2-15}
& (2,2) & 3	& 3	& 3	& 3 & 3	& 3	& 3	& 5	& 5	& 5	& 3	& 4 & 3 (8) \\
\cline{1-15} \cline{1-15}

\multirow{4}{*}{SpeakEasy} & (1,1) & 0.1	& 0.9	& 0.65	& 0.65	& 0.05	& 0.9	& 0.05	& 1	& 0.95	& 0.9	& 0.9	& 0.9 & \textcolor{red}{\textbf{\emph{0.9} (\emph{5})}} \\
\cline{2-15}
& (1,2) & 0.55	& 0.9	& 0.65	& 0.35	& 0.35	& 0.9	& 0.9	& 1	& 0.95	& 0.9	& 0.9	& 0.35 & \textcolor{red}{\textbf{\emph{0.9} (\emph{5})}} \\
\cline{2-15}
& (2,1) & 0.1	& 0.9	& 0.1	& 0.65	& 0.05	& 0.9	& 0.05	& 1	& 0.95	& 0.9	& 0.9	& 0.65 & 0.9 (4) \\
\cline{2-15}
& (2,2) & 0.1	& 0.9	& 0.1	& 0.65	& 0.35	& 0.9	& 0.9	& 0.65	& 0.65	& 1	& 0.9	& 0.65 & \{0.65,0.9\} (4) \\
\hline \hline
\end{tabular}
\end{table}

\vspace{1.5em}
\subsection{Protein-protein Interaction Networks}
We consider eight protein-protein interaction networks in the experiments. \textbf{Collins\_cyc}, \textbf{Collins\_cyc\_w}, \textbf{Collins\_mips}, \textbf{Collins\_sgd}, \textbf{Gavin\_cyc}, \textbf{Gavin\_cyc\_w}, \textbf{Gavin\_mips}, and \textbf{Gavin\_sgd} are two kinds (referred as Collins \cite{Collins} and Gavin \cite{Gavin} here) of popular high throughput protein-protein interaction networks derived from measurements obtained by affinity purification and mass spectrometry (AP-MS) techniques \cite{SpeakEasy}. These two kinds of networks are further refined with three gold-standards for protein complexes, including the classic Munich Information Center for Protein Sequences (MIPS) \cite{MIPS} and the more recent Saccharomyces Genome Database (SGD) \cite{SGD}. The complete MIPS dataset as well as partial information from SGD are incorporated into a third protein complex list known as CYC2008 \cite{CYC}. Thus, we have \textbf{Collins\_cyc}, \textbf{Collins\_mips}, \textbf{Collins\_sgd}, \textbf{Gavin\_cyc}, \textbf{Gavin\_mips}, and \textbf{Gavin\_sgd}, respectively. \textbf{Collins\_cyc\_w} and \textbf{Gavin\_cyc\_w} are respectively the weighted versions of \textbf{Collins\_cyc} and \textbf{Gavin\_cyc}, in which the weight is proportional to the probability a given interaction pair truly exists.

\begin{table}
\centering
\caption{The best value of threshold $r$ for SLPA, the best value of parameter $k$ for CFinder, and the best value of threshold $tr$ for SpeakEasy determined by the twelve community quality metrics with four possible combinations of the two versions of belonging coefficient and two version of belonging function on \textbf{Collins\_cyc}. The best value for subcolumn of the last column is marked by red italic font.}
\label{tab:collins_cyc:global}       
\vspace{-0.8em}
\setlength{\tabcolsep}{3pt}
\begin{tabular}{c||c|c|c|c|c|c|c|c|c|c|c|c|c||c}
\hline \hline
Algorithm & (BC,BF) & $Q_{ov}$ & $NQ_{ov}$ & $Q_{ov}^L$ & $Q_{ds}^{ov}$ & $IE$ & $ID$ & $CNT$ & $BE$ & $EXP$ & $CND$ & $F$ & $D$ & \\
\hline
\multirow{4}{*}{SLPA} & (1,1) & 0.05	& 0.5	& 0.5	& 0.4	& 0.1	& 0.5	& 0.05	& 0.5	& 0.5	& 0.5	& 0.5	& 0.5 & \textcolor{red}{\textbf{\emph{0.5} (\emph{8})}} \\
\cline{2-15}
& (1,2) & 0.5	& 0.45	& 0.5	& 0.5	& 0.2	& 0.5	& 0.2	& 0.5	& 0.4	& 0.2	& 0.2	& 0.4 & 0.5 (5) \\
\cline{2-15}
& (2,1) & 0.05	& 0.5	& 0.5	& 0.4	& 0.1	& 0.5	& 0.05	& 0.5	& 0.5	& 0.5	& 0.5	& 0.5 & \textcolor{red}{\textbf{\emph{0.5} (\emph{8})}} \\
\cline{2-15}
& (2,2) & 0.5	& 0.5	& 0.5	& 0.5	& 0.2	& 0.5	& 0.2	& 0.5	& 0.4	& 0.2	& 0.2	& 0.4 & 0.5 (6) \\
\cline{1-15} \cline{1-15}

\multirow{4}{*}{CFinder} & (1,1) & 3	& 3	& 3	& 5	& 3	& 3	& 3	& 3	& 3	& 3	& 3	& 3 & \textcolor{red}{\textbf{\emph{3} (\emph{11})}} \\
\cline{2-15}
& (1,2) & 3	& 3	& 3	& 5	& 3	& 3	& 3	& 3	& 3	& 3	& 3	& 3 & \textcolor{red}{\textbf{\emph{3} (\emph{11})}} \\
\cline{2-15}
& (2,1) & 3	& 9	& 3	& 6	& 3	& 3	& 3	& 18	& 18	& 20	& 3	& 3 & 3 (7) \\
\cline{2-15}
& (2,2) & 3	& 3	& 3	& 6	& 3	& 3	& 3	& 17-20	& 17-20	& 17-20	& 3	& 3 & 3 (8) \\
\cline{1-15} \cline{1-15}

\multirow{4}{*}{SpeakEasy} & (1,1) & 0.05	& 0.9	& 0.9	& 1	& 0.9	& 0.9	& 0.05	& 0.8	& 0.9	& 0.9	& 0.9	& 0.9 & 0.9 (8) \\
\cline{2-15}
& (1,2) & 0.9	& 0.9	& 0.9	& 1	& 0.9	& 0.9	& 0.9	& 0.8	& 0.9	& 0.9	& 0.9	& 0.9 & \textcolor{red}{\textbf{\emph{0.9} (\emph{10})}} \\
\cline{2-15}
& (2,1) & 0.05	& 0.55	& 0.9	& 0.95	& 0.05	& 0.9	& 0.05	& 0.8	& 0.95	& 0.9	& 0.9	& 0.95 & 0.9 (4) \\
\cline{2-15}
& (2,2) & 0.9	& 0.7	& 0.9	& 0.95	& 0.9	& 0.9	& 0.9	& 0.8	& 0.25	& 0.25	& 0.9	& 0.95 & 0.9 (6) \\
\hline \hline
\end{tabular}
\end{table}

\begin{table}
\centering
\caption{The best value of threshold $r$ for SLPA and the best value of threshold $tr$ for SpeakEasy determined by the twelve community quality metrics with four possible combinations of the two versions of belonging coefficient and two version of belonging function on \textbf{Collins\_cyc\_w}. The best value for subcolumn of the last column is marked by red italic font.}
\label{tab:collins_cyc_w:global}       
\vspace{-0.8em}
\setlength{\tabcolsep}{2.5pt}
\begin{tabular}{c||c|c|c|c|c|c|c|c|c|c|c|c|c||c}
\hline \hline
Algorithm & (BC,BF) & $Q_{ov}$ & $NQ_{ov}$ & $Q_{ov}^L$ & $Q_{ds}^{ov}$ & $IE$ & $ID$ & $CNT$ & $BE$ & $EXP$ & $CND$ & $F$ & $D$ & \\
\hline
\multirow{4}{*}{SLPA} & (1,1) & 0.05	& 0.35	& 0.05	& 0.35	& 0.05	& 0.25	& 0.05	& 0.5	& 0.05	& 0.05	& 0.05	& 0.2 & 0.05 (7) \\
\cline{2-15}
& (1,2) & 0.05	& 0.35	& 0.05	& 0.35	& 0.05	& 0.25	& 0.4	& 0.05	& 0.05	& 0.05	& 0.05	& 0.35 & 0.05 (7) \\
\cline{2-15}
& (2,1) & 0.05	& 0.35	& 0.05	& 0.35	& 0.05	& 0.25	& 0.05	& 0.5	& 0.5	& 0.05	& 0.05	& 0.2 & 0.05 (6) \\
\cline{2-15}
& (2,2) & 0.05	& 0.5	& 0.05	& 0.35	& 0.05	& 0.25	& 0.05	& 0.05	& 0.05	& 0.05	& 0.05	& 0.2 & \textcolor{red}{\textbf{\emph{0.05} (\emph{8})}} \\
\cline{1-15} \cline{1-15}

\multirow{4}{*}{SpeakEasy} & (1,1) & 0.1	& 0.55	& 0.55	& 0.9	& 0.1	& 0.65	& 0.1	& 0.95	& 0.55	& 0.55	& 0.55	& 0.55 & 0.55 (6) \\
\cline{2-15}
& (1,2) & 0.55	& 0.55	& 0.55	& 0.9	& 0.55	& 0.65	& 0.55	& 0.95	& 0.55	& 0.55	& 0.55	& 0.55 & \textcolor{red}{\textbf{\emph{0.55} (\emph{9})}} \\
\cline{2-15}
& (2,1) & 0.1	& 0.4	& 0.55	& 0.9	& 0.1	& 0.65	& 0.1	& 0.95	& 0.8	& 0.55	& 0.55	& 0.45 & \{0.1,0.55\} (3) \\
\cline{2-15}
& (2,2) & 0.55	& 0.4	& 0.55	& 0.9	& 0.55	& 0.65	& 0.55	& 0.8	& 0.8	& 0.8	& 0.7	& 0.45 & 0.55 (4) \\
\hline \hline
\end{tabular}
\end{table}

\begin{table}
\centering
\caption{The best value of threshold $r$ for SLPA, the best value of parameter $k$ for CFinder, and the best value of threshold $tr$ for SpeakEasy determined by the twelve community quality metrics with four possible combinations of the two versions of belonging coefficient and two version of belonging function on \textbf{Collins\_mips}. The best value for subcolumn of the last column is marked by red italic font.}
\label{tab:collins_mips:global}       
\vspace{-0.8em}
\setlength{\tabcolsep}{1.5pt}
\begin{tabular}{c||c|c|c|c|c|c|c|c|c|c|c|c|c||c}
\hline \hline
Algorithm & (BC,BF) & $Q_{ov}$ & $NQ_{ov}$ & $Q_{ov}^L$ & $Q_{ds}^{ov}$ & $IE$ & $ID$ & $CNT$ & $BE$ & $EXP$ & $CND$ & $F$ & $D$ & \\
\hline
\multirow{4}{*}{SLPA} & (1,1) & 0.15 & 0.45	& 0.45	& 0.45	& 0.05	& 0.5	& 0.05	& 0.5	& 0.45	& 0.15	& 0.05	& 0.45 & 0.45 (5) \\
\cline{2-15}
& (1,2) & 0.45	& 0.45	& 0.45	& 0.45	& 0.05	& 0.5	& 0.15	& 0.5	& 0.45	& 0.05	& 0.05	& 0.45 & \textcolor{red}{\textbf{\emph{0.45} (\emph{6})}} \\
\cline{2-15}
& (2,1) & 0.15	& 0.45	& 0.05	& 0.45	& 0.05	& 0.5	& 0.05	& 0.5	& 0.5	& 0.45	& 0.05	& 0.45 & \{0.05,0.45\} {4} \\
\cline{2-15}
& (2,2) & 0.45	& 0.5	& 0.05	& 0.45	& 0.05	& 0.5	& 0.15	& 0.5	& 0.45	& 0.05	& 0.05	& 0.45 & \{0.05,0.45\} (4) \\
\cline{1-15} \cline{1-15}

\multirow{4}{*}{CFinder} & (1,1) & 4	& 3	& 3	& 5	& 3	& 3	& 3	& 3	& 3	& 3	& 3	& 3 & \textcolor{red}{\textbf{\emph{3} (\emph{10})}} \\
\cline{2-15}
& (1,2) & 4	& 3	& 3	& 5	& 3	& 3	& 3	& 3	& 3	& 3	& 3	& 3 & \textcolor{red}{\textbf{\emph{3} (\emph{10})}} \\
\cline{2-15}
& (2,1) & 4	& 7	& 3	& 7	& 3	& 3	& 3	& 19	& 19	& 20	& 3	& 3 & 3 (6) \\
\cline{2-15}
& (2,2) & 4	& 7	& 3	& 7	& 3	& 3	& 3	& 16-20	& 16-20	& 16-20	& 3	& 3 & 3 (6) \\
\cline{1-15} \cline{1-15}

\multirow{4}{*}{SpeakEasy} & (1,1) & 0.05	& 0.4	& 0.4	& 1	& 0.4	& 0.4	& 0.35	& 0.6	& 0.7	& 0.4	& 0.4	& 0.4 & 0.4 (7) \\
\cline{2-15}
& (1,2) & 0.4	& 0.4	& 0.4	& 1	& 0.4	& 0.4	& 0.4	& 0.6	& 0.4	& 0.4	& 0.4	& 0.4 & \textcolor{red}{\textbf{\emph{0.4} (\emph{10})}} \\
\cline{2-15}
& (2,1) & 0.05	& 0.4	& 0.25	& 1	& 0.4	& 0.4	& 0.35	& 0.6	& 0.75	& 0.4	& 0.4	& 0.75 & 0.4 (5) \\
\cline{2-15}
& (2,2) & 0.25	& 0.95	& 0.25	& 1	& 0.4	& 0.4	& 0.4	& 0.6	& 0.6	& 0.6	& 0.25	& 0.75 & \{0.25,0.4,0.6\} (3) \\
\hline \hline
\end{tabular}
\end{table}

\begin{table}
\centering
\caption{The best value of threshold $r$ for SLPA, the best value of parameter $k$ for CFinder, and the best value of threshold $tr$ for SpeakEasy determined by the twelve community quality metrics with four possible combinations of the two versions of belonging coefficient and two version of belonging function on \textbf{Collins\_sgd}. The best value for subcolumn of the last column is marked by red italic font.}
\label{tab:collins_sgd:global}       
\vspace{-0.8em}
\setlength{\tabcolsep}{2pt}
\begin{tabular}{c||c|c|c|c|c|c|c|c|c|c|c|c|c||c}
\hline \hline
Algorithm & (BC,BF) & $Q_{ov}$ & $NQ_{ov}$ & $Q_{ov}^L$ & $Q_{ds}^{ov}$ & $IE$ & $ID$ & $CNT$ & $BE$ & $EXP$ & $CND$ & $F$ & $D$ & \\
\hline
\multirow{4}{*}{SLPA} & (1,1) & 0.05	& 0.5	& 0.35	& 0.45	& 0.05	& 0.5	& 0.05	& 0.5	& 0.5	& 0.35	& 0.35	& 0.45 & 0.5 (4) \\
\cline{2-15}
& (1,2) & 0.35	& 0.45	& 0.35	& 0.45	& 0.1	& 0.5	& 0.1	& 0.5	& 0.35	& 0.1	& 0.1	& 0.45 & 0.1 (4) \\
\cline{2-15}
& (2,1) & 0.05	& 0.5	& 0.1	& 0.45	& 0.05	& 0.5	& 0.05	& 0.5	& 0.5	& 0.35	& 0.35	& 0.45 & 0.5 (4) \\
\cline{2-15}
& (2,2) & 0.35	& 0.5	& 0.1	& 0.45	& 0.1	& 0.3	& 0.1	& 0.5	& 0.1	& 0.1	& 0.1	& 0.4 & \textcolor{red}{\textbf{\emph{0.1} (\emph{6})}} \\
\cline{1-15} \cline{1-15}

\multirow{4}{*}{CFinder} & (1,1) & 3	& 3	& 3	& 3	& 3	& 3	& 3	& 3	& 3	& 3	& 3	& 3 & \textcolor{red}{\textbf{\emph{3} (\emph{12})}} \\
\cline{2-15}
& (1,2) & 3	& 3	& 3	& 3	& 3	& 3	& 3	& 3	& 3	& 3	& 3	& 3 & \textcolor{red}{\textbf{\emph{3} (\emph{12})}} \\
\cline{2-15}
& (2,1) & 3	& 3	& 3	& 6	& 3	& 3	& 3	& 16-20	& 16-20	& 16-20	& 3	& 3 & 3 (8) \\
\cline{2-15}
& (2,2) & 3	& 3	& 3	& 6	& 3	& 3	& 3	& 16-20	& 16-20	& 16-20	& 3	& 3 & 3 (8) \\
\cline{1-15} \cline{1-15}

\multirow{4}{*}{SpeakEasy} & (1,1) & 0.1 & 0.8	& 0.8	& 0.75	& 0.05	& 0.8	& 0.05	& 0.5	& 0.8	& 0.8	& 0.8	& 0.8 & 0.8 (7) \\
\cline{2-15}
& (1,2) & 0.8	& 0.8	& 0.8	& 0.75	& 0.8	& 0.8	& 0.8	& 0.35	& 0.8	& 0.8	& 0.8	& 0.8 & \textcolor{red}{\textbf{\emph{0.8} (\emph{10})}} \\
\cline{2-15}
& (2,1) & 0.05	& 0.8	& 0.35	& 0.5	& 0.05	& 0.8	& 0.15	& 0.5	& 0.5	& 0.8	& 0.8	& 0.5 & \{0.5,0.8\} (4) \\
\cline{2-15}
& (2,2) & 0.35	& 0.8	& 0.35	& 0.5	& 0.8	& 0.8	& 0.8	& 0.5	& 0.5	& 0.5	& 0.8	& 0.5 & \{0.5,0.8\} (5) \\
\hline \hline
\end{tabular}
\end{table}

\begin{table}
\centering
\caption{The best value of threshold $r$ for SLPA, the best value of parameter $k$ for CFinder, and the best value of threshold $tr$ for SpeakEasy determined by the twelve community quality metrics with four possible combinations of the two versions of belonging coefficient and two version of belonging function on \textbf{Gavin\_cyc}. The best value for subcolumn of the last column is marked by red italic font.}
\label{tab:gavin_cyc:global}       
\vspace{-0.8em}
\setlength{\tabcolsep}{2.7pt}
\begin{tabular}{c||c|c|c|c|c|c|c|c|c|c|c|c|c||c}
\hline \hline
Algorithm & (BC,BF) & $Q_{ov}$ & $NQ_{ov}$ & $Q_{ov}^L$ & $Q_{ds}^{ov}$ & $IE$ & $ID$ & $CNT$ & $BE$ & $EXP$ & $CND$ & $F$ & $D$ & \\
\hline
\multirow{4}{*}{SLPA} & (1,1) & 0.05	& 0.45	& 0.5	& 0.45	& 0.05	& 0.5	& 0.05	& 0.5	& 0.5	& 0.45	& 0.45	& 0.45 & 0.45 (5) \\
\cline{2-15}
& (1,2) & 0.5	& 0.45	& 0.5	& 0.45	& 0.15	& 0.5	& 0.2	& 0.5	& 0.45	& 0.45	& 0.45	& 0.45 & 0.45 (6) \\
\cline{2-15}
& (2,1) & 0.05	& 0.45	& 0.2	& 0.45	& 0.05	& 0.5	& 0.05	& 0.5	& 0.5	& 0.45	& 0.45	& 0.45 & 0.45 (5) \\
\cline{2-15}
& (2,2) & 0.45	& 0.45	& 0.2	& 0.45	& 0.15	& 0.5	& 0.2	& 0.5	& 0.45	& 0.45	& 0.45	& 0.45 & \textcolor{red}{\textbf{\emph{0.45} (\emph{7})}} \\
\cline{1-15} \cline{1-15}

\multirow{4}{*}{CFinder} & (1,1) & 3	& 3	& 3	& 3	& 3	& 3	& 3	& 4	& 3	& 3	& 3	& 3 & 3 (11) \\
\cline{2-15}
& (1,2) & 3	& 3	& 3	& 3	& 3	& 3	& 3	& 3	& 3	& 3	& 3	& 3 & \textcolor{red}{\textbf{\emph{3} (\emph{12})}} \\
\cline{2-15}
& (2,1) & 3	& 4	& 3	& 5	& 3	& 3	& 3	& 13	& 13	& 18-20	& 3	& 4 & 3 (6) \\
\cline{2-15}
& (2,2) & 3	& 4	& 3	& 5	& 3	& 3	& 3	& 11-20	& 11-20	& 11-20	& 3	& 4 & 3 (6) \\
\cline{1-15} \cline{1-15}

\multirow{4}{*}{SpeakEasy} & (1,1) & 0.25	& 0.95	& 0.7	& 0.85	& 0.05	& 1	& 0.05	& 0.7	& 0.7	& 0.7	& 0.7	& 0.7 & 0.7 (6) \\
\cline{2-15}
& (1,2) & 0.7	& 0.95	& 0.7	& 0.85	& 0.7	& 1	& 1	& 0.7	& 0.7	& 0.7	& 0.7	& 0.7 & \textcolor{red}{\textbf{\emph{0.7} (\emph{8})}} \\
\cline{2-15}
& (2,1) & 0.1	& 0.95	& 0.7	& 0.85	& 0.05	& 1	& 0.05	& 1	& 0.7	& 0.7	& 0.7	& 0.7 & 0.7 (5) \\
\cline{2-15}
& (2,2) & 0.7	& 0.95	& 0.7	& 0.85	& 0.7	& 1	& 1	& 0.7	& 0.7	& 0.85	& 0.7	& 0.85 & 0.7 (6) \\
\hline \hline
\end{tabular}
\end{table}

\begin{table}
\centering
\caption{The best value of $r$ for SLPA and the best value of $tr$ for SpeakEasy determined by the twelve community quality metrics with four combinations of belonging coefficient and belonging function on \textbf{Gavin\_cyc\_w}. The best value for subcolumn of the last column is marked by red italic font.}
\label{tab:gavin_cyc_w:global}       
\vspace{-0.8em}
\setlength{\tabcolsep}{1pt}
\begin{tabular}{c||c|c|c|c|c|c|c|c|c|c|c|c|c||c}
\hline \hline
Algorithm & (BC,BF) & $Q_{ov}$ & $NQ_{ov}$ & $Q_{ov}^L$ & $Q_{ds}^{ov}$ & $IE$ & $ID$ & $CNT$ & $BE$ & $EXP$ & $CND$ & $F$ & $D$ & \\
\hline
\multirow{4}{*}{SLPA} & (1,1) & 0.05	& 0.4	& 0.35	& 0.3	& 0.05	& 0.35	& 0.05	& 0.5	& 0.45	& 0.35	& 0.35	& 0.35 & 0.35 (5) \\
\cline{2-15}
& (1,2) & 0.35	& 0.4	& 0.35	& 0.3	& 0.05	& 0.35	& 0.3	& 0.45	& 0.3	& 0.3	& 0.3	& 0.3 & \textcolor{red}{\textbf{\emph{0.3} (\emph{6})}} \\
\cline{2-15}
& (2,1) & 0.05	& 0.45	& 0.05	& 0.3	& 0.05	& 0.35	& 0.05	& 0.5	& 0.5	& 0.35	& 0.35	& 0.45 & 0.05 (4) \\
\cline{2-15}
& (2,2) & 0.3	& 0.45	& 0.05	& 0.3	& 0.05	& 0.35	& 0.3	& 0.45	& 0.5	& 0.5	& 0.3	& 0.3 & 0.3 (5) \\
\cline{1-15} \cline{1-15}

\multirow{4}{*}{SpeakEasy} & (1,1) & 0.05	& 0.4	& 0.45	& 0.65	& 0.05	& 0.65	& 0.1	& 0.95	& 0.7	& 0.7	& 0.7	& 0.7 & 0.7 (4) \\
\cline{2-15}
& (1,2) & 0.45	& 0.4	& 0.45	& 0.65	& 0.7	& 0.65	& 0.7	& 0.95	& 0.45	& 0.7	& 0.7	& 0.7 & \textcolor{red}{\textbf{\emph{0.7} (\emph{5})}} \\
\cline{2-15}
& (2,1) & 0.05	& 0.7	& 0.45	& 0.5	& 0.05	& 0.65	& 0.1	& 0.95	& 0.95	& 0.5	& 0.7	& 1 & \{0.05,0.5,0.7,0.95\} (2) \\
\cline{2-15}
& (2,2) & 0.45	& 0.7	& 0.45	& 0.65	& 0.7	& 0.65	& 0.7	& 0.95	& 0.95	& 0.9	& 0.4	& 1 & 0.7 (3) \\
\hline \hline
\end{tabular}
\end{table}

Tables~\ref{tab:collins_cyc:global}-\ref{tab:gavin_sgd:global} show the best value of threshold $r$ for SLPA, the best value of parameter $k$ for CFinder, and the best value of threshold $tr$ for SpeakEasy determined by the twelve community quality metrics with four possible combinations of the two versions of belonging coefficient and two version of belonging function on the eight protein-protein interaction networks. Results for CFinder on \textbf{Collins\_cyc\_w} and \textbf{Gavin\_cyc\_w} are not provided because it has not finished running on \textbf{Collins\_cyc\_w} and \textbf{Gavin\_cyc\_w} for more than two months processing many potential $k$-cliques associated with intensity larger than the intensity threshold \cite{CPMw}.

From Table~\ref{tab:collins_cyc:global} we could see that on \textbf{Collins\_cyc} SLPA implies that the first version of belonging function is better than the second version, CFinder indicates that the first version of belonging coefficient is better than the second one, and SpeakEasy demonstrates that (BC,BF)=(1,2) is the best among all four combinations. In summary, there are two out of three algorithms show that (BC,BF)=(1,2) is the best on \textbf{Collins\_cyc}. Also, SLPA in Table~\ref{tab:collins_cyc_w:global} implies that (BC,BF)=(2,2) is the best, while SpeakEasy shows that (BC,BF)=(1,2) is the best on \textbf{Collins\_cyc\_w}. It can be seen from Table~\ref{tab:collins_mips:global} that all three algorithms support the conclusion that (BC,BF)=(1,2) is the best among the four combinations on \textbf{Collins\_mips}. In addition on \textbf{Collins\_sgd} (Table~\ref{tab:collins_sgd:global}), SLPA shows that (BC,BF)=(2,2) is the best, CFinder implies that the first version of belonging coefficient is better than the second version, and SpeakEasy indicates that (BC,BF)=(1,2) is the best among the four combinations of belonging coefficient and belonging function. Thus, two out of three algorithms conclude that (BC,BF)=(1,2) is the best among the four combinations on \textbf{Collins\_sgd}.

It can be observed from Table~\ref{tab:gavin_cyc:global} that on \textbf{Ganvin\_cyc} SLPA shows that (BC,BF)=(2,2) is the best, while CFinder and SpeakEasy indicate that (BC,BF)=(1,2) is the best. Hence, two out of three algorithms support that (BC,BF)=(1,2) is the best among the four combinations on \textbf{Ganvin\_cyc}. From Tables~\ref{tab:gavin_cyc_w:global}-\ref{tab:gavin_sgd:global}, we could learn that all three algorithms show that (BC,BF)=(1,2) is the best among the four combinations on \textbf{Ganvin\_cyc\_w}, \textbf{Ganvin\_mips}, and \textbf{Ganvin\_sgd}.

From the analysis above, we could conclude that (BC,BF)=(1,2) is the best among the four possible combinations of the two versions of belonging coefficient and two versions of belonging function on protein-protein interaction networks.

\begin{table}
\vspace{1em}
\centering
\caption{The best value of threshold $r$ for SLPA, the best value of parameter $k$ for CFinder, and the best value of threshold $tr$ for SpeakEasy determined by the twelve community quality metrics with four possible combinations of the two versions of belonging coefficient and two version of belonging function on \textbf{Gavin\_mips}. The best value for subcolumn of the last column is marked by red italic font.}
\label{tab:gavin_mips:global}       
\vspace{-0.8em}
\setlength{\tabcolsep}{2.8pt}
\begin{tabular}{c||c|c|c|c|c|c|c|c|c|c|c|c|c||c}
\hline \hline
Algorithm & (BC,BF) & $Q_{ov}$ & $NQ_{ov}$ & $Q_{ov}^L$ & $Q_{ds}^{ov}$ & $IE$ & $ID$ & $CNT$ & $BE$ & $EXP$ & $CND$ & $F$ & $D$ & \\
\hline
\multirow{4}{*}{SLPA} & (1,1) & 0.05	& 0.5	& 0.5	& 0.5	& 0.05	& 0.5	& 0.05	& 0.5	& 0.5	& 0.5	& 0.5	& 0.5 & 0.5 (9) \\
\cline{2-15}
& (1,2) & 0.5	& 0.5	& 0.5	& 0.5	& 0.1	& 0.5	& 0.5	& 0.5	& 0.5	& 0.5	& 0.5	& 0.5 & \textcolor{red}{\textbf{\emph{0.5} (\emph{11})}} \\
\cline{2-15}
& (2,1) & 0.05	& 0.5	& 0.25	& 0.5	& 0.05	& 0.5	& 0.05	& 0.5	& 0.5	& 0.5	& 0.5	& 0.5 & 0.5 (8) \\
\cline{2-15}
& (2,2) & 0.5	& 0.45	& 0.25	& 0.5	& 0.1	& 0.4	& 0.3	& 0.5	& 0.35	& 0.5	& 0.5	& 0.5 & 0.5 (6) \\
\cline{1-15} \cline{1-15}

\multirow{4}{*}{CFinder} & (1,1) & 3	& 3	& 3	& 3	& 3	& 3	& 3	& 3	& 3	& 3	& 3	& 3 & \textcolor{red}{\textbf{\emph{3} (\emph{12})}} \\
\cline{2-15}
& (1,2) & 3	& 3	& 3	& 3	& 3	& 3	& 3	& 3	& 3	& 3	& 3	& 3 & \textcolor{red}{\textbf{\emph{3} (\emph{12})}} \\
\cline{2-15}
& (2,1) & 3	& 3	& 3	& 4	& 3	& 3	& 3	& 12	& 12	& 18-20	& 3	& 4 & 3 (7) \\
\cline{2-15}
& (2,2) & 3	& 4	& 3	& 4	& 3	& 3	& 3	& 11-20	& 11-20	& 11-20	& 3	& 4 & 3 (6) \\
\cline{1-15} \cline{1-15}

\multirow{4}{*}{SpeakEasy} & (1,1) & 0.05	& 0.4	& 0.9	& 0.7	& 0.05	& 0.95	& 0.1	& 0.9	& 0.9	& 0.9	& 0.9	& 0.9 & 0.9 (6) \\
\cline{2-15}
& (1,2) & 0.9	& 0.4	& 0.9	& 0.7	& 0.9	& 0.95	& 0.9	& 0.4	& 0.9	& 0.9	& 0.9	& 0.9 & \textcolor{red}{\textbf{\emph{0.9} (\emph{8})}} \\
\cline{2-15}
& (2,1) & 0.05	& 0.95	& 0.9	& 0.7	& 0.05	& 0.95	& 0.1	& 0.9	& 0.9	& 0.9	& 0.9	& 0.4 & 0.9 (5) \\
\cline{2-15}
& (2,2) & 0.9	& 0.95	& 0.9	& 0.7	& 0.9	& 0.35	& 0.9	& 0.4	& 0.4	& 0.4	& 0.9	& 0.4 & 0.9 (5) \\
\hline \hline
\end{tabular}
\end{table}

\begin{table}
\centering
\caption{The best value of threshold $r$ for SLPA, the best value of parameter $k$ for CFinder, and the best value of threshold $tr$ for SpeakEasy determined by the twelve community quality metrics with four possible combinations of the two versions of belonging coefficient and two version of belonging function on \textbf{Gavin\_sgd}. The best value for subcolumn of the last column is marked by red italic font.}
\label{tab:gavin_sgd:global}       
\vspace{-0.8em}
\setlength{\tabcolsep}{3pt}
\begin{tabular}{c||c|c|c|c|c|c|c|c|c|c|c|c|c||c}
\hline \hline
Algorithm & (BC,BF) & $Q_{ov}$ & $NQ_{ov}$ & $Q_{ov}^L$ & $Q_{ds}^{ov}$ & $IE$ & $ID$ & $CNT$ & $BE$ & $EXP$ & $CND$ & $F$ & $D$ & \\
\hline
\multirow{4}{*}{SLPA} & (1,1) & 0.05	& 0.5	& 0.5	& 0.5	& 0.1	& 0.35	& 0.1	& 0.5	& 0.5	& 0.5	& 0.5	& 0.5 & 0.5 (8) \\
\cline{2-15}
& (1,2) & 0.5	& 0.5	& 0.5	& 0.5	& 0.1	& 0.35	& 0.5	& 0.35	& 0.5	& 0.5	& 0.5	& 0.5 & \textcolor{red}{\textbf{\emph{0.5} (\emph{9})}} \\
\cline{2-15}
& (2,1) & 0.05	& 0.5	& 0.15	& 0.5	& 0.1	& 0.35	& 0.05	& 0.5	& 0.5	& 0.5	& 0.5	& 0.5 & 0.5 (7) \\
\cline{2-15}
& (2,2) & 0.5	& 0.45	& 0.15	& 0.5	& 0.1	& 0.35	& 0.5	& 0.35	& 0.5	& 0.5	& 0.5	& 0.5 & 0.5 (7) \\
\cline{1-15} \cline{1-15}

\multirow{4}{*}{CFinder} & (1,1) & 3	& 3	& 3	& 4	& 3	& 3	& 3	& 3	& 3	& 3	& 3	& 3 & \textcolor{red}{\textbf{\emph{3} (\emph{11})}} \\
\cline{2-15}
& (1,2) & 3	& 3	& 3	& 4	& 3	& 3	& 3	& 3	& 3	& 3	& 3	& 3 & \textcolor{red}{\textbf{\emph{3} (\emph{11})}} \\
\cline{2-15}
& (2,1) & 3	& 4	& 3	& 5	& 3	& 3	& 3	& 13	& 13	& 13	& 3	& 4 & 3 (6) \\
\cline{2-15}
& (2,2) & 3	& 4	& 3	& 5	& 3	& 3	& 3	& 11-17	& 11-17	& 11-17	& 3	& 4 & 3 (6) \\
\cline{1-15} \cline{1-15}

\multirow{4}{*}{SpeakEasy} & (1,1) & 0.05	& 0.8	& 0.45	& 0.5	& 0.8	& 0.8	& 0.15	& 0.7	& 0.45	& 0.8	& 0.8	& 0.45 & 0.8 (5) \\
\cline{2-15}
& (1,2) & 0.45	& 0.8	& 0.45	& 0.5	& 0.8	& 0.8	& 0.8	& 0.7	& 0.45	& 0.8	& 0.8	& 0.45 & \textcolor{red}{\textbf{\emph{0.8} (\emph{6})}} \\
\cline{2-15}
& (2,1) & 0.05	& 0.8	& 0.8	& 0.5	& 0.8	& 0.8	& 0.15	& 0.7	& 0.45	& 0.8	& 0.8	& 0.45 & \textcolor{red}{\textbf{\emph{0.8} (\emph{6})}} \\
\cline{2-15}
& (2,2) & 0.45	& 0.8	& 0.8	& 0.5	& 0.8	& 0.8	& 0.8	& 0.7	& 0.45	& 0.45	& 0.8	& 0.45 & \textcolor{red}{\textbf{\emph{0.8} (\emph{6})}} \\
\hline \hline
\end{tabular}
\end{table}

\vspace{1em}
\subsection{Amazon Product Network}
This is a product co-purchased network of the Amazon website \cite{Leskovec_dataset}. If a product $p_i$ is frequently co-purchased with product $p_j$, the graph contains an undirected edge from $p_i$ to $p_j$. There are 319948 nodes and 880215 edges. Table~\ref{tab:amazon:global} shows the best value of threshold $r$ for SLPA, the best value of parameter $k$ for CFinder, and the best value of threshold $tr$ for SpeakEasy determined by the twelve community quality metrics with four possible combinations of the two versions of belonging coefficient and two version of belonging function on Amazon product network. We can see that all three algorithms show that (BC,BF)=(1,2) is the best among the four combinations of belonging coefficient and belonging function.

\begin{table}
\vspace{1em}
\centering
\caption{The best value of threshold $r$ for SLPA, the best value of parameter $k$ for CFinder, and the best value of threshold $tr$ for SpeakEasy determined by the twelve quality metrics with four combinations of the two versions of belonging coefficient and two version of belonging function on Amazon product network. The best value for subcolumn of the last column is marked by red italic font.}
\label{tab:amazon:global}       
\vspace{-0.8em}
\setlength{\tabcolsep}{3pt}
\begin{tabular}{c||c|c|c|c|c|c|c|c|c|c|c|c|c||c}
\hline \hline
Algorithm & (BC,BF) & $Q_{ov}$ & $NQ_{ov}$ & $Q_{ov}^L$ & $Q_{ds}^{ov}$ & $IE$ & $ID$ & $CNT$ & $BE$ & $EXP$ & $CND$ & $F$ & $D$ & \\
\hline
\multirow{4}{*}{SLPA} & (1,1) & 0.05	& 0.5	& 0.5	& 0.5	& 0.1	& 0.45	& 0.05	& 0.5	& 0.5	& 0.5	& 0.5	& 0.5 & 0.5 (8) \\
\cline{2-15}
& (1,2) & 0.5	& 0.5	& 0.5	& 0.5	& 0.5	& 0.4	& 0.5	& 0.45	& 0.5	& 0.5	& 0.5	& 0.5 & \textcolor{red}{\textbf{\emph{0.5} (\emph{10})}} \\
\cline{2-15}
& (2,1) & 0.05	& 0.5	& 0.3	& 0.45	& 0.05	& 0.45	& 0.05	& 0.5	& 0.5	& 0.5	& 0.5	& 0.5 & 0.5 (6) \\
\cline{2-15}
& (2,2) & 0.5	& 0.5	& 0.3	& 0.5	& 0.5	& 0.4	& 0.5	& 0.4	& 0.5	& 0.5	& 0.5	& 0.5 & 0.5 (9) \\
\cline{1-15} \cline{1-15}

\multirow{4}{*}{CFinder} & (1,1) & 3	& 3	& 3	& 3	& 3	& 3	& 3	& 7	& 3	& 3	& 3	& 3 & \textcolor{red}{\textbf{\emph{3} (\emph{11})}} \\
\cline{2-15}
& (1,2) & 3	& 3	& 3	& 3	& 3	& 3	& 3	& 5	& 3	& 3	& 3	& 3 & \textcolor{red}{\textbf{\emph{3} (\emph{11})}} \\
\cline{2-15}
& (2,1) & 3	& 4	& 3	& 4	& 3	& 3	& 3	& 7	& 7	& 7	& 3	& 5 & 3 (6) \\
\cline{2-15}
& (2,2) & 3	& 4	& 3	& 4	& 3	& 3	& 3	& 7	& 7	& 7	& 3	& 4 & 3 (6) \\
\cline{1-15} \cline{1-15}

\multirow{4}{*}{SpeakEasy} & (1,1) & 0.1	& 1	& 1	& 1	& 0.1	& 1	& 0.1	& 1	& 1	& 1	& 1	& 1 & 1 (9) \\
\cline{2-15}
& (1,2) & 1	& 1	& 1	& 1	& 1	& 1	& 1	& 1	& 1	& 0.95	& 1	& 1 & \textcolor{red}{\textbf{\emph{1} (\emph{11})}} \\
\cline{2-15}
& (2,1) & 0.1	& 1	& 0.95	& 1	& 0.1	& 1	& 0.1	& 1	& 1	& 1	& 1	& 1 & 1 (8) \\
\cline{2-15}
& (2,2) & 1	& 1	& 0.95	& 1	& 1	& 1	& 1	& 1	& 1	& 1	& 1	& 1 & \textcolor{red}{\textbf{\emph{1} (\emph{11})}} \\
\hline \hline
\end{tabular}
\end{table}

\vspace{1em}
\subsection{DBLP Collaboration Network}
The DBLP computer science bibliography provides a comprehensive list of research papers in computer science. In this DBLP co-authorship network, two authors are connected if they publish at least one paper together \cite{Leskovec_dataset}. It has 260998 nodes and 950059 edges in total. Table~\ref{tab:dblp:global} shows the best value of threshold $r$ for SLPA, the best value of parameter $k$ for CFinder, and the best value of threshold $tr$ for SpeakEasy determined by the twelve community quality metrics with four possible combinations of the two versions of belonging coefficient and two version of belonging function on DBLP collaboration network. We can see that all three algorithms show that (BC,BF)=(1,2) is the best among the four combinations of belonging coefficient and belonging function.

\begin{table}
\vspace{1em}
\centering
\caption{The best value of threshold $r$ for SLPA, the best value of parameter $k$ for CFinder, and the best value of threshold $tr$ for SpeakEasy determined by the twelve quality metrics with four combinations of the two versions of belonging coefficient and two version of belonging function on DBLP collaboration network. The best value for subcolumn of the last column is marked by red italic font.}
\label{tab:dblp:global}       
\vspace{-0.8em}
\setlength{\tabcolsep}{3pt}
\begin{tabular}{c||c|c|c|c|c|c|c|c|c|c|c|c|c||c}
\hline \hline
Algorithm & (BC,BF) & $Q_{ov}$ & $NQ_{ov}$ & $Q_{ov}^L$ & $Q_{ds}^{ov}$ & $IE$ & $ID$ & $CNT$ & $BE$ & $EXP$ & $CND$ & $F$ & $D$ & \\
\hline
\multirow{4}{*}{SLPA} & (1,1) & 0.05	& 0.5	& 0.5	& 0.45	& 0.05	& 0.45	& 0.05	& 0.5	& 0.5	& 0.5	& 0.5	& 0.5 & 0.5 (7) \\
\cline{2-15}
& (1,2) & 0.5	& 0.5	& 0.5	& 0.5	& 0.5	& 0.45	& 0.5	& 0.45	& 0.5	& 0.5	& 0.5	& 0.5 & \textcolor{red}{\textbf{\emph{0.5} (\emph{10})}} \\
\cline{2-15}
& (2,1) & 0.05	& 0.5	& 0.3	& 0.45	& 0.05	& 0.45	& 0.05	& 0.5	& 0.5	& 0.5	& 0.5	& 0.5 & 0.5 (6) \\
\cline{2-15}
& (2,2) & 0.5	& 0.5	& 0.3	& 0.5	& 0.15	& 0.45	& 0.5	& 0.45	& 0.5	& 0.5	& 0.5	& 0.5 & 0.5 (8) \\
\cline{1-15} \cline{1-15}

\multirow{4}{*}{CFinder} & (1,1) & 4 &	3 &	3	& 3	& 3	& 3	& 3	& 18 & 7	& 3	& 3	& 3 & 3 (9) \\
\cline{2-15}
& (1,2) & 4	& 3	& 3	& 3	& 3	& 3	& 3	& 9	& 3	& 3	& 3	& 3  & \textcolor{red}{\textbf{\emph{3} (\emph{10})}} \\
\cline{2-15}
& (2,1) & 4	& 5	& 3	& 6	& 3	& 3	& 3	& 20	& 19	& 20	& 3	& 16  & 3 (5) \\
\cline{2-15}
& (2,2) & 4	& 5	& 3	& 5	& 3	& 3	& 3	& 20	& 19	& 20	& 3	& 9 & 3 (5) \\
\cline{1-15} \cline{1-15}

\multirow{4}{*}{SpeakEasy} & (1,1) & 0.1	& 1	& 1	& 1	& 0.1	& 1	& 0.1	& 1	& 1	& 1	& 1	& 1 & 1 (9) \\
\cline{2-15}
& (1,2) & 1	& 0.95	& 1	& 1	& 1	& 1	& 0.95	& 1	& 1	& 1	& 1	& 1 & \textcolor{red}{\textbf{\emph{1} (\emph{10})}} \\
\cline{2-15}
& (2,1) & 0.1	& 1	& 0.95	& 1	& 0.1	& 1	& 0.1	& 1	& 1	& 1	& 1	& 1 & 1 (8) \\
\cline{2-15}
& (2,2) & 1	& 1	& 0.95	& 1	& 1	& 1	& 0.95	& 1	& 1	& 1	& 1	& 1 & \textcolor{red}{\textbf{\emph{1} (\emph{10})}} \\
\hline \hline
\end{tabular}
\end{table}

\vspace{1em}
\section{LFR Benchmark Networks}

LFR (named after the initials of names of authors) benchmark networks \cite{LFROv} have become a standard in the evaluation of the performance of community detection algorithms. The LFR benchmark network that we used here has $1000$ nodes with average degree $15$ and maximum degree $50$. The exponent $\gamma$ for the degree sequence varies from $2$ to $3$. The exponent $\beta$ for the community size distribution ranges from $1$ to $2$. Then, four pairs of the exponents $(\gamma, \beta) = (2, 1), (2, 2), (3, 1), \text{and}~(3, 2)$ are chosen in order to explore the widest spectrum of graph structures. The mixing parameter $\mu$ is varied from $0.05$ to $0.95$. It means that each node shares a fraction $(1- \mu)$ of its edges with the other nodes in its community and shares a fraction $\mu$ of its edges with the nodes outside its community. Thus, low mixing parameters indicate strong community structure. The degree of overlap is determined by two parameters. $O_n$ is the number of overlapping nodes, and $O_m$ is the number of communities to which each overlapping node belongs. $O_n$ here is set to 10\% of the total number of nodes. Instead of fixing $O_m$, we allow it to vary from 1 to 8 indicating the overlapping diversity of overlapping nodes. By increasing the value of $O_m$, we create harder detection tasks. Also, we generate $10$ network instances for each configuration of these parameters. Hence, each metric value for a certain configuration of LFR represents the average metric values of all $10$ instances. Since the experimental results are similar for all four pairs of exponents $(\gamma, \beta) = (2, 1), (2, 2), (3, 1), \text{and}~(3, 2)$, for the sake of brevity, we only present the results for $(\gamma, \beta) = (2, 1)$ here. In addition, there results are similar for different values of $\mu$, so here we only show the results for $\mu=0.3,0.35,\text{and~} 0.4$. We choose $\mu=0.3,0.35,\text{and~} 0.4$ to better illustrate the results since with $\mu=0.3,0.35,\text{and~} 0.4$ the community structures generated by LFR are around the boundary of well-separated communities and well-connected communities. For each node, $\mu=0.5$ means that the number of its edges with other nodes in its communities is equal to the number of its edges with nodes outside its community, which makes the community structure difficult to discover.

\begin{table}
\centering
\caption{The best value of threshold $r$ for SLPA determined by the twelve community quality metrics with four possible combinations of the two versions of belonging coefficient and two version of belonging function on LFR benchmark networks with $(\alpha,\beta)=(1,2)$ and $\mu=0.3,0.35,0.4$. The best value for subcolumn of the last column is marked by red italic font.}
\label{tab:lfr:slpa}       
\vspace{-1em}
\setlength{\tabcolsep}{1.5pt}
\begin{tabular}{c|c||c|c|c|c|c|c|c|c|c|c|c|c|c||c}
\hline \hline
$\mu$ & $O_m$ & (BC,BF) & $Q_{ov}$ & $NQ_{ov}$ & $Q_{ov}^L$ & $Q_{ds}^{ov}$ & $IE$ & $ID$ & $CNT$ & $BE$ & $EXP$ & $CND$ & $F$ & $D$ & \\
\hline
\multirow{20}{*}{0.3} & \multirow{5}{*}{1} & (1,1) & 0.3 &	0.5 &	0.5	& 0.5	& 0.05	& 0.45	& 0.05	& 0.45	& 0.5	& 0.5	& 0.5	& 0.5	& 0.5 (7)
 \\
\cline{3-16}
& & (1,2) & 0.5	& 0.5	& 0.5	& 0.5	& 0.05	& 0.45	& 0.2	& 0.45	& 0.5	& 0.5	& 0.5	& 0.5	& \textcolor{red}{\textbf{\emph{0.5} (\emph{8})}} \\
\cline{3-16}
& & (2,1) & 0.3	& 0.5	& 0.2	& 0.5	& 0.05	& 0.5	& 0.05	& 0.45	& 0.5	& 0.5	& 0.5	& 0.5	& 0.5 (7) \\
\cline{3-16}
& & (2,2) & 0.5	& 0.5	& 0.2	& 0.5	& 0.05	& 0.5	& 0.2	& 0.45	& 0.5	& 0.2	& 0.2	& 0.5	& 0.5 (6) \\
\cline{2-16} \cline{2-16}

& \multirow{5}{*}{2} & (1,1) & 0.05	& 0.5	& 0.5	& 0.25	& 0.05	& 0.5	& 0.05	& 0.5	& 0.5	& 0.5	& 0.5	& 0.5	& \textcolor{red}{\textbf{\emph{0.5} (\emph{8})}} \\
\cline{3-16}
& & (1,2) & 0.5	& 0.5	& 0.5	& 0.5	& 0.3	& 0.5	& 0.3	& 0.45	& 0.4	& 0.3	& 0.3	& 0.4	& 0.5 (5) \\
\cline{3-16}
& & (2,1) & 0.05	& 0.5	& 0.3	& 0.25	& 0.05	& 0.5	& 0.05	& 0.5	& 0.5	& 0.5	& 0.5	& 0.5	& 0.5 (7) \\
\cline{3-16}
& & (2,2) & 0.4	& 0.5	& 0.3	& 0.4	& 0.15	& 0.5	& 0.3	& 0.45	& 0.35	& 0.3	& 0.3	& 0.35	& 0.3 (4) \\
\cline{2-16} \cline{2-16}

& \multirow{5}{*}{4} & (1,1) & 0.05	& 0.5	& 0.5	& 0.5	& 0.05	& 0.5	& 0.05	& 0.5	& 0.5	& 0.5	& 0.5	& 0.5	& 0.5 (9) \\
\cline{3-16}
& & (1,2) & 0.5	& 0.5	& 0.5	& 0.5	& 0.15	& 0.5	& 0.5	& 0.5	& 0.4	& 0.5	& 0.5	& 0.5	& \textcolor{red}{\textbf{\emph{0.5} (\emph{10})}} \\
\cline{3-16}
& & (2,1) & 0.05	& 0.5	& 0.3	& 0.5	& 0.05	& 0.5	& 0.05	& 0.5	& 0.5	& 0.5	& 0.5	& 0.5	& 0.5 (8) \\
\cline{3-16}
& & (2,2) & 0.45	& 0.5	& 0.3	& 0.5	& 0.15	& 0.5	& 0.35	& 0.5	& 0.4	& 0.4	& 0.5	& 0.4	& 0.5 (5) \\
\cline{2-16} \cline{2-16}

& \multirow{5}{*}{6} & (1,1) & 0.05	& 0.5	& 0.5	& 0.5	& 0.05	& 0.5	& 0.05	& 0.5	& 0.5	& 0.5	& 0.5	& 0.5	& 0.5 (9) \\
\cline{3-16}
& & (1,2) & 0.5	& 0.5	& 0.5	& 0.5	& 0.1	& 0.5	& 0.5	& 0.45	& 0.5	& 0.5	& 0.5	& 0.5	& \textcolor{red}{\textbf{\emph{0.5} (\emph{10})}} \\
\cline{3-16}
& & (2,1) & 0.05	& 0.5	& 0.3	& 0.5	& 0.05	& 0.5	& 0.05	& 0.5	& 0.5	& 0.5	& 0.5	& 0.5	& 0.5 (8) \\
\cline{3-16}
& & (2,2) & 0.4	& 0.5	& 0.3	& 0.45	& 0.1	& 0.5	& 0.2	& 0.45	& 0.5	& 0.5	& 0.5	& 0.5	& 0.5 (6) \\
\cline{2-16} \cline{2-16}

& \multirow{5}{*}{8} & (1,1) & 0.1	& 0.5	& 0.5	& 0.5	& 0.05	& 0.5	& 0.05	& 0.5	& 0.5	& 0.5	& 0.5	& 0.5	& 0.5 (9) \\
\cline{3-16}
& & (1,2) & 0.5	& 0.5	& 0.5	& 0.5	& 0.1	& 0.5	& 0.5	& 0.45	& 0.5	& 0.5	& 0.5	& 0.5	& \textcolor{red}{\textbf{\emph{0.5} (\emph{10})}} \\
\cline{3-16}
& & (2,1) & 0.1	& 0.5	& 0.2	& 0.5	& 0.05	& 0.5	& 0.05	& 0.5	& 0.5	& 0.5	& 0.5	& 0.5	& 0.5 (8) \\
\cline{3-16}
& & (2,2) & 0.5	& 0.5	& 0.2	& 0.5	& 0.1	& 0.5	& 0.5	& 0.45	& 0.5	& 0.5	& 0.5	& 0.5	& 0.5 (9) \\
\cline{1-16} \cline{1-16}

\multirow{20}{*}{0.35} & \multirow{5}{*}{1} & (1,1) & 0.25	& 0.5	& 0.25	& 0.4	& 0.05	& 0.5	& 0.15	& 0.5	& 0.5	& 0.25	& 0.25	& 0.25	& 0.25 (5) \\
\cline{3-16}
& & (1,2) & 0.25	& 0.5	& 0.25	& 0.4	& 0.1	& 0.45	& 0.15	& 0.45	& 0.25	& 0.25	& 0.25	& 0.25	& \textcolor{red}{\textbf{\emph{0.25} (\emph{6})}} \\
\cline{3-16}
& & (2,1) & 0.2	& 0.5	& 0.15	& 0.4	& 0.05	& 0.45	& 0.05	& 0.5	& 0.25	& 0.25	& 0.25	& 0.25	& 0.25 (4) \\
\cline{3-16}
& & (2,2) & 0.25	& 0.5	& 0.15	& 0.4	& 0.05	& 0.45	& 0.15	& 0.45	& 0.25	& 0.25	& 0.25	& 0.25	& 0.25 (5) \\
\cline{2-16} \cline{2-16}

& \multirow{5}{*}{2} & (1,1) & 0.05	& 0.5	& 0.45	& 0.3	& 0.05	& 0.5	& 0.05	& 0.5	& 0.5	& 0.5	& 0.5	& 0.5	& \textcolor{red}{\textbf{\emph{0.5} (\emph{7})}} \\
\cline{3-16}
& & (1,2) & 0.45	& 0.5	& 0.45	& 0.45	& 0.15	& 0.45	& 0.3	& 0.45	& 0.45	& 0.3	& 0.3	& 0.45	& \textcolor{red}{\textbf{\emph{0.45} (\emph{7})}} \\
\cline{3-16}
& & (2,1) & 0.05	& 0.5	& 0.3	& 0.25	& 0.05	& 0.5	& 0.05	& 0.5	& 0.5	& 0.5	& 0.5	& 0.5	& \textcolor{red}{\textbf{\emph{0.5} (\emph{7})}} \\
\cline{3-16}
& & (2,2) & 0.45	& 0.5	& 0.3	& 0.45	& 0.15	& 0.45	& 0.15	& 0.45	& 0.45	& 0.3	& 0.3	& 0.45	& 0.45 (6) \\
\cline{2-16} \cline{2-16}

& \multirow{5}{*}{4} & (1,1) & 0.05	& 0.5	& 0.5	& 0.5	& 0.05	& 0.45	&0.05	& 0.5	& 0.5	& 0.5	& 0.5	& 0.5	& 0.5 (8) \\
\cline{3-16}
& & (1,2) & 0.5	& 0.5	& 0.5	& 0.5	& 0.15	& 0.45	& 0.5	& 0.45	& 0.5	& 0.5	& 0.5	& 0.5	& \textcolor{red}{\textbf{\emph{0.5} (\emph{9})}} \\
\cline{3-16}
& & (2,1) & 0.05	& 0.5	& 0.3	& 0.5	& 0.05	& 0.45	& 0.05	& 0.5	& 0.5	& 0.5	& 0.5	& 0.5	& 0.5 (7) \\
\cline{3-16}
& & (2,2) & 0.5	& 0.5	& 0.3	& 0.45	& 0.1	& 0.45	& 0.5	& 0.45	& 0.45	& 0.5	& 0.5	& 0.5	& 0.5 (6) \\
\cline{2-16} \cline{2-16}

& \multirow{5}{*}{6} & (1,1) & 0.05	& 0.5	& 0.5	& 0.45	& 0.05	& 0.5	& 0.05	& 0.5	& 0.5	& 0.5	& 0.5	& 0.5	& 0.5 (8) \\
\cline{3-16}
& & (1,2) & 0.5	& 0.5	& 0.5	& 0.5	& 0.1	& 0.45	& 0.5 & 0.45	& 0.5	& 0.5	& 0.5	& 0.5	& \textcolor{red}{\textbf{\emph{0.5} (\emph{9})}} \\
\cline{3-16}
& & (2,1) & 0.05	& 0.5	& 0.25	& 0.45	& 0.05	& 0.5	&0.05	& 0.5	& 0.5	& 0.5	& 0.5	& 0.5	& 0.5 (7) \\
\cline{3-16}
& & (2,2) & 0.5	& 0.5	& 0.25	& 0.5	& 0.1	& 0.45	 & 0.5	& 0.45	& 0.5	& 0.5	& 0.5	& 0.5	& 0.5 (8) \\
\cline{2-16} \cline{2-16}

& \multirow{5}{*}{8} & (1,1) & 0.35	& 0.5	& 0.5	& 0.45	& 0.05	& 0.45	& 0.05	& 0.5	& 0.5	& 0.5 &	0.5	& 0.5	& 0.5 (7) \\
\cline{3-16}
& & (1,2) & 0.5	& 0.5	& 0.5	& 0.5	& 0.05	& 0.45	& 0.5	& 0.45	& 0.5	& 0.5	& 0.5	& 0.5	& \textcolor{red}{\textbf{\emph{0.5} (\emph{9})}} \\
\cline{3-16}
& & (2,1) & 0.4	& 0.5	& 0.2	& 0.45	& 0.05	& 0.45	& 0.05	& 0.5	& 0.5	& 0.5	& 0.5	& 0.5	& 0.5 (6) \\
\cline{3-16}
& & (2,2) & 0.5	& 0.5	& 0.2	& 0.5	& 0.05	& 0.45	& 0.5	& 0.45	& 0.5	& 0.5	& 0.5	& 0.5	& 0.5 (8) \\
\cline{1-16} \cline{1-16}

\multirow{20}{*}{0.4} & \multirow{5}{*}{1} & (1,1) & 0.25	& 0.5	& 0.5	& 0.45	& 0.05	& 0.45	& 0.1	& 0.5	& 0.5	& 0.25	& 0.5	& 0.5 &	\textcolor{red}{\textbf{\emph{0.5} (\emph{6})}} \\
\cline{3-16}
& & (1,2) & 0.5	& 0.45	& 0.5	& 0.45	& 0.1	& 0.4	& 0.25	& 0.4	& 0.25	& 0.25	& 0.25	& 0.25	& 0.25 (5) \\
\cline{3-16}
& & (2,1) & 0.2	& 0.5	& 0.2	& 0.45	& 0.05	& 0.45	& 0.05	& 0.5	& 0.5	& 0.25	& 0.25	& 0.5	& 0.5 (4) \\
\cline{3-16}
& & (2,2) & 0.45	& 0.45	& 0.2	& 0.45	& 0.1	& 0.4	& 0.2	& 0.4	& 0.25	& 0.2	& 0.25	& 0.25	& \{0.2,0.25,0.45\} (3) \\
\cline{2-16} \cline{2-16}

& \multirow{5}{*}{2} & (1,1) & 0.05	& 0.5	& 0.5	& 0.3	& 0.05	& 0.5	& 0.05	& 0.5	& 0.5	& 0.5	& 0.5	& 0.5	& 0.5 (8) \\
\cline{3-16}
& & (1,2) & 0.5	& 0.5	& 0.5	& 0.5	& 0.15	& 0.5	& 0.5	& 0.4	& 0.45	& 0.5	& 0.5	& 0.5	& \textcolor{red}{\textbf{\emph{0.5} (\emph{9})}} \\
\cline{3-16}
& & (2,1) & 0.05	& 0.5	& 0.3	& 0.3	& 0.05	& 0.5	& 0.05	& 0.5	& 0.5	& 0.5	& 0.5	& 0.5	& 0.5 (7) \\
\cline{3-16}
& & (2,2) & 0.45	& 0.5	& 0.3	& 0.5	& 0.15	& 0.4	& 0.5	& 0.35	& 0.45	& 0.5	& 0.5	& 0.45	& 0.5 (5) \\
\cline{2-16} \cline{2-16}

& \multirow{5}{*}{4} & (1,1) & 0.1	& 0.5	& 0.5	& 0.5	& 0.05	& 0.5	& 0.05	& 0.5	& 0.5	& 0.5	& 0.5	& 0.5	& 0.5 (9) \\
\cline{3-16}
& & (1,2) & 0.5	& 0.5	& 0.5	& 0.5	& 0.2	& 0.5	& 0.5	& 0.4	& 0.5	& 0.5	& 0.5	& 0.5	& \textcolor{red}{\textbf{\emph{0.5} (\emph{10})}} \\
\cline{3-16}
& & (2,1) & 0.1	& 0.5	& 0.3	& 0.5	& 0.05	& 0.5	& 0.05	& 0.5	& 0.5	& 0.5	& 0.5	& 0.5	& 0.5 (8) \\
\cline{3-16}
& & (2,2) & 0.5	& 0.5	& 0.3	& 0.5	& 0.15	& 0.5	& 0.5	& 0.4	& 0.45	& 0.5	& 0.5	& 0.5	& 0.5 (8) \\
\cline{2-16} \cline{2-16}

& \multirow{5}{*}{6} & (1,1) & 0.5	& 0.5	& 0.5	& 0.45	& 0.05	& 0.5	& 0.05	& 0.5	& 0.5	& 0.5	& 0.5	& 0.5	& 0.5 (9) \\
\cline{3-16}
& & (1,2) & 0.5	& 0.5	& 0.5	& 0.5	& 0.05	& 0.5	& 0.5	& 0.45	& 0.5	& 0.5	& 0.5	& 0.5	& \textcolor{red}{\textbf{\emph{0.5} (\emph{10})}} \\
\cline{3-16}
& & (2,1) & 0.5	& 0.5	& 0.5	& 0.45	& 0.05	& 0.5	& 0.05	& 0.5	& 0.5	& 0.5	& 0.5	& 0.5	& 0.5 (9) \\
\cline{3-16}
& & (2,2) & 0.5	& 0.5	& 0.5	& 0.5	& 0.05	& 0.45	& 0.5	& 0.4	& 0.5	& 0.5	& 0.5	& 0.5	& 0.5 (9) \\
\cline{2-16} \cline{2-16}

& \multirow{5}{*}{8} & (1,1) & 0.5	& 0.5	& 0.5	& 0.4	& 0.05	& 0.45	& 0.05	& 0.5	& 0.5	& 0.5	& 0.5	& 0.5	& 0.5 (8) \\
\cline{3-16}
& & (1,2) & 0.5	& 0.5	& 0.5	& 0.5	& 0.1	& 0.45	& 0.5	& 0.4	& 0.5	& 0.5	& 0.5	& 0.5	& \textcolor{red}{\textbf{\emph{0.5} (\emph{9})}} \\
\cline{3-16}
& & (2,1) & 0.5	& 0.5	& 0.25	& 0.4	& 0.05	& 0.45	& 0.05	& 0.5	& 0.5	& 0.5	& 0.5	& 0.5	& 0.5 (7) \\
\cline{3-16}
& & (2,2) & 0.5	& 0.5	& 0.25	& 0.5	& 0.05	& 0.4	& 0.5	& 0.4	& 0.5	& 0.5	& 0.5	& 0.5	& 0.5 (8) \\
\hline \hline
\end{tabular}
\end{table}

\begin{table}
\centering
\caption{The best value of parameter $k$ for CFinder determined by the twelve community quality metrics with four possible combinations of the two versions of belonging coefficient and two version of belonging function on LFR benchmark networks with $(\alpha,\beta)=(1,2)$ and $\mu=0.3,0.35,0.4$. The best value for subcolumn of the last column is marked by red italic font.}
\label{tab:lfr:cfinder}       
\vspace{-1em}
\setlength{\tabcolsep}{3.5pt}
\begin{tabular}{c|c||c|c|c|c|c|c|c|c|c|c|c|c|c||c}
\hline \hline
$\mu$ & $O_m$ & (BC,BF) & $Q_{ov}$ & $NQ_{ov}$ & $Q_{ov}^L$ & $Q_{ds}^{ov}$ & $IE$ & $ID$ & $CNT$ & $BE$ & $EXP$ & $CND$ & $F$ & $D$ & \\
\hline
\multirow{20}{*}{0.3} & \multirow{5}{*}{1} & (1,1) & 4	& 4	& 4	& 4	& 3	& 3	& 4	& 10	& 4	& 4	& 4	& 4	& \textcolor{red}{\textbf{\emph{4} (\emph{9})}} \\
\cline{3-16}
& & (1,2) & 4	& 4	& 4	& 4	& 3	& 3	& 4	& 10	& 4	& 4	& 4	& 4	& \textcolor{red}{\textbf{\emph{4} (\emph{9})}} \\
\cline{3-16}
& & (2,1) & 4	& 5	& 4	& 4	& 3	& 3	& 4	& 10	& 10	& 4	& 4	& 4	& 4 (7) \\
\cline{3-16}
& & (2,2) & 4	& 5	& 4	& 4	& 3	& 3	& 4	& 10	& 10	& 10	& 4	& 4	& 4 (6) \\
\cline{2-16} \cline{2-16}

& \multirow{5}{*}{2} & (1,1) & 4	& 5	& 4	& 4	& 3	& 3	& 3	& 9	& 4	& 3	& 3	& 4	& \{3,4\} (5) \\
\cline{3-16}
& & (1,2) & 4	& 5	& 4	& 4	& 3	& 3	& 4	& 9	& 4	& 3	& 3	& 4	& \textcolor{red}{\textbf{\emph{4} (\emph{6})}} \\
\cline{3-16}
& & (2,1) & 4	& 5	& 4	& 4	& 3	& 3	& 3	& 9	& 9	& 3	& 3	& 4	& 3 (5) \\
\cline{3-16}
& & (2,2) & 4	& 6	& 4	& 4	& 3	& 3	& 4	& 9	& 9	& 9	& 3	& 4	& 4 (5) \\
\cline{2-16} \cline{2-16}

& \multirow{5}{*}{4} & (1,1) & 4	& 5	& 4	& 4	& 3	& 3	& 3	& 12	& 4	& 3	& 3	& 3	& \textcolor{red}{\textbf{\emph{3} (\emph{6})}} \\
\cline{3-16}
& & (1,2) & 4	& 5	& 4	& 4	& 3	& 3	& 3	& 12	& 4	& 3	& 3	& 3	& \textcolor{red}{\textbf{\emph{3} (\emph{6})}} \\
\cline{3-16}
& & (2,1) & 4	& 6	& 4	& 4	& 3	& 3	& 3	& 12	& 12	& 12	& 3	& 9	& 3 (4) \\
\cline{3-16}
& & (2,2) & 4	& 6	& 4	& 4	& 3	& 3	& 3	& 12	& 12	& 12	& 3	& 7	& 3 (4) \\
\cline{2-16} \cline{2-16}

& \multirow{5}{*}{6} & (1,1) & 4	& 5	& 4	& 4	& 3	& 3	& 4	& 11	& 5	& 3	& 3	& 3	& 3 (5) \\
\cline{3-16}
& & (1,2) & 4	& 5	& 4	& 4	& 3	& 3	& 4	& 11	& 4	& 4	& 4	& 3	& \textcolor{red}{\textbf{\emph{4} (\emph{7})}} \\
\cline{3-16}
& & (2,1) & 4	& 7	& 4	& 5	& 3	& 3	& 4	& 11	& 11	& 11	& 3	& 7	& \{3,4,11\} (3) \\
\cline{3-16}
& & (2,2) & 4	& 7	& 4	& 5	& 3	& 3	& 4	& 11	& 11	& 11	& 4	& 7	& 4 (4) \\
\cline{2-16} \cline{2-16}

& \multirow{5}{*}{8} & (1,1) & 4	& 5	& 4	& 4	& 3	& 3	& 4	& 12	& 6	& 3	& 3	& 3	& 3 (5) \\
\cline{3-16}
& & (1,2) & 4	& 5	& 4	& 4	& 3	& 3	& 4	& 12	& 4	& 4	& 4	& 3	& \textcolor{red}{\textbf{\emph{4} (\emph{7})}} \\
\cline{3-16}
& & (2,1) & 4	& 6	& 4	& 5	& 3	& 3	& 4	& 12	& 12	& 12	& 3	& 7	& \{3,4,12\} (3) \\
\cline{3-16}
& & (2,2) & 4	& 6	& 4	& 5	& 3	& 3	& 4	& 12	& 12	& 12	& 4	& 6	& 4 (4) \\
\cline{1-16} \cline{1-16}

\multirow{20}{*}{0.35} & \multirow{5}{*}{1} & (1,1) & 4	& 4	& 4	& 4	& 3	& 3	& 3	& 10	& 4	& 4	& 4	& 4	& \textcolor{red}{\textbf{\emph{4} (\emph{8})}} \\
\cline{3-16}
& & (1,2) & 4	& 4	& 4	& 4	& 3	& 3	& 4	& 10	& 4	& 3	& 4	& 4	& \textcolor{red}{\textbf{\emph{4} (\emph{8})}} \\
\cline{3-16}
& & (2,1) & 4	& 4	& 4	& 4	& 3	& 3	& 3	& 10	& 10	& 10	& 4	& 4	& 4 (6) \\
\cline{3-16}
& & (2,2) & 4	& 7	& 4	& 4	& 3	& 3	& 4	& 10	& 10	& 10	& 3	& 4	& 4 (5) \\
\cline{2-16} \cline{2-16}

& \multirow{5}{*}{2} & (1,1) & 4	& 4	& 4	& 4	& 3	& 3	& 3	& 8	& 4	& 3	& 3	& 3	& \textcolor{red}{\textbf{\emph{3} (\emph{6})}} \\
\cline{3-16}
& & (1,2) & 4	& 4	& 4	& 4	& 3	& 3	& 3	& 8	& 4	& 3	& 3	& 3	& \textcolor{red}{\textbf{\emph{3} (\emph{6})}} \\
\cline{3-16}
& & (2,1) & 4	& 5	& 4	& 4	& 3	& 3	& 3	& 8	& 8	& 8	& 3	& 8	& \{3,8\} (4) \\
\cline{3-16}
& & (2,2) & 4	& 7	& 4	& 4	& 3	& 3	& 4	& 8	& 8	& 8	& 3	& 7	& 4 (4) \\
\cline{2-16} \cline{2-16}

& \multirow{5}{*}{4} & (1,1) & 4	& 5	& 4	& 4	& 3	& 3	& 3	& 10	& 4	& 3	& 3	& 3	& \textcolor{red}{\textbf{\emph{3} (\emph{6})}} \\
\cline{3-16}
& & (1,2) & 4	& 4	& 4	& 4	& 3	& 3	& 4	& 10	& 4	& 3	& 3	& 3	& \textcolor{red}{\textbf{\emph{4} (\emph{6})}} \\
\cline{3-16}
& & (2,1) & 4	& 6	& 4	& 4	& 3	& 3	& 3	& 10	& 10	& 10	& 3	& 9	& 3 (4) \\
\cline{3-16}
& & (2,2) & 4	& 8	& 4	& 4	& 3	& 3	& 4	& 10	& 10	& 10	& 3	& 7	& 4 (4) \\
\cline{2-16} \cline{2-16}

& \multirow{5}{*}{6} & (1,1) & 4	& 5	& 4	& 4	& 3	& 3	& 4	& 11	& 5	& 4	& 3	& 3	& 4 (5) \\
\cline{3-16}
& & (1,2) & 4	& 5	& 4	& 4	& 3	& 3	& 4	& 11	& 4	& 4	& 4	& 3	& \textcolor{red}{\textbf{\emph{4} (\emph{7})}} \\
\cline{3-16}
& & (2,1) & 4	& 6	& 4	& 4	& 3	& 3	& 4	& 11	& 11	& 11	& 4	& 8	& 4 (5) \\
\cline{3-16}
& & (2,2) & 4	& 7	& 4	& 4	& 3	& 3	& 4	& 11	& 11	& 11	& 4	& 7	& 4 (5) \\
\cline{2-16} \cline{2-16}

& \multirow{5}{*}{8} & (1,1) & 4	& 5	& 4	& 4	& 3	& 3	& 4	& 11	& 6	& 4	& 4	& 3	& 4 (6) \\
\cline{3-16}
& & (1,2) & 4	& 5	& 4	& 4	& 3	& 3	& 4	& 11	& 4	& 4	& 4	& 3	& \textcolor{red}{\textbf{\emph{4} (\emph{7})}} \\
\cline{3-16}
& & (2,1) & 4	& 6	& 4	& 5	& 3	& 3	& 4	& 11	& 11	& 11	& 4	& 6	& 4 (4) \\
\cline{3-16}
& & (2,2) & 4	& 6	& 4	& 4	& 3	& 3	& 4	& 11	& 11	& 11	& 4	& 6	& 4 (5) \\
\cline{1-16} \cline{1-16}

\multirow{20}{*}{0.4} & \multirow{5}{*}{1} & (1,1) & 4	& 4	& 4	& 4	& 3	& 3	& 3	& 8	& 4	& 3	& 3	& 4	& \textcolor{red}{\textbf{\emph{4} (\emph{6})}} \\
\cline{3-16}
& & (1,2) & 4	& 4	& 4	& 4	& 3	& 3	& 3	& 8	& 4	& 3	& 3	& 3	& \textcolor{red}{\textbf{\emph{3} (\emph{6})}} \\
\cline{3-16}
& & (2,1) & 4	& 4	& 4	& 4	& 3	& 3	& 3	& 8	& 8	& 8	& 3	& 8	& \{3,4,8\} (4) \\
\cline{3-16}
& & (2,2) & 4	& 7	& 4	& 4	& 3	& 3	& 4	& 8	& 8	& 8	& 3	& 7	& 4 (4) \\
\cline{2-16} \cline{2-16}

& \multirow{5}{*}{2} & (1,1) & 4	& 4	& 4	& 4	& 3	& 3	& 3	& 9	& 4	& 3	& 3	& 3	& \textcolor{red}{\textbf{\emph{3} (\emph{6})}} \\
\cline{3-16}
& & (1,2) & 4	& 4	& 4	& 4	& 3	& 3	& 3	& 9	& 4	& 3	& 3	& 3	& \textcolor{red}{\textbf{\emph{3} (\emph{6})}} \\
\cline{3-16}
& & (2,1) & 4	& 5	& 4	& 4	& 3	& 3	& 3	& 9	& 9	& 9	& 3	& 9	& \{3,9\} (4) \\
\cline{3-16}
& & (2,2) & 4	& 7	& 4	& 4	& 3	& 3	& 4	& 9	& 9	& 9	& 3	& 7	& 4 (4) \\
\cline{2-16} \cline{2-16}

& \multirow{5}{*}{4} & (1,1) & 4	& 4	& 4	& 4	& 3	& 3	& 3	& 9	& 5	& 3	& 3	& 3	& \textcolor{red}{\textbf{\emph{3} (\emph{6})}} \\
\cline{3-16}
& & (1,2) & 4	& 4	& 4	& 4	& 3	& 3	& 4	& 9	& 4	& 3	& 3	& 3	& \textcolor{red}{\textbf{\emph{4} (\emph{6})}} \\
\cline{3-16}
& & (2,1) & 4	& 5 &	4	& 4	& 3	& 3	& 3	& 9	& 9	& 9	& 3	& 9	& \{3,9\} (4) \\
\cline{3-16}
& & (2,2) & 4	& 7	& 4	& 4	& 3	& 3	& 4	& 9	& 9	& 9	& 3	& 7	& 4 (4) \\
\cline{2-16} \cline{2-16}

& \multirow{5}{*}{6} & (1,1) & 4	& 4	& 4	& 4	& 3	& 3	& 4	& 10	& 5	& 4	& 4	& 3	& 4 (7) \\
\cline{3-16}
& & (1,2) & 4	& 4	& 4	& 4	& 3	& 3	& 4	& 10	& 4	& 4	& 4	& 3	& \textcolor{red}{\textbf{\emph{4} (\emph{8})}} \\
\cline{3-16}
& & (2,1) & 4	& 5	& 4	& 4	& 3	& 3	& 4	& 10	& 10	& 10	& 4	& 10	& 4 (5) \\
\cline{3-16}
& & (2,2) & 4	& 7	& 4	& 4	& 3	& 3	& 4	& 10	& 10	& 10	& 4	& 7	& 4 (5) \\
\cline{2-16} \cline{2-16}

& \multirow{5}{*}{8} & (1,1) & 4	& 5	& 4	& 4	& 3	& 3	& 4	& 10	& 5	& 4	& 4	& 3	& 4 (6) \\
\cline{3-16}
& & (1,2) & 4	& 5	& 4	& 4	& 3	& 3	& 4	& 10	& 4	& 4	& 4	& 3	& \textcolor{red}{\textbf{\emph{4} (\emph{7})}} \\
\cline{3-16}
& & (2,1) & 4	& 5	& 4	& 4	& 3	& 3	& 4	& 10	& 10	& 10	& 4	& 9	& 4 (5) \\
\cline{3-16}
& & (2,2) & 4	& 6	& 4	& 4	& 3	& 3	& 4	& 10	& 10	& 10	& 4	& 6	& 4 (5) \\
\hline \hline
\end{tabular}
\end{table}

\begin{table}
\centering
\caption{The best value of threshold $tr$ for SpeakEasy determined by the twelve community quality metrics with four possible combinations of the two versions of belonging coefficient and two version of belonging function on LFR benchmark networks with $(\alpha,\beta)=(1,2)$ and $\mu=0.3,0.35,0.4$. The best value for subcolumn of the last column is marked by red italic font.}
\label{tab:lfr:speak_easy}       
\vspace{-1em}
\setlength{\tabcolsep}{0.5pt}
\begin{tabular}{c|c||c|c|c|c|c|c|c|c|c|c|c|c|c||c}
\hline \hline
$\mu$ & $O_m$ & (BC,BF) & $Q_{ov}$ & $NQ_{ov}$ & $Q_{ov}^L$ & $Q_{ds}^{ov}$ & $IE$ & $ID$ & $CNT$ & $BE$ & $EXP$ & $CND$ & $F$ & $D$ & \\
\hline
\multirow{20}{*}{0.3} & \multirow{5}{*}{1} & (1,1) & 0.75	& 0.2	& 0.75	& 0.75	& 0.75	& 0.8	& 0.75	& 0.8	& 0.75	& 0.75	& 0.75	& 0.75	& \textcolor{red}{\textbf{\emph{0.75} (\emph{9})}} \\
\cline{3-16}
& & (1,2) & 0.75	& 0.2	& 0.75	& 0.75	& 0.75	& 0.8	& 0.75	& 0.8	& 0.75	& 0.75	& 0.75	& 0.75	& \textcolor{red}{\textbf{\emph{0.75} (\emph{9})}} \\
\cline{3-16}
& & (2,1) & 0.75	& 0.2	& 0.75	& 0.75	& 0.75	& 0.8	& 0.75	& 0.8	& 0.75	& 0.75	& 0.75	& 0.75	& \textcolor{red}{\textbf{\emph{0.75} (\emph{9})}} \\
\cline{3-16}
& & (2,2) & 0.75	& 0.2	& 0.75	& 0.75	& 0.75	& 0.8	& 0.75	& 0.8	& 0.75	& 0.75	& 0.75	& 0.75	& \textcolor{red}{\textbf{\emph{0.75} (\emph{9})}} \\
\cline{2-16} \cline{2-16}

& \multirow{5}{*}{2} & (1,1) & 0.05	& 0.85	& 0.8	& 0.6	& 0.8	& 0.4	& 0.8	& 0.9	& 0.8	& 0.8	& 0.8	& 0.8	& 0.8 (7) \\
\cline{3-16}
& & (1,2) & 0.8	& 0.85	& 0.8	& 0.6	& 0.8	& 0.25	& 0.8	& 0.4	& 0.8	& 0.8	& 0.8	& 0.8	& \textcolor{red}{\textbf{\emph{0.8} (\emph{8})}} \\
\cline{3-16}
& & (2,1) & 0.05	& 0.85	& 0.8	& 0.6	& 0.8	& 0.4	& 0.8	& 0.9	& 1	& 0.8	& 0.8	& 0.8	& 0.8 (6) \\
\cline{3-16}
& & (2,2) & 0.8	& 0.85	& 0.8	& 0.6	& 0.8	& 0.25	& 0.8	& 0.4	& 1	& 0.8	& 0.8	& 0.8	& 0.8 (7) \\
\cline{2-16} \cline{2-16}

& \multirow{5}{*}{4} & (1,1) & 0.05	& 1	& 0.7	& 0.95	& 0.05	& 1	& 0.05	& 1	& 1	& 0.8	& 0.8	& 0.95	& \textcolor{red}{\textbf{\emph{1} (\emph{4})}} \\
\cline{3-16}
& & (1,2) & 0.7	& 1	& 0.7	& 0.95	& 0.2	& 1	& 0.6	& 1	& 1	& 0.65	& 0.65	& 0.6	& \textcolor{red}{\textbf{\emph{1} (\emph{4})}} \\
\cline{3-16}
& & (2,1) & 0.05	& 1	& 0.7	& 0.95	& 0.05	& 1	& 0.05	& 1	& 1	& 0.8	& 0.8	& 0.95	& \textcolor{red}{\textbf{\emph{1} (\emph{4})}} \\
\cline{3-16}
& & (2,2) & 0.25	& 1	& 0.7	& 0.45	& 0.2	& 1	& 0.6	& 1	& 1	& 0.45	& 0.65	& 0.2	& \textcolor{red}{\textbf{\emph{1} (\emph{4})}} \\
\cline{2-16} \cline{2-16}

& \multirow{5}{*}{6} & (1,1) & 0.05	& 0.85	& 0.5	& 1	& 0.05	& 0.65	& 0.05	& 0.7	& 0.95	& 0.95	& 0.95	& 0.95	& \textcolor{red}{\textbf{\emph{0.95} (\emph{4})}} \\
\cline{3-16}
& & (1,2) & 0.35	& 0.85	& 0.5	& 0.5	& 0.35	& 0.65	& 0.35	& 0.6	& 0.95	& 0.95	& 0.95	& 0.95	& \textcolor{red}{\textbf{\emph{0.95} (\emph{4})}} \\
\cline{3-16}
& & (2,1) & 0.05	& 0.85	& 0.35	& 1	& 0.05	& 0.65	& 0.05	& 0.7	& 0.7	& 0.95	& 0.95	& 1	& 0.05 (3) \\
\cline{3-16}
& & (2,2) & 0.35	& 0.85	& 0.35	& 0.5	& 0.35	& 0.65	& 0.35	& 0.7	& 0.7	& 0.7	& 0.95	& 0.7	& \textcolor{red}{\textbf{\{\emph{0.35},\emph{0.7}\} (\emph{4})}} \\
\cline{2-16} \cline{2-16}

& \multirow{5}{*}{8} & (1,1) & 0.05	& 1	& 0.4	& 0.85	& 0.05	& 0.95	& 0.05	& 0.9	& 0.9	& 0.6	& 0.75	& 1	& 0.05 (3) \\
\cline{3-16}
& & (1,2) & 0.4	& 1	& 0.4	& 0.85	& 0.4	& 0.95	& 0.4	& 0.35	& 0.6	& 0.6	& 0.6	& 0.6	& \textcolor{red}{\textbf{\{\emph{0.4},\emph{0.6}\} (\emph{4})}} \\
\cline{3-16}
& & (2,1) & 0.05	& 1	& 0.4	& 0.85	& 0.05	& 0.95	& 0.05	& 0.9	& 0.85	& 0.6	& 0.75	& 0.85	& \{0.05,0.85\} (3) \\
\cline{3-16}
& & (2,2) & 0.35	& 1	& 0.4	& 0.85	& 0.4	& 0.95	& 0.4	& 0.35	& 0.85	& 0.7	& 0.6	& 0.85	& \{0.4,0.85\} (3) \\
\cline{1-16} \cline{1-16}

\multirow{20}{*}{0.35} & \multirow{5}{*}{1} & (1,1) & 0.8,0.9	& 0.75	& 0.8	& 0.8,0.9	& 0.8,0.9	& 0.45	& 0.8,0.9	& 0.45	& 0.8,0.9	& 0.8,0.9 &	0.8,0.9	& 0.8,0.9	& \textcolor{red}{\textbf{\emph{0.8} (\emph{9})}} \\
\cline{3-16}
& & (1,2) & 0.8,0.9	& 0.75	& 0.8	& 0.8,0.9	& 0.8,0.9	& 0.45	& 0.8,0.9	& 0.45	& 0.8,0.9	& 0.8,0.9	& 0.8,0.9	& 0.8,0.9	& \textcolor{red}{\textbf{\emph{0.8} (\emph{9})}} \\
\cline{3-16}
& & (2,1) & 0.8,0.9	& 0.75	& 0.8	& 0.8,0.9	& 0.8,0.9	& 0.45	& 0.8,0.9	& 0.45	& 0.8,0.9	& 0.8,0.9	& 0.8,0.9	& 0.8,0.9	& \textcolor{red}{\textbf{\emph{0.8} (\emph{9})}} \\
\cline{3-16}
& & (2,2) & 0.8,0.9	& 0.75	& 0.8	& 0.8,0.9	& 0.8,0.9	& 0.45	& 0.8,0.9	& 0.45	& 0.8,0.9	& 0.8,0.9	& 0.8,0.9	& 0.8,0.9	& \textcolor{red}{\textbf{\emph{0.8} (\emph{9})}} \\
\cline{2-16} \cline{2-16}

& \multirow{5}{*}{2} & (1,1) & 0.05	& 0.85	& 0.95	& 1	& 0.05	& 0.9	& 0.05	& 0.9	& 0.95	& 0.95	& 0.95	& 0.95	& 0.95 (5) \\
\cline{3-16}
& & (1,2) & 0.95	& 0.85	& 0.95	& 1	& 0.35	& 0.9	& 0.95	& 0.5	& 0.75	& 0.95	& 0.95	& 0.95	& \textcolor{red}{\textbf{\emph{0.95} (\emph{6})}} \\
\cline{3-16}
& & (2,1) & 0.05	& 0.85	& 0.35	& 1	& 0.05	& 0.9	& 0.05	& 0.9	& 0.95	& 0.95	& 0.95	& 0.95	& 0.95 (4) \\
\cline{3-16}
& & (2,2) & 0.35	& 0.85	& 0.35	& 1	& 0.35	& 0.9	& 0.95	& 0.5	& 0.75	& 0.75	& 0.95	& 0.95	& \{0.35,0.95\} (3) \\
\cline{2-16} \cline{2-16}

& \multirow{5}{*}{4} & (1,1) & 0.05	& 0.95	& 0.9	& 1	& 0.05	& 0.95	& 0.05	& 0.95	& 0.9	& 0.75	& 0.75	& 1	& \{0.05,0.95\} (3) \\
\cline{3-16}
& & (1,2) & 0.9	& 0.95	& 0.9	& 1	& 0.65	& 0.95	& 0.75	& 0.95	& 0.9	& 0.75	& 0.75	& 0.75	& \textcolor{red}{\textbf{\emph{0.75} (\emph{4})}} \\
\cline{3-16}
& & (2,1) & 0.05	& 0.95	& 0.65	& 1	& 0.05	& 0.95	& 0.05	& 0.95	& 0.9	& 0.75	& 0.75	& 0.9	& \{0.05,0.95\} (3) \\
\cline{3-16}
& & (2,2) & 0.9	& 0.95	& 0.65	& 1	& 0.65	& 0.95	& 0.75	& 0.95	& 0.6	& 0.6	& 0.75	& 0.9	& 0.95 (3) \\
\cline{2-16} \cline{2-16}

& \multirow{5}{*}{6} & (1,1) & 0.05	& 1	& 0.8	& 0.85	& 0.05	& 1	& 0.05	& 1	& 0.85	& 0.85	& 0.85	& 0.85	& \textcolor{red}{\textbf{\emph{0.85} (\emph{5})}} \\
\cline{3-16}
& & (1,2) & 0.8	& 0.95	& 0.8	& 0.45	& 0.3	& 1	& 0.3	& 0.1	& 0.85	& 0.85	& 0.85	& 0.85	& 0.85 (4) \\
\cline{3-16}
& & (2,1) & 0.05	& 1	& 0.45	& 0.85	& 0.05	& 1	& 0.05	& 1	& 0.9	& 0.85	& 0.85	& 0.85	& 0.85 (4) \\
\cline{3-16}
& & (2,2) & 0.45	& 0.95	& 0.45	& 0.45	& 0.3	& 1	& 0.3	& 0.35	& 0.9	& 0.9	& 0.85	& 0.85	& 0.45 (3) \\
\cline{2-16} \cline{2-16}

& \multirow{5}{*}{8} & (1,1) & 0.05	& 1	& 0.9	& 0.8	& 0.05	& 0.85	& 0.05	& 1	& 0.95	& 0.85	& 0.85	& 0.9	& \{0.05,0.85\} (3) \\
\cline{3-16}
& & (1,2) & 0.9	& 1	& 0.9	& 0.8	& 0.9	& 0.85	& 0.85	& 0.3	& 0.9	& 0.85	& 0.85	& 0.9	& \textcolor{red}{\textbf{\emph{0.9} (\emph{5})}} \\
\cline{3-16}
& & (2,1) & 0.05	& 1	& 0.9	& 0.8	& 0.05	& 0.85	& 0.05	& 1	& 1	& 0.85	& 0.85	& 0.95	& \{0.05,0.85,1\} (3) \\
\cline{3-16}
& & (2,2) & 0.9	& 1	& 0.9	& 0.8	& 0.9	& 0.85	& 0.85	& 0.3	& 0.2	& 1	& 0.85	& 0.95	& \{0.85,0.9\} (3) \\
\cline{1-16} \cline{1-16}

\multirow{20}{*}{0.4} & \multirow{5}{*}{1} & (1,1) & 0.15	& 0.75	& 0.15	& 0.15	& 0.15	& 0.75	& 0.15	& 0.75	& 0.15	& 0.15	& 0.15	& 0.15	& \textcolor{red}{\textbf{\emph{0.15} (\emph{9})}} \\
\cline{3-16}
& & (1,2) & 0.15	& 0.2	& 0.15	& 0.15	& 0.15	& 0.75	& 0.15	& 0.75	& 0.15	& 0.15	& 0.15	& 0.15	& \textcolor{red}{\textbf{\emph{0.15} (\emph{9})}} \\
\cline{3-16}
& & (2,1) & 0.15	& 0.75	& 0.15	& 0.15	& 0.15	& 0.75	& 0.15	& 0.75	& 0.15	& 0.15	& 0.15	& 0.15	& \textcolor{red}{\textbf{\emph{0.15} (\emph{9})}} \\
\cline{3-16}
& & (2,2) & 0.15	& 0.2	& 0.15	& 0.15	& 0.15	& 0.75	& 0.15	& 0.75	& 0.15	& 0.15	& 0.15	& 0.15	& \textcolor{red}{\textbf{\emph{0.15} (\emph{9})}} \\
\cline{2-16} \cline{2-16}

& \multirow{5}{*}{2} & (1,1) & 0.05	& 0.85	& 0.95	& 0.8	& 0.1	& 0.85	& 0.05	& 0.85	& 0.95	& 0.8	& 0.8	& 0.95	& \{0.8,0.85,0.95\} (3) \\
\cline{3-16}
& & (1,2) & 0.95	& 0.85	& 0.95	& 0.75	& 0.8	& 0.85	& 0.75	& 0.85	& 0.95	& 0.8	& 0.8	& 0.95	& 0.95 (4) \\
\cline{3-16}
& & (2,1) & 0.05	& 0.85	& 0.95	& 0.75	& 0.05	& 0.85	& 0.05	& 0.85	& 0.95	& 0.8	& 0.8	& 0.95	& \{0.05,0.85,0.95\} (3) \\
\cline{3-16}
& & (2,2) & 0.95	& 0.85	& 0.95	& 0.75	& 0.8	& 0.85	& 0.75	& 0.6	& 0.95	& 0.95	& 0.8	& 0.95	& \textcolor{red}{\textbf{\emph{0.95} (\emph{5})}} \\
\cline{2-16} \cline{2-16}

& \multirow{5}{*}{4} & (1,1) & 0.05	& 1	& 0.95	& 0.9	& 0.15	& 1	& 0.05	& 0.9	& 0.95	& 0.95	& 0.95	& 0.95	& 0.95 (5) \\
\cline{3-16}
& & (1,2) & 0.95	& 1	& 0.95	& 0.9	& 0.5	& 1	& 0.95	& 0.9	& 0.75	& 0.95	& 0.95	& 0.95	& \textcolor{red}{\textbf{\emph{0.95} (\emph{6})}} \\
\cline{3-16}
& & (2,1) & 0.05	& 1	& 0.95	& 0.9	& 0.05	& 1	& 0.05	& 0.9	& 0.9	& 0.95	& 0.95	& 0.95	& 0.95 (4) \\
\cline{3-16}
& & (2,2) & 0.95	& 0.9	& 0.95	& 0.9	& 0.5	& 1	& 0.95	& 0.9	& 0.7	& 0.7	& 0.95	& 0.7	& 0.95 (4) \\
\cline{2-16} \cline{2-16}

& \multirow{5}{*}{6} & (1,1) & 0.1	& 0.95	& 0.75	& 0.75	& 0.15	& 0.95	& 0.1	& 0.95	& 0.95	& 0.9	& 0.9	& 0.8	& 0.95 (4) \\
\cline{3-16}
& & (1,2) & 0.75	& 0.95	& 0.75	& 0.75	& 0.15	& 0.95	& 0.65	& 0.45	& 0.8	& 0.65	& 0.65	& 0.8	& \{0.65,0.75\} (3) \\
\cline{3-16}
& & (2,1) & 0.05	& 0.95	& 0.75	& 0.75	& 0.05	& 0.95	& 0.1	& 0.95	& 0.75	& 0.9	& 0.9	& 0.75	& 0.75 (4) \\
\cline{3-16}
& & (2,2) & 0.75	& 0.95	& 0.75	& 0.75	& 0.15	& 0.95	& 0.65	& 0.45	& 0.75	& 0.75	& 0.65	& 0.75	& \textcolor{red}{\textbf{\emph{0.75} (\emph{6})}} \\
\cline{2-16} \cline{2-16}

& \multirow{5}{*}{8} & (1,1) & 0.05	& 1	& 0.9	& 0.8	& 0.05	& 1	& 0.05	& 1	& 0.9	& 0.95	& 0.95	& 0.9	& \{0.05,0.9,1\} (3) \\
\cline{3-16}
& & (1,2) & 0.9	& 1	& 0.9	& 0.8	& 0.4	& 1	& 0.4	& 0.45	& 0.9	& 0.4	& 0.4	& 0.4	& \textcolor{red}{\textbf{\emph{0.4} (\emph{5})}} \\
\cline{3-16}
& & (2,1) & 0.05	& 1	& 0.9	& 0.8	& 0.05	& 1	& 0.05	& 1	& 0.8	& 0.95	& 0.95	& 0.95	& \{0.05,0.95,1\} (3) \\
\cline{3-16}
& & (2,2) & 0.9	& 1	& 0.9	& 0.8	& 0.4	& 1	& 0.4	& 0.45	& 0.5	& 0.5	& 0.4	& 0.5	& \{0.4,0.5\} (3) \\
\hline \hline
\end{tabular}
\end{table}

Tables~\ref{tab:lfr:slpa}-\ref{tab:lfr:speak_easy} respectively show the best value of threshold $r$ for SLPA, the best value of parameter $k$ for CFinder, and the best value of threshold $tr$ for SpeakEasy determined by the twelve community quality metrics with four possible combinations of the two versions of belonging coefficient and two version of belonging function on LFR benchmark networks with $(\alpha,\beta)=(1,2)$ and $\mu=0.3,0.35,0.4$. Table~\ref{tab:lfr:slpa} implies that (BC,BF)=(1,2) is the best among the four possible combinations on all configurations of LFR benchmark networks except $\mu=0.3,O_m=2$ and $\mu=0.4,O_m=1$ when using SLPA. Table~\ref{tab:lfr:cfinder} demonstrates that (BC,BF)=(1,2) is the best on all configurations of LFR benchmark networks when using CFinder. Table~\ref{tab:lfr:speak_easy} indicates that (BC,BF)=(1,2) is the best among the four combinations on all configurations of LFR benchmark networks except $\mu=0.35,O_m=6$ and $\mu=0.4,O_m=2,4$ when using SpeakEasy. Consequently, we could conclude that the overlapping community quality metrics with the first version of belonging coefficient and the second version of the belonging function are the best among the four possible combinations on LFR benchmark networks.

\vspace{2em}